\documentclass{article}

\usepackage{arxiv}
\usepackage[utf8]{inputenc} % allow utf-8 input
\usepackage[T1]{fontenc}    % use 8-bit T1 fonts
\usepackage{hyperref}       % hyperlinks
\usepackage{url}            % simple URL typesetting
\usepackage{booktabs}       % professional-quality tables
\usepackage{amsfonts}       % blackboard math symbols
\usepackage{nicefrac}       % compact symbols for 1/2, etc.
\usepackage{microtype}      % microtypography
\usepackage{graphicx}
\usepackage{natbib}
\usepackage{doi}
\usepackage{bm}
\usepackage{amsmath}
\usepackage{multirow}
\usepackage{subcaption}
\usepackage{comment}

\usepackage{fancyhdr}
\newcommand{\bftheta}{\mbox{\boldmath $\theta$}}

\newcommand{\bfy}{\mathbf{y}}

\newcommand{\bfh}{\mathbf{h}}

\title{Stochastic Volatility in Mean Models with Heavy Tails: A Fast Approximate Bayesian Inference Using Hidden Markov Models}

\author{
		Bruno E. Holtz\thanks{University of S\~ao Paulo, Instituto de Ciências Matemáticas e de Computação}\\
        bruno.holtz@usp.br.\\
        \and
		Carlos A. Abanto-Valle\thanks{Federal University of Rio de Janeiro, Department of Statistics} \\
        cabantovalle@im.ufrj.br.\\
		\and 
		Ricardo S. Ehlers\footnotemark[1]\\
        ehlers@icmc.usp.br.\\
		\and 
        Gabriel Rodríguez \thanks{Pontificia Universidad Católica del Perú, Department of Economics}\\
        gabriel.rodriguez@pucp.edu.pe
}
	
\date{This version: June, 2026}

\hypersetup{
            pdftitle={A template for the arxiv style},
            pdfsubject={q-bio.NC, q-bio.QM},
            pdfauthor={David S.~Hippocampus, Elias D.~Striatum},
            pdfkeywords={First keyword, Second keyword, More},
}

\begin{document}
\maketitle

%\pagestyle{myheadings}
%\markboth{}{}

\begin{abstract}
	This paper extends the approximate Bayesian estimation framework for Stochastic Volatility in Mean (SVM) models to accommodate heavy-tailed distributions from the Scale Mixture of Normals (SMN) family. To overcome the computational challenges arising from these models, we propose a numerically stable estimation procedure that exploits special functions to eliminate the need for direct numerical integration. Furthermore, the implementation incorporates parallel computing strategies that substantially reduce computational costs. Simulation studies and empirical applications demonstrate that the proposed approach delivers accurate inference while achieving computational times that are approximately an order of magnitude smaller than those required by conventional Markov chain Monte Carlo (MCMC) methods.
\end{abstract}

% keywords can be removed

\section{Introduction}
\label{sec-intro}

Stochastic volatility (SV) models play a central role in the econometric analysis of financial time series, particularly in markets characterized by persistent and substantial uncertainty. Since the seminal work of \citet{taylor1986}, these models have provided a flexible and theoretically grounded framework for describing the dynamic evolution of volatility. Such a representation is essential for effective decision-making by both policymakers and market participants. Moreover, SV models are firmly rooted in the principles of financial economics \citep{Melino1990} and have proved capable of reproducing important stylized features of financial return data, including volatility clustering and leptokurtosis \citep{Carnero2004}.
	
Volatility asymmetry is a well-documented characteristic of financial time series, whereby the impact of shocks on volatility depends on the direction of returns. In particular, negative shocks tend to generate substantially larger increases in volatility than positive shocks of the same magnitude. One of the principal theoretical explanations for this phenomenon is the volatility feedback theory proposed in \cite{French1987}.
	
Since their original formulation, SV models have been extended in several directions to better capture the empirical dynamics of financial markets. In particular, they have been widely employed to investigate asymmetric volatility behavior and to model the dynamics of daily stock return volatility. Early implementations of SV models, however, often required extensive pre-modeling procedures in which the conditional mean and conditional variance were estimated separately. To overcome the limitations of this two-step estimation approach, \citet{Koopman2002} introduced the stochastic volatility-in-mean (SVM) specification, in which volatility is incorporated as a regressor in the conditional mean equation for returns. This formulation was designed to capture the volatility feedback effect. Furthermore, \citet{AbantoValle2012} extended the SVM framework by assuming that the observational errors follow the scale mixture of normal (SMN) family, thereby accommodating heavy-tailed distributions. Such departures from normality in financial returns are well documented in the literature \citep{Mandelbrot1963, Liesenfeld2000, Chib2002, jacquier2004bayesian}.
	
It is worth noting that, in any SV model, the likelihood function involves high-dimensional integration over the latent volatility process. MCMC methods circumvent the direct evaluation of these integrals by jointly sampling the latent volatilities and model parameters. However, the implementation of these methods may entail substantial computational costs and potential convergence issues in the simulated chains. From a practical perspective, however, practitioners are primarily interested in obtaining reliable results within a reasonable computational time using readily available software tools.
	
In this context, one of the main contributions of this article is to extend the approximate Bayesian inference framework introduced by \cite{Abanto2024}. In that work, the likelihood function of the SVM model under the assumption of Gaussian returns is numerically approximated by discretizing the continuous latent volatility process and subsequently evaluated using the hidden Markov model (HMM) machinery. The underlying idea of approximating the likelihood through a finite-state representation of the latent process was originally proposed by \cite{Langrock2012} for maximum likelihood estimation of stochastic volatility models and later adapted by \cite{Abanto2017} to maximum likelihood inference for SVM models with heavy-tailed distributions.

Building on these developments, the present paper makes two further contributions. First, we extend the HMM approximation to the broad class of Scale Mixtures of Normals (SMN), thereby encompassing several important heavy-tailed models within a unified framework. Second, we embed the resulting approximate likelihood into a Bayesian inferential procedure based on importance sampling, allowing efficient approximation of posterior quantities without resorting to computationally intensive MCMC algorithms. The resulting methodology combines the modeling flexibility of SMN distributions with the computational efficiency of HMM-based likelihood evaluation, providing a practical alternative for Bayesian analysis of SVM models.

To address the associated computational challenges, we develop an algorithm that exploits special functions in order to avoid direct numerical integration. The proposed methods were implemented using the \texttt{Rcpp} and \texttt{RcppArmadillo} packages \citep{Eddelbuettel2014}, which provide an efficient interface between \texttt{R} and \texttt{C++}. In addition, we present empirical evidence demonstrating that the use of parallel computing techniques leads to substantial gains in computational performance.
	
The remainder of this paper is organized as follows. Section \ref{SMNdist} provides a brief introduction to SMN distributions. Section \ref{SVMmod} outlines the proposed SVM model and the corresponding Bayesian estimation procedure, while the results of the simulation studies are presented in Section \ref{sim}. In Section \ref{Appl}, we report an empirical application to international market indexes. Finally, Section \ref{discuss} concludes the paper with some final remarks and directions for future research.

\section{Scale Mixture of Normals Distributions}\label{SMNdist}
	
The SMN framework provides a flexible class of models for a wide range of symmetric and unimodal distributions, particularly those exhibiting heavier tails than the normal distribution. As defined in \cite{lange1993normal} and \cite{Choy2008}, a random variable $Y$ is said to belong to the SMN family if it admits the stochastic representation
\[
Y = \mu + \lambda^{-1/2} Z,
\]
where $\mu$ denotes a location parameter, $Z \sim N(0,\sigma^2)$ represents the Gaussian component, and $\lambda$ is a positive mixing random variable. The variable $\lambda$ is assumed to be independent of $Z$ and to have probability density function $h(\lambda \mid \nu)$, where the parameter $\nu > 0$ controls the tail thickness of the resulting distribution. In this case, we write
\[
    Y \sim \mathcal{SMN}(\mu,\sigma^2,\nu).
\]
	
The marginal density of $Y$ is obtained by integrating out the mixing variable $\lambda$, yielding
	\begin{equation}
		f(y \mid \mu, \sigma^2, \nu)
		=
		\int_0^\infty
		\mathcal{N}\left(y \mid \mu, \sigma^2 / \lambda\right)
		h(\lambda \mid \nu)
		\, d\lambda,
		\label{eq:smn_density}
	\end{equation}
	where the choice of $h(\lambda \mid \nu)$ determines the tail behavior of the distribution. In this study, we consider three specific mixing distributions: the Gamma distribution, $\lambda \sim \mathcal{G}(\nu/2,\nu/2)$; the Beta distribution, $\lambda \sim \mathcal{B}e(\nu,1)$; and the Inverse Gamma distribution, $\lambda \sim \mathcal{IG}(\nu/2,\nu/2)$. These choices correspond, respectively, to the Student-\textit{t}, Slash, and Variance Gamma distributions. Finally, note that the standard normal distribution is recovered as a special case when $\lambda$ degenerates to 1.

\section{The Stochastic Volatility in Mean (SVM) Model}\label{SVMmod}
	
	The SVM model provides a convenient framework for studying the interaction between financial returns and time-varying volatility by explicitly incorporating volatility into the conditional mean equation. The model is specified as
	\begin{subequations}
		\begin{eqnarray}
			y_{t} &=& \beta_{0} + \beta_{1} y_{t-1}
			+ \beta_{2} e^{h_{t}}
			+ e^{\frac{h_{t}}{2}}\epsilon_{t},
			\label{EqModSMNret} \\
			h_{t+1} &=& \mu + \phi(h_{t}-\mu)
			+ \sigma_{\eta}\eta_{t},
			\qquad
			\eta_{t}\sim\mathcal{N}(0,1),
			\label{EqModSMNstat}
		\end{eqnarray}
	\end{subequations}
	where $y_t$ denotes the compounded return at time $t$, while $h_t$ represents the latent log-volatility process, for $t=1,\ldots,T$. To guarantee stationarity of the volatility dynamics, we assume that $|\phi|<1$. Under this condition, the latent process admits the stationary distribution $h_1 \sim \mathcal{N}\left(\mu,\frac{\sigma_{\eta}^{2}}{1-\phi^{2}}\right).$
	In addition, we impose $|\beta_1|<1$ in order to ensure stationarity of the conditional mean process. The innovation terms $\epsilon_t$ and $\eta_t$ are assumed to be serially independent and mutually independent. Unlike the standard Gaussian SVM specification, we allow the innovation term $\epsilon_t$ to follow a Scale Mixture of Normals distribution, namely $\epsilon_t \sim \mathcal{SMN}(0,1,\nu),$
	as introduced in Section~\ref{SMNdist}. This extension substantially increases the flexibility of the model by allowing for heavier tails and greater robustness against extreme observations. Depending on the choice of the mixing distribution, the resulting models correspond to the SVM with Student-\textit{t} errors (SVM-\textit{t}), Slash errors (SVM-S), and Variance Gamma errors (SVM-VG). These formulations generalize the classical Gaussian SVM model, hereafter denoted by SVM-N, introduced by \citet{Koopman2002}.
	
	An important feature of the SVM framework is that the latent volatility enters directly into the mean equation through the coefficient $\beta_2$. This parameter captures both the ex-ante relationship between returns and volatility and the volatility feedback effect. In particular, negative values of $\beta_2$ are consistent with the volatility feedback hypothesis, according to which increases in expected future volatility are associated with lower contemporaneous returns. Further theoretical details regarding the volatility feedback effect and the SVM framework are provided in the Supplementary Material.

	\subsection{Likelihood evaluation}\label{Likel}
	
	Let $\bfy_T=(y_1,\dots,y_T)$ and $\bfh_T=(h_1,\dots,h_T)$ denote, respectively, the observed returns and the latent log-volatility process in the SVM model defined by equations \eqref{EqModSMNret} and \eqref{EqModSMNstat}. For a parameter vector $\bftheta$, the likelihood function is given by
	\begin{eqnarray*}
		\mathcal{L}(\bftheta)
		&=&
		\int p(\bfy_T,\bfh_T \mid y_0,\bftheta)\, d\bfh_T \\
		&=&
		\int
		p(\bfy_T \mid y_0,\bfh_T,\bftheta)
		\, p(\bfh_T \mid \bftheta)
		\, d\bfh_T,
	\end{eqnarray*}
	which corresponds to a $T$-dimensional integral. By exploiting the conditional dependence structure of the model, we have
	\[
	p(\bfy_T \mid y_0,\bfh_T,\bftheta)
	=
	p(y_1 \mid y_0,h_1)
	\prod_{t=2}^{T}
	p(y_t \mid y_{t-1},h_t),
	\]
	and
	\[
	p(\bfh_T \mid \bftheta)
	=
	p(h_1 \mid \bftheta)
	\prod_{t=2}^{T}
	p(h_t \mid h_{t-1}).
	\]
	
	In the SVM models, the likelihood function is analytically intractable due to the presence of the high-dimensional integral over the latent volatilities. MCMC methods circumvent this difficulty by jointly sampling the latent states and model parameters. However, such methods may involve substantial computational costs and potential convergence issues in the simulated chains.
	
	To overcome these limitations, we build upon the approaches proposed by \cite{Langrock2012}, \cite{Abanto2017} and \cite{Abanto2024}, who introduced efficient likelihood approximation methods for SV and SVM models based on the hidden Markov model (HMM) framework. In the present work, we extend these ideas to the more flexible class of Scale Mixtures of Normals distributions while preserving the computational efficiency of likelihood-based estimation relative to conventional MCMC methods.
	
	The key idea is to approximate the likelihood function by discretizing the continuous latent volatility process. This is accomplished through a rectangular quadrature rule based on $m$ equidistant intervals, $B_i=(b_{i-1},b_i),$ $ 
	i=1,\dots,m,$
	with midpoint $b_i^{*}$ and interval length $b$. Under this discretization scheme, the likelihood function can be approximated by
	\begin{eqnarray}
		\nonumber
		\mathcal{L}(\bftheta)
		&\approx&
		b^T
		\sum_{i_1=1}^{m}
		\cdots
		\sum_{i_T=1}^{m}
		p(y_1 \mid y_0,h_1=b_{i_1}^{*})\, p(h_1=b_{i_1}^{*})
		\\
		&&
		\times
		\prod_{t=2}^{T}
		p(y_t \mid y_{t-1},h_t=b_{i_t}^{*})
		\,
		p(h_t=b_{i_t}^{*} \mid h_{t-1}=b_{i_{t-1}}^{*})
		\nonumber\\
		&&
		=
		\mathcal{L}_{\text{approx}},
		\label{approxL}
	\end{eqnarray}
	which transforms the original continuous-state likelihood into a discrete approximation that can be efficiently evaluated using HMM techniques.
	
	The direct numerical evaluation of the approximate likelihood in \eqref{approxL} is generally computationally infeasible, since it involves $m^T$ summands. However, the SVM model can be represented as a state space model (SSM) and subsequently approximated by a discrete-state SSM, which is equivalent to a hidden Markov model (HMM). These models preserve the same dependence structure as the original SVM specification while operating on a finite state space \citep{Langrock2011,Langrock2012}. Under this representation, the approximate likelihood can be written as
	\begin{equation}
		\mathcal{L}_{\text{approx}}
		=
		\bm{\delta}
		\bm{P}(y_1)
		\bm{\Gamma}
		\bm{P}(y_2)
		\cdots
		\bm{\Gamma}
		\bm{P}(y_T)
		\bm{1}',
		\label{approxLHMM}
	\end{equation}
	where $\bm{\delta}$ denotes the vector associated with the stationary distribution of the latent log-volatility process, evaluated at the discretized states $b_i^{*}$ and multiplied by the interval length $b$, that is,
	\[
	\bm{\delta}
	=
	\mathcal{N}\left(
	\mu,
	\frac{\sigma_{\eta}^{2}}{1-\phi^{2}}
	\right)b.
	\]
	Furthermore,
	\[
	\bm{P}(y_t)
	=
	\text{diag}
	\left\{
	p(y_t \mid y_{t-1}, h_t=b_i^{*})
	\right\},
	\]
	is an $m\times m$ diagonal matrix containing the conditional densities of the observations, while
	\[
	\bm{\Gamma}
	=
	(\gamma_{ij}),
	\qquad
	\gamma_{ij}
	=
	p(h_t=b_j^{*}\mid h_{t-1}=b_i^{*}),
	\]
	is the transition probability matrix associated with the discretized latent volatility process. Finally, $\bm{1}'$ denotes an $m\times1$ vector of ones. All these quantities are completely determined by equations \eqref{EqModSMNret} and \eqref{EqModSMNstat}.
	
	This representation allows the approximate likelihood in \eqref{approxLHMM} to be evaluated efficiently through recursive HMM algorithms. Although the approximation accuracy can be improved by increasing the number of discretization points $m$ and, if necessary, enlarging the interval $[b_0,b_m]$, these choices must be made carefully. In particular, larger values of $m$ increase the dimension of the transition matrix $\bm{\Gamma}$ and consequently raise the computational cost of the likelihood evaluation.
	
	An additional advantage of the HMM-based approximation is that it naturally facilitates the computation of predictive distributions, which are useful for forecast evaluation and model diagnostics. Since these procedures closely follow those described in \citet{Abanto2024}, further details are provided in the Supplementary Material.
	
	\subsection{Bayesian Inference for the SVM model}\label{BSVM}
	
	To complete the Bayesian specification of the model, prior distributions must be assigned to the parameters. However, some parameters of the SVM model are constrained, namely
	\[
	|\beta_1|<1,
	\qquad
	|\phi|<1,
	\qquad
	\sigma_{\eta}>0.
	\]
	To avoid working directly within constrained parameter spaces, we consider the following reparameterizations:
	\[
	\gamma
	=
	\log\left(
	\frac{1+\beta_1}{1-\beta_1}
	\right),
	\qquad
	\psi
	=
	\log\left(
	\frac{1+\phi}{1-\phi}
	\right),
	\qquad
	\omega
	=
	\log(\sigma_{\eta}).
	\]
	These transformations map the original constrained spaces onto the real line, thereby simplifying the computational implementation. In particular, because the posterior approximation relies on multivariate normal proposals, it is convenient to define the prior distributions on this unconstrained scale.
	
	Throughout this paper, independent normal prior distributions are assigned to all transformed parameters, with hyperparameters selected according to the corresponding modeling context. Let
	\[
	\bftheta
	=
	(\beta_0,\gamma,\beta_2,\mu,\psi,\omega)',
	\]
	and denote by $p(\bftheta)$ the prior density of $\bftheta$. Since the transformations are one-to-one, the likelihood function remains invariant under reparameterization. Consequently, from equation \eqref{approxL}, the approximate posterior density is given by
	\begin{equation}
		p(\bftheta \mid y_0,\bfy_T)
		\propto
		p(\bftheta)\,
		\mathcal{L}_{\text{approx}}(\bftheta),
		\label{MAP}
	\end{equation}
	where $\bfy_T=(y_1,\ldots,y_T)'$.
	
	Posterior inference is based on expectations of the form
	\[
	E\!\left[
	h(\bftheta)\mid y_0,\bfy_T
	\right],
	\]
	which are approximated using Importance Sampling. Let $q(\bftheta)$ denote the importance density. The corresponding importance weights are defined as
	\[
	\omega(\bftheta)
	=
	\frac{
		p(\bftheta\mid y_0,\bfy_T)
	}{
		q(\bftheta)
	}.
	\]
	Thus, given independent draws
	\[
	\bftheta_1,\ldots,\bftheta_m
	\sim
	q(\bftheta),
	\]
	the posterior expectation can be estimated by
	\begin{equation}
		\bar h
		=
		\frac{
			\sum_{i=1}^{m}
			h(\bftheta_i)\,
			\omega(\bftheta_i)
		}{
			\sum_{i=1}^{m}
			\omega(\bftheta_i)
		}.
		\label{MCest}
	\end{equation}
	
	In this work, the importance density $q(\bftheta)$ is specified as a multivariate normal distribution centered at the posterior mode,
	\[
	\hat{\bftheta}_{MAP},
	\]
	with covariance matrix given by the inverse of the Hessian matrix evaluated at $\hat{\bftheta}_{MAP}$. These quantities are obtained through numerical maximization of the posterior density in \eqref{MAP}.

\section{Simulation Study}\label{sim}

We now illustrate the methodology described in Sections~\ref{Likel} and~\ref{BSVM}. The objectives of this simulation study are twofold. First, we investigate the finite-sample performance of the proposed approximate Bayesian procedure in terms of parameter recovery and uncertainty quantification. Second, we assess the numerical stability of the HMM approximation by examining the effect of the number of discretization points, $m$, on the resulting posterior summaries and computational cost. In particular, we seek to identify values of $m$ for which the approximation becomes sufficiently stable while maintaining the computational advantages of the proposed methodology.
	
	The computational implementation was developed using the \texttt{Rcpp} and \texttt{RcppArmadillo} packages, which provide an efficient interface between \texttt{R} and \texttt{C++}. From equation~(\ref{EqModSMNret}), the conditional densities
	\[
	p(y_t \mid y_{t-1}, h_t=b_i^*)
	\quad \text{and} \quad
	p(y_t \mid y_{t-1}, h_t=b_j^*)
	\]
	can be evaluated independently for all $i \neq j$, $i,j \in \{1,\ldots,m\}$. As a result, the computation of the diagonal matrix $\mathbf{P}(y_t)$ can be parallelized for each $t=1,\ldots,T$. To exploit this feature, we employed the \texttt{RcppParallel} package.
	
	In addition, the evaluation of the marginal densities associated with the Slash and Variance Gamma distributions was optimized using specialized functions available through the \texttt{RcppGSL} package. Since the integral in equation~(\ref{eq:smn_density}) does not admit a closed-form expression for these distributions, direct numerical quadrature may become computationally expensive and potentially unstable. Instead, by expressing the densities in terms of lower incomplete gamma and modified Bessel functions, we obtained a numerically stable implementation together with a substantial reduction in computational time.
	
	Similarly to the Student-\textit{t} distribution, whose density can be evaluated through the Gamma function, the Slash and Variance Gamma densities can also be computed efficiently using highly optimized special functions from the \textit{GNU Scientific Library} (GSL). The resulting implementation enables fast and stable likelihood evaluations. Details of the analytical derivations are provided in the Supplementary Material.
	
	As a first experiment, we generated synthetic data from the SVM-\textit{t}, SVM-S, and SVM-VG models with a sample size $T=6000$. Using subsamples of sizes $T=1500$ and $T=3000$, we investigated the estimation accuracy as a function of the number of discretization points $m$ over the interval $(-2.5,2.5)$ while recording the corresponding computational times (in minutes).\footnote{All analyses were performed on a notebook equipped with an Intel Core i5-13450HX processor and 16 GB of RAM, running \texttt{R} 4.5.0 on Ubuntu 24.04.2 LTS. The results are reported in Tables~\ref{tab1}, \ref{tab2}, and \ref{tab3}.}
	
	% sim1
	% t-Student
	\begin{table}[h!]
		\caption{SVM-\textit{t} Model estimation results using HMM approach ($b_{m}=-b_{0}=2.5$). First row: Posterior mean. Second row: $95 \%$ credibility interval. True value parameters: $\bm{\theta} = (\beta_0, \beta_1, \beta_2, \mu, \phi, \sigma, \nu)'=(0.2, 0.07, -0.18, 0.1, 0.98, 0.1, 10)'$.}
		\centering
		\scalebox{0.65}{
			\begin{tabular}{l l c c c c c c c c}
				\toprule
				$T$ & $m$ & $\beta_{0}$ & $\beta_{1}$ & $\beta_{2}$ & $\mu$ & $\phi$ & $\sigma$ &  $\nu$ & Time (min) \\  
				\midrule
				\multirow{8}{*}{1500}   &\multirow{2}{*}{50}   & 0.2790           & 0.0767           & -0.2200          & 0.4674           & 0.9702           & 0.0892            & 17.7890 & 0.11\\
				&                      & (0.0359, 0.5899) & (0.0304, 0.1243) & (-0.4185, -0.0569) & (0.2644, 0.6504) & (0.9265, 0.9937) & (0.0427, 0.1637) & (8.1842, 30.2036) & -\\                                      
				\addlinespace
				&\multirow{2}{*}{100}  & 0.2790           & 0.0759           & -0.2129          & 0.5065           & 0.9567           & 0.1005            & 22.7853 & 0.31\\
				&                      & (0.0515, 0.6129) & (0.0343, 0.1182) & (-0.4433, -0.0803) & (0.2479, 0.6724) & (0.9267, 0.9941) & (0.0388, 0.1398) & (8.4719, 33.3950) & -\\                                      
				\addlinespace
				&\multirow{2}{*}{150}  & 0.2381           & 0.0777           & -0.1968          & 0.4871           & 0.9646           & 0.0987            & 21.7996 & 0.54\\
				&                      & (0.0518, 0.5461) & (0.0401, 0.1222) & (-0.3984, -0.0805) & (0.2875, 0.6636) & (0.9396, 0.9940) & (0.0335, 0.1531) & (8.3053, 30.9576) & -\\
				\addlinespace
				&\multirow{2}{*}{200}  & 0.2577           & 0.0784           & -0.2054          & 0.4951           & 0.9731           & 0.0805            & 18.5696 & 0.74\\
				&                      & (0.0139, 0.5558) & (0.0286, 0.1241) & (-0.3924, -0.0573) & (0.2482, 0.7633) & (0.9344, 0.9960) & (0.0319, 0.1448) & (8.1633, 31.8923) & -\\
				\midrule
				\multirow{8}{*}{3000}   &\multirow{2}{*}{50}   & 0.2773           & 0.0725           & -0.2246          & 0.2925           & 0.9778           & 0.0966            & 12.6272 & 0.17\\
				&                      & (0.1547, 0.4080) & (0.0355, 0.1078) & (-0.3285, -0.1262) & (0.0797, 0.5072) & (0.9577, 0.9908) & (0.0628, 0.1438) & (8.3228, 19.5050) & -\\                                      
				\addlinespace
				&\multirow{2}{*}{100}  & 0.2764           & 0.0787           & -0.2251          & 0.2767           & 0.9794           & 0.0939            & 13.0465 & 0.87\\
				&                      & (0.1567, 0.4070) & (0.0377, 0.1146) & (-0.3249, -0.1363) & (0.1259, 0.4516) & (0.9612, 0.9934) & (0.0643, 0.1369) & (8.6110, 19.8466) & -\\                                      
				\addlinespace
				&\multirow{2}{*}{150}  & 0.2724           & 0.0728           & -0.2223          & 0.2864           & 0.9767           & 0.0994            & 12.6525 & 1.16\\
				&                      & (0.1299, 0.4078) & (0.0413, 0.1064) & (-0.3310, -0.1152) & (0.0699, 0.4465) & (0.9574, 0.9896) & (0.0674, 0.1472) & (8.0818, 19.1827) & -\\
				\addlinespace
				&\multirow{2}{*}{200}  & 0.2765           & 0.0723           & -0.2256          & 0.2910           & 0.9769           & 0.0981            & 12.8578 & 1.50\\
				&                      & (0.1289, 0.4116) & (0.0400, 0.1086) & (-0.3337, -0.1209) & (0.1311, 0.4551) & (0.9591, 0.9884) & (0.0678, 0.1398) & (8.3503, 19.9422) & -\\
				\midrule
				\multirow{8}{*}{6000}   &\multirow{2}{*}{50}   & 0.2345           & 0.0756           & -0.2095          & 0.1987           & 0.9774           & 0.0924            & 10.1381 & 0.37\\
				&                      & (0.1345, 0.3417) & (0.0508, 0.1002) & (-0.2971, -0.1336) & (0.0858, 0.3226) & (0.9613, 0.9861) & (0.0711, 0.1245) & (7.9085, 13.5205) & -\\                                      
				\addlinespace
				&\multirow{2}{*}{100}  & 0.2353           & 0.0755           & -0.2110          & 0.1954           & 0.9776           & 0.0914            & 10.0962 & 1.34\\
				&                      & (0.1384, 0.3383) & (0.0498, 0.0992) & (-0.2966, -0.1289) & (0.0787, 0.3062) & (0.9652, 0.9873) & (0.0688, 0.1181) & (7.8442, 13.5025) & -\\                                      
				\addlinespace
				&\multirow{2}{*}{150}  & 0.2317           & 0.0771           & -0.2079          & 0.1992           & 0.9772           & 0.0924            & 10.1289 & 1.79\\
				&                      & (0.1300, 0.3232) & (0.0511, 0.1022) & (-0.2879, -0.1237) & (0.0788, 0.3287) & (0.9650, 0.9867) & (0.0682, 0.1180) & (7.8713, 13.5882) & -\\
				\addlinespace
				&\multirow{2}{*}{200}  & 0.2362           & 0.0762           & -0.2118          & 0.1980           & 0.9773           & 0.0916            & 10.0737 & 2.29\\
				&                      & (0.1478, 0.3278) & (0.0515, 0.1002) & (-0.2950, -0.1366) & (0.0762, 0.3236) & (0.9656, 0.9869) & (0.0691, 0.1166) & (7.7266, 13.5836) & -\\
				\bottomrule
			\end{tabular}
		}
		\label{tab1}
	\end{table}
	% slash
	\begin{table}[h!]
		\caption{SVM-S Model estimation results using HMM approach ($b_{m}=-b_{0}=2.5$). First row: Posterior mean. Second row: $95 \%$ credibility interval. True value parameters: $\bm{\theta} = (\beta_0, \beta_1, \beta_2, \mu, \phi, \sigma, \nu)'=(0.2, 0.07, -0.18, 0.1, 0.98, 0.1, 2)'$.}
		\centering
		\scalebox{0.65}{
			\begin{tabular}{l l c c c c c c c c}
				\toprule
				$T$ & $m$ & $\beta_{0}$ & $\beta_{1}$ & $\beta_{2}$ & $\mu$ & $\phi$ & $\sigma$ &  $\nu$ & Time (min) \\  
				\midrule
				\multirow{8}{*}{1500}   &\multirow{2}{*}{50}   & 0.1870           & 0.0291           & -0.1780          & 0.1210           & 0.9640           & 0.1450            & 2.2100 & 0.10\\
				&                      & (-0.0054, 0.3920) & (-0.0225, 0.0818) & (-0.3530, -0.0290) & (-0.1380, 0.3810) & (0.9210, 0.9880) & (0.0816, 0.2410) & (1.6800, 2.9700) & -\\                                      
				\addlinespace
				&\multirow{2}{*}{100}  & 0.1860           & 0.0271           & -0.1790          & 0.1270           & 0.9650           & 0.1450            & 2.2000 & 0.48\\
				&                      & (-0.0039, 0.3780) & (-0.0261, 0.0794) & (-0.3660, -0.0138) & (-0.1340, 0.3990) & (0.9200, 0.9880) & (0.0798, 0.2430) & (1.7100, 2.9200) & -\\                                      
				\addlinespace
				&\multirow{2}{*}{150}  & 0.1780           & 0.0307           & -0.1710          & 0.1310           & 0.9640           & 0.1450            & 2.2200 & 0.69\\
				&                      & (0.0101, 0.3960) & (-0.0178, 0.0780) & (-0.3730, -0.0280) & (-0.1600, 0.5950) & (0.9200, 0.9900) & (0.0753, 0.2370) & (1.7100, 3.2400) & -\\
				\addlinespace
				&\multirow{2}{*}{200}  & 0.1850           & 0.0292           & -0.1780          & 0.1350           & 0.9630           & 0.1480            & 2.2100 & 0.91\\
				&                      & (0.0116, 0.4000) & (-0.0179, 0.0783) & (-0.3700, -0.0273) & (-0.1020, 0.4150) & (0.9070, 0.9880) & (0.0832, 0.2660) & (1.6800, 2.9600) & -\\
				\midrule
				\multirow{8}{*}{3000}   &\multirow{2}{*}{50}   & 0.1770           & 0.0379           & -0.1800          & 0.0865           & 0.9620           & 0.1240            & 2.0600 & 0.19\\
				&                      & (0.0305, 0.3370) & (0.0035, 0.0745) & (-0.3280, -0.0534) & (-0.0783, 0.2480) & (0.9280, 0.9820) & (0.0797, 0.1830) & (1.7500, 2.4700) & -\\                                      
				\addlinespace
				&\multirow{2}{*}{100}  & 0.1740           & 0.0376           & -0.1790          & 0.0882           & 0.9630           & 0.1220            & 2.0700 & 0.94\\
				&                      & (0.0122, 0.3440) & (-0.0005, 0.0733) & (-0.3390, -0.0420) & (-0.1010, 0.2540) & (0.9350, 0.9820) & (0.0834, 0.1800) & (1.7300, 2.4800) & -\\                                      
				\addlinespace
				&\multirow{2}{*}{150}  & 0.1720           & 0.0365           & -0.1780          & 0.0996           & 0.9630           & 0.1230            & 2.0700 & 1.34\\
				&                      & (0.0278, 0.3630) & (0.0022, 0.0726) & (-0.3490, -0.0463) & (-0.0817, 0.3340) & (0.9270, 0.9820) & (0.0806, 0.1860) & (1.7700, 2.4700) & -\\
				\addlinespace
				&\multirow{2}{*}{200}  & 0.1710           & 0.0386           & -0.1750          & 0.0921           & 0.9630           & 0.1230            & 2.0700 & 1.68\\
				&                      & (0.0158, 0.3380) & (0.0046, 0.0715) & (-0.3200, -0.0389) & (-0.0809, 0.2450) & (0.9290, 0.9810) & (0.0790, 0.1860) & (1.7300, 2.5200) & -\\
				\midrule
				\multirow{8}{*}{6000}   &\multirow{2}{*}{50}   & 0.1530           & 0.0625           & -0.1640          & 0.1280           & 0.9720           & 0.1150            & 2.1200 & 0.27\\
				&                      & (0.0589, 0.2560) & (0.0371, 0.0889) & (-0.2500, -0.0798) & (-0.0121, 0.2790) & (0.9580, 0.9830) & (0.0892, 0.1460) & (1.8800, 2.4500) & -\\                                      
				\addlinespace
				&\multirow{2}{*}{100}  & 0.1540           & 0.0623           & -0.1640          & 0.1290           & 0.9730           & 0.1150            & 2.1300 & 1.50\\
				&                      & (0.0606, 0.2460) & (0.0360, 0.0865) & (-0.2500, -0.0836) & (-0.0119, 0.2740) & (0.9590, 0.9830) & (0.0895, 0.1460) & (1.8800, 2.4300) & -\\                                      
				\addlinespace
				&\multirow{2}{*}{150}  & 0.1500           & 0.0619           & -0.1620          & 0.1290           & 0.9730           & 0.1140            & 2.1200 & 1.92\\
				&                      & (0.0552, 0.2400) & (0.0387, 0.0850) & (-0.2440, -0.0878) & (-0.0089, 0.2620) & (0.9590, 0.9840) & (0.0847, 0.1450) & (1.8700, 2.4300) & -\\
				\addlinespace
				&\multirow{2}{*}{200}  & 0.1520           & 0.0620           & -0.1620          & 0.1270           & 0.9730           & 0.1150            & 2.1200 & 2.50\\
				&                      & (0.0593, 0.2420) & (0.0384, 0.0873) & (-0.2480, -0.0810) & (-0.0084, 0.2550) & (0.9580, 0.9830) & (0.0877, 0.1450) & (1.8800, 2.4000) & -\\
				
				\bottomrule
			\end{tabular}
		}
		\label{tab2}
	\end{table}
	% vg
	\begin{table}[h!]
		\caption{SVM-VG Model estimation results using HMM approach ($b_{m}=-b_{0}=2.5$). First row: Posterior mean. Second row: $95 \%$ credibility interval. True value parameters: $\bm{\theta} = (\beta_0, \beta_1, \beta_2, \mu, \phi, \sigma, \nu)'=(0.2, 0.07, -0.18, 0.1, 0.98, 0.1, 10)'$.}
		\centering
		\scalebox{0.65}{
			\begin{tabular}{l l c c c c c c c c}
				\toprule
				$T$ & $m$ & $\beta_{0}$ & $\beta_{1}$ & $\beta_{2}$ & $\mu$ & $\phi$ & $\sigma$ &  $\nu$ & Time (min) \\  
				\midrule
				\multirow{8}{*}{1500}   &\multirow{2}{*}{20}   & 0.0495           & 0.0817           & -0.0103          & -0.0326          & 0.9826           & 0.0694            & 8.0632  & 0.12\\
				&                      & (-0.0987, 0.2514) & (0.0348, 0.1278) & (-0.2087, 0.1764) & (-0.3052, 0.2909) & (0.9446, 0.9961) & (0.0392, 0.1356) & (4.4965, 15.2174) & -\\
				\addlinespace
				&\multirow{2}{*}{100}  & 0.0515           & 0.0846           & -0.0130          & -0.0516          & 0.9861           & 0.0601            & 8.1849  & 0.72\\
				&                      & (-0.1276, 0.2289) & (0.0326, 0.1271) & (-0.2121, 0.1842) & (-0.2649, 0.2001) & (0.9472, 0.9984) & (0.0259, 0.1311) & (4.7431, 14.4914) & -\\
				\addlinespace
				&\multirow{2}{*}{150}  & 0.0694           & 0.0801           & -0.0337          & -0.0382          & 0.9881           & 0.0563            & 7.8975  & 1.09\\
				&                      & (-0.1042, 0.2544) & (0.0314, 0.1275) & (-0.2355, 0.1461) & (-0.2944, 0.2222) & (0.9552, 0.9981) & (0.0239, 0.1230) & (4.8587, 14.2783) & -\\
				\addlinespace
				&\multirow{2}{*}{200}  & 0.0534           & 0.0811           & -0.0159          & -0.0468          & 0.9862           & 0.0595            & 8.1122  & 1.33\\
				&                      & (-0.1304, 0.2673) & (0.0323, 0.1305) & (-0.2259, 0.1548) & (-0.3202, 0.2659) & (0.9513, 0.9976) & (0.0216, 0.1267) & (4.7400, 17.3844) & -\\
				\midrule
				\multirow{8}{*}{3000}   &\multirow{2}{*}{50}   & 0.2173           & 0.0959           & -0.1865          & 0.1206           & 0.9733           & 0.1146            & 10.6950 & 0.12\\
				&                      & (0.1222, 0.3278) & (0.0601, 0.1285) & (-0.2787, -0.1037) & (-0.0406, 0.3044) & (0.9547, 0.9868) & (0.0789, 0.1581) & (6.5310, 17.9871) & -\\                                      
				\addlinespace
				&\multirow{2}{*}{100}  & 0.2131           & 0.0937           & -0.1832          & 0.1214           & 0.9737           & 0.1135            & 10.2522 & 0.72\\
				&                      & (0.1145, 0.3223) & (0.0586, 0.1292) & (-0.2763, -0.0992) & (-0.0672, 0.3047) & (0.9545, 0.9871) & (0.0793, 0.1537) & (6.3973, 16.0696) & -\\                               
				\addlinespace
				&\multirow{2}{*}{150}  & 0.2165           & 0.0957           & -0.1842          & 0.1288           & 0.9743           & 0.1129            & 10.4172 & 1.09\\
				&                      & (0.1190, 0.3144) & (0.0604, 0.1320) & (-0.2666, -0.0980) & (-0.0677, 0.2942) & (0.9538, 0.9878) & (0.0794, 0.1539) & (6.4775, 17.2436) & -\\
				\addlinespace
				&\multirow{2}{*}{200}  & 0.2145           & 0.0937           & -0.1849          & 0.1228           & 0.9737           & 0.1129            & 10.9670 & 1.33\\
				&                      & (0.1148, 0.3105) & (0.0505, 0.1331) & (-0.2789, -0.0972) & (-0.0521, 0.3108) & (0.9535, 0.9862) & (0.0769, 0.1603) & (6.4543, 21.1818) & -\\
				
				\midrule
				\multirow{8}{*}{6000}   &\multirow{2}{*}{50}   & 0.2198           & 0.0767           & -0.1973          & 0.0947           & 0.9778           & 0.1029            & 11.5695 & 0.44\\
				&                      & (0.1542, 0.2942) & (0.0512, 0.1025) & (-0.2575, -0.1378) & (-0.0214, 0.2329) & (0.9664, 0.9871) & (0.0785, 0.1294) & (7.9558, 16.8269) & -\\    
				\addlinespace
				&\multirow{2}{*}{100}  & 0.2214           & 0.0781           & -0.1997          & 0.0957           & 0.9777           & 0.1030            & 11.7157 & 1.81\\
				&                      & (0.1560, 0.2899) & (0.0503, 0.1027) & (-0.2646, -0.1417) & (-0.0196, 0.2447) & (0.9665, 0.9868) & (0.0795, 0.1304) & (7.9809, 17.9708) & -\\                      
				\addlinespace
				&\multirow{2}{*}{150}  & 0.2136           & 0.0780           & -0.1922          & 0.0873           & 0.9781           & 0.1034            & 11.5569 & 2.40\\
				&                      & (0.1346, 0.2912) & (0.0542, 0.1043) & (-0.2568, -0.1338) & (-0.0793, 0.2244) & (0.9659, 0.9865) & (0.0807, 0.1309) & (8.0835, 17.2979) & -\\
				\addlinespace
				&\multirow{2}{*}{200}  & 0.2213           & 0.0784           & -0.1988          & 0.0933           & 0.9775           & 0.1032            & 11.5230 & 2.98\\
				&                      & (0.1514, 0.2942) & (0.0506, 0.1028) & (-0.2626, -0.1361) & (-0.0323, 0.2124) & (0.9655, 0.9863) & (0.0799, 0.1315) & (7.8730, 17.0545) & -\\
				\bottomrule
			\end{tabular}
		}
		\label{tab3}
	\end{table} 
    
	The results show that, as the sample size increases, the posterior means converge toward the true parameter values and the corresponding $95\%$ credible intervals consistently contain the true values. Moreover, even under the most computationally demanding scenario ($T=6000$ and $m=200$), the proposed methodology remained highly efficient, requiring less than three minutes to complete the estimation procedure.
	
	As a second experiment, we conducted a Monte Carlo study to evaluate the finite-sample performance of the proposed estimators. Specifically, we generated $n=300$ datasets from the SVM-\textit{t}, SVM-S, and SVM-VG models and fitted them using $m=50,100,150,$ and $200$ discretization points, with $b_m=-b_0=2.5$, for sample sizes $T=1500,3000,$ and $6000$. For each scenario, we computed the posterior mean estimates, the mean relative bias (MRB), the mean relative absolute deviation (MRAD), and the relative mean squared error (RMSE). In addition, we evaluated the coverage rates (CR) of the associated $95\%$ credible intervals. The results are summarized in Tables~\ref{tab4}, \ref{tab5}, and \ref{tab6}.
	
	% t
	\begin{table}[h!]
		\centering
		\caption{SVM-\textit{t} model estimation results using HMM approach ($b_{m}=-b_{0}=2.5$) based on $n=300$ replicates of $T=1500, 3000, 6000$.}
		\setlength{\tabcolsep}{2.5pt} % Ajuste fino do espaço entre colunas
		\scalebox{0.80}{
			\begin{tabular}{l c r *{5}{c} c *{5}{c} c *{5}{c}} 
				\toprule
				$m$ & Param. & {True} & \multicolumn{5}{c}{$T=1500$} & & \multicolumn{5}{c}{$T=3000$} & & \multicolumn{5}{c}{$T=6000$} \\
				\cmidrule(lr){4-8} \cmidrule(lr){10-14} \cmidrule(lr){16-20}
				& & {Value} & {Mean} & {MRB} & {MRAD} & {RMSE} & {CR} & & {Mean} & {MRB} & {MRAD} & {RMSE} & {CR} & & {Mean} & {MRB} & {MRAD} & {RMSE} & {CR} \\
				\midrule
				
				% Bloco m=50
				\multirow{7}{*}{50} 
				& $\beta_{0}$ & 0.20  & 0.2181 & 0.0904 & 0.3379 & 0.1978 & 0.93 && 0.2054 & 0.0270 & 0.2509 & 0.1083 & 0.94 && 0.2035 & 0.0175 & 0.1610 & 0.0405 & 0.94 \\
				& $\beta_{1}$ & 0.07  & 0.0700 & 0.0002 & 0.3127 & 0.1481 & 0.93 && 0.0681 & -0.0267 & 0.2015 & 0.0659 & 0.94 && 0.0692 & -0.0113 & 0.1486 & 0.0328 & 0.96 \\
				& $\beta_{2}$ & -0.18 & -0.1977 & 0.0982 & 0.3551 & 0.2255 & 0.94 && -0.1846 & 0.0256 & 0.2523 & 0.1092 & 0.93 && -0.1825 & 0.0140 & 0.1622 & 0.0415 & 0.95 \\
				& $\mu$        & 0.10  & 0.1317 & 0.3169 & 1.2464 & 2.3543 & 0.92 && 0.1166 & 0.1664 & 0.8218 & 1.0469 & 0.93 && 0.1042 & 0.0421 & 0.5884 & 0.5646 & 0.96 \\
				& $\phi$       & 0.98  & 0.9744 & -0.0057 & 0.0115 & 0.0004 & 0.92 && 0.9777 & -0.0023 & 0.0066 & 0.0001 & 0.94 && 0.9796 & -0.0004 & 0.0042 & 0.0000 & 0.94 \\
				& $\sigma$     & 0.10  & 0.1057 & 0.0565 & 0.1989 & 0.0744 & 0.94 && 0.1013 & 0.0132 & 0.1424 & 0.0337 & 0.94 && 0.0999 & -0.0014 & 0.0992 & 0.0173 & 0.95 \\
				& $\nu$        & 10.00 & 13.5817 & 0.3582 & 0.4519 & 0.4824 & 0.89 && 11.3893 & 0.1389 & 0.2241 & 0.1102 & 0.93 && 10.5115 & 0.0512 & 0.1248 & 0.0368 & 0.92 \\
				\midrule
				
				% Bloco m=100
				\multirow{7}{*}{100} 
				& $\beta_{0}$ & 0.20  & 0.2182 & 0.0910 & 0.3713 & 0.2287 & 0.93 && 0.2070 & 0.0351 & 0.2340 & 0.0923 & 0.93 && 0.2013 & 0.0066 & 0.1580 & 0.0384 & 0.95 \\
				& $\beta_{1}$ & 0.07  & 0.0681 & -0.0264 & 0.2984 & 0.1416 & 0.94 && 0.0694 & -0.0084 & 0.2088 & 0.0687 & 0.94 && 0.0693 & -0.0105 & 0.1537 & 0.0366 & 0.94 \\
				& $\beta_{2}$ & -0.18 & -0.2001 & 0.1119 & 0.3859 & 0.2576 & 0.93 && -0.1880 & 0.0447 & 0.2300 & 0.0942 & 0.92 && -0.1822 & 0.0124 & 0.1625 & 0.0405 & 0.96 \\
				& $\mu$        & 0.10  & 0.1011 & 0.0106 & 1.1180 & 2.1134 & 0.94 && 0.1072 & 0.0721 & 0.7394 & 0.9451 & 0.94 && 0.1014 & 0.0141 & 0.6135 & 0.5967 & 0.95 \\
				& $\phi$       & 0.98  & 0.9759 & -0.0042 & 0.0105 & 0.0002 & 0.93 && 0.9790 & -0.0010 & 0.0062 & 0.0001 & 0.95 && 0.9794 & -0.0007 & 0.0041 & 0.0000 & 0.96 \\
				& $\sigma$     & 0.10  & 0.1033 & 0.0334 & 0.2035 & 0.0711 & 0.94 && 0.1004 & 0.0042 & 0.1496 & 0.0350 & 0.96 && 0.1010 & 0.0097 & 0.0999 & 0.0159 & 0.94 \\
				& $\nu$        & 10.00 & 13.0695 & 0.3069 & 0.4281 & 0.4572 & 0.91 && 11.3675 & 0.1368 & 0.2291 & 0.1095 & 0.94 && 10.6026 & 0.0603 & 0.1364 & 0.0381 & 0.92 \\
				\midrule
				
				% Bloco m=150
				\multirow{7}{*}{150} 
				& $\beta_{0}$ & 0.20  & 0.2193 & 0.0965 & 0.3874 & 0.2631 & 0.91 && 0.2040 & 0.0201 & 0.2325 & 0.0901 & 0.94 && 0.1998 & -0.0008 & 0.1492 & 0.0359 & 0.94 \\
				& $\beta_{1}$ & 0.07  & 0.0656 & -0.0622 & 0.2994 & 0.1396 & 0.94 && 0.0689 & -0.0162 & 0.2151 & 0.0699 & 0.95 && 0.0700 & 0.0004 & 0.1406 & 0.0327 & 0.94 \\
				& $\beta_{2}$ & -0.18 & -0.1977 & 0.0981 & 0.4012 & 0.2862 & 0.89 && -0.1844 & 0.0247 & 0.2390 & 0.0961 & 0.93 && -0.1813 & 0.0070 & 0.1454 & 0.0340 & 0.96 \\
				& $\mu$        & 0.10  & 0.1114 & 0.1138 & 1.1619 & 2.3759 & 0.94 && 0.1017 & 0.0168 & 0.7997 & 1.0296 & 0.97 && 0.1022 & 0.0217 & 0.5823 & 0.5340 & 0.95 \\
				& $\phi$       & 0.98  & 0.9753 & -0.0048 & 0.0113 & 0.0003 & 0.93 && 0.9788 & -0.0012 & 0.0063 & 0.0001 & 0.95 && 0.9795 & -0.0006 & 0.0043 & 0.0000 & 0.94 \\
				& $\sigma$     & 0.10  & 0.1037 & 0.0375 & 0.2390 & 0.0950 & 0.94 && 0.1008 & 0.0083 & 0.1411 & 0.0311 & 0.95 && 0.1007 & 0.0069 & 0.1068 & 0.0179 & 0.95 \\
				& $\nu$        & 10.00 & 12.8654 & 0.2865 & 0.3791 & 0.3110 & 0.95 && 11.5119 & 0.1512 & 0.2342 & 0.1404 & 0.95 && 10.4845 & 0.0484 & 0.1262 & 0.0282 & 0.94 \\
				\midrule
				
				% Bloco m=200
				\multirow{7}{*}{200} 
				& $\beta_{0}$ & 0.20  & 0.2145 & 0.0727 & 0.3605 & 0.2157 & 0.94 && 0.2018 & 0.0090 & 0.2389 & 0.0934 & 0.92 && 0.2008 & 0.0042 & 0.1405 & 0.0305 & 0.96 \\
				& $\beta_{1}$ & 0.07  & 0.0645 & -0.0783 & 0.3089 & 0.1426 & 0.96 && 0.0689 & -0.0162 & 0.2088 & 0.0685 & 0.94 && 0.0706 & 0.0090 & 0.1497 & 0.0345 & 0.95 \\
				& $\beta_{2}$ & -0.18 & -0.1915 & 0.0636 & 0.3763 & 0.2379 & 0.93 && -0.1831 & 0.0170 & 0.2452 & 0.1012 & 0.90 && -0.1808 & 0.0042 & 0.1463 & 0.0337 & 0.96 \\
				& $\mu$        & 0.10  & 0.1051 & 0.0513 & 1.1291 & 2.0130 & 0.96 && 0.0969 & -0.0313 & 0.8135 & 1.1073 & 0.93 && 0.0969 & -0.0306 & 0.5578 & 0.4820 & 0.96 \\
				& $\phi$       & 0.98  & 0.9754 & -0.0047 & 0.0113 & 0.0003 & 0.92 && 0.9785 & -0.0016 & 0.0064 & 0.0001 & 0.91 && 0.9796 & -0.0004 & 0.0038 & 0.0000 & 0.94 \\
				& $\sigma$     & 0.10  & 0.1025 & 0.0246 & 0.2346 & 0.0903 & 0.93 && 0.1007 & 0.0070 & 0.1563 & 0.0398 & 0.93 && 0.1003 & 0.0032 & 0.0948 & 0.0144 & 0.96 \\
				& $\nu$        & 10.00 & 12.9523 & 0.2952 & 0.3903 & 0.3505 & 0.93 && 11.0530 & 0.1053 & 0.2006 & 0.0810 & 0.95 && 10.4509 & 0.0451 & 0.1115 & 0.0237 & 0.95 \\
				\bottomrule
			\end{tabular}
		}
		\label{tab4}
	\end{table}
	% slash
	\begin{table}[h!]
		\centering
		\caption{SVM-S model estimation results using HMM approach ($b_{m}=-b_{0}=2.5$) based on $n=300$ replicates of $T=1500, 3000, 6000$.}
		\setlength{\tabcolsep}{2.5pt} % Ajuste fino do espaço entre colunas
		\scalebox{0.8}{
			\begin{tabular}{l c r *{5}{c} c *{5}{c} c *{5}{c}} 
				\toprule
				$m$ & Param. & {True} & \multicolumn{5}{c}{$T=1500$} & & \multicolumn{5}{c}{$T=3000$} & & \multicolumn{5}{c}{$T=6000$} \\
				\cmidrule(lr){4-8} \cmidrule(lr){10-14} \cmidrule(lr){16-20}
				& & {Value} & {Mean} & {MRB} & {MRAD} & {RMSE} & {CR} & & {Mean} & {MRB} & {MRAD} & {RMSE} & {CR} & & {Mean} & {MRB} & {MRAD} & {RMSE} & {CR} \\
				\midrule
				
				% Bloco m=50
				\multirow{7}{*}{50} 
				& $\beta_{0}$ & 0.20  & 0.2136 & 0.0680 & 0.4636 & 0.3471 & 0.92 && 0.2051 & 0.0255 & 0.2807 & 0.1249 & 0.94 && 0.2048 & 0.0239 & 0.1766 & 0.0534 & 0.96 \\
				& $\beta_{1}$ & 0.07  & 0.0696 & -0.0056 & 0.2933 & 0.1327 & 0.94 && 0.0697 & -0.0039 & 0.1976 & 0.0594 & 0.96 && 0.0707 & 0.0107 & 0.1466 & 0.0324 & 0.94 \\
				& $\beta_{2}$ & -0.18 & -0.1886 & 0.0478 & 0.4706 & 0.3737 & 0.92 && -0.1870 & 0.0391 & 0.2749 & 0.1166 & 0.96 && -0.1852 & 0.0291 & 0.1829 & 0.0589 & 0.95 \\
				& $\mu$        & 0.10  & 0.1199 & 0.1987 & 1.2874 & 2.5611 & 0.93 && 0.0986 & -0.0142 & 0.8810 & 1.2287 & 0.94 && 0.0993 & -0.0074 & 0.5771 & 0.5157 & 0.96 \\
				& $\phi$       & 0.98  & 0.9739 & -0.0062 & 0.0116 & 0.0004 & 0.92 && 0.9784 & -0.0016 & 0.0061 & 0.0001 & 0.96 && 0.9791 & -0.0009 & 0.0044 & 0.0000 & 0.93 \\
				& $\sigma$     & 0.10  & 0.1044 & 0.0435 & 0.2216 & 0.0872 & 0.94 && 0.1012 & 0.0124 & 0.1421 & 0.0311 & 0.96 && 0.1009 & 0.0088 & 0.1141 & 0.0211 & 0.94 \\
				& $\nu$        & 2.00  & 2.1309 & 0.0654 & 0.1263 & 0.0305 & 0.91 && 2.0291 & 0.0145 & 0.0650 & 0.0069 & 0.97 && 2.0159 & 0.0080 & 0.0491 & 0.0039 & 0.94 \\
				\midrule
				
				% Bloco m=100
				\multirow{7}{*}{100} 
				& $\beta_{0}$ & 0.20  & 0.2063 & 0.0314 & 0.4632 & 0.3912 & 0.92 && 0.2026 & 0.0129 & 0.2672 & 0.1163 & 0.94 && 0.2039 & 0.0193 & 0.1923 & 0.0573 & 0.94 \\
				& $\beta_{1}$ & 0.07  & 0.0688 & -0.0166 & 0.2870 & 0.1339 & 0.93 && 0.0698 & -0.0026 & 0.2034 & 0.0665 & 0.92 && 0.0704 & 0.0058 & 0.1338 & 0.0305 & 0.95 \\
				& $\beta_{2}$ & -0.18 & -0.1933 & 0.0737 & 0.4973 & 0.4603 & 0.91 && -0.1826 & 0.0146 & 0.2758 & 0.1209 & 0.95 && -0.1841 & 0.0228 & 0.2029 & 0.0629 & 0.95 \\
				& $\mu$        & 0.10  & 0.0888 & -0.1123 & 1.2205 & 2.4484 & 0.93 && 0.1038 & 0.0375 & 0.8429 & 1.1408 & 0.95 && 0.0945 & -0.0552 & 0.6301 & 0.6071 & 0.94 \\
				& $\phi$       & 0.98  & 0.9756 & -0.0045 & 0.0104 & 0.0003 & 0.94 && 0.9778 & -0.0022 & 0.0067 & 0.0001 & 0.91 && 0.9792 & -0.0008 & 0.0041 & 0.0000 & 0.96 \\
				& $\sigma$     & 0.10  & 0.1030 & 0.0299 & 0.2173 & 0.0777 & 0.94 && 0.1038 & 0.0380 & 0.1559 & 0.0401 & 0.94 && 0.0994 & -0.0064 & 0.1023 & 0.0165 & 0.96 \\
				& $\nu$        & 2.00  & 2.0864 & 0.0432 & 0.1170 & 0.0272 & 0.92 && 2.0547 & 0.0273 & 0.0784 & 0.0108 & 0.92 && 2.0091 & 0.0045 & 0.0512 & 0.0044 & 0.93 \\
				\midrule
				
				% Bloco m=150
				\multirow{7}{*}{150} 
				& $\beta_{0}$ & 0.20  & 0.2089 & 0.0443 & 0.4654 & 0.3650 & 0.94 && 0.2059 & 0.0296 & 0.2514 & 0.1099 & 0.95 && 0.2060 & 0.0302 & 0.1761 & 0.0502 & 0.96 \\
				& $\beta_{1}$ & 0.07  & 0.0681 & -0.0268 & 0.2840 & 0.1235 & 0.94 && 0.0711 & 0.0154 & 0.2007 & 0.0638 & 0.92 && 0.0690 & -0.0136 & 0.1368 & 0.0299 & 0.96 \\
				& $\beta_{2}$ & -0.18 & -0.1902 & 0.0564 & 0.4720 & 0.3934 & 0.93 && -0.1860 & 0.0335 & 0.2594 & 0.1121 & 0.95 && -0.1863 & 0.0352 & 0.1814 & 0.0545 & 0.96 \\
				& $\mu$        & 0.10  & 0.1026 & 0.0263 & 1.2585 & 2.6216 & 0.95 && 0.1065 & 0.0649 & 0.9350 & 1.3081 & 0.96 && 0.0978 & -0.0220 & 0.5921 & 0.5751 & 0.94 \\
				& $\phi$       & 0.98  & 0.9757 & -0.0044 & 0.0108 & 0.0003 & 0.92 && 0.9790 & -0.0010 & 0.0071 & 0.0001 & 0.90 && 0.9796 & -0.0004 & 0.0041 & 0.0000 & 0.95 \\
				& $\sigma$     & 0.10  & 0.1012 & 0.0116 & 0.2373 & 0.0921 & 0.91 && 0.1007 & 0.0074 & 0.1685 & 0.0477 & 0.90 && 0.0989 & -0.0114 & 0.1038 & 0.0172 & 0.95 \\
				& $\nu$        & 2.00  & 2.0977 & 0.0489 & 0.1261 & 0.0305 & 0.92 && 2.0561 & 0.0280 & 0.0776 & 0.0104 & 0.92 && 2.0150 & 0.0075 & 0.0477 & 0.0037 & 0.96 \\
				\midrule
				
				% Bloco m=200
				\multirow{7}{*}{200} 
				& $\beta_{0}$ & 0.20  & 0.2210 & 0.1048 & 0.4643 & 0.3970 & 0.92 && 0.2084 & 0.0420 & 0.2755 & 0.1188 & 0.93 && 0.2023 & 0.0117 & 0.1842 & 0.0522 & 0.96 \\
				& $\beta_{1}$ & 0.07  & 0.0681 & -0.0272 & 0.2785 & 0.1177 & 0.95 && 0.0700 & -0.0001 & 0.2156 & 0.0715 & 0.93 && 0.0705 & 0.0074 & 0.1422 & 0.0316 & 0.95 \\
				& $\beta_{2}$ & -0.18 & -0.1994 & 0.1080 & 0.4739 & 0.4493 & 0.91 && -0.1872 & 0.0399 & 0.2739 & 0.1208 & 0.93 && -0.1841 & 0.0226 & 0.1840 & 0.0522 & 0.96 \\
				& $\mu$        & 0.10  & 0.1101 & 0.1013 & 1.2828 & 2.6064 & 0.93 && 0.1029 & 0.0288 & 0.7918 & 1.0156 & 0.97 && 0.1012 & 0.0117 & 0.6621 & 0.6423 & 0.95 \\
				& $\phi$       & 0.98  & 0.9758 & -0.0043 & 0.0107 & 0.0003 & 0.93 && 0.9787 & -0.0013 & 0.0063 & 0.0001 & 0.93 && 0.9792 & -0.0009 & 0.0045 & 0.0000 & 0.94 \\
				& $\sigma$     & 0.10  & 0.1031 & 0.0312 & 0.2302 & 0.0904 & 0.92 && 0.1008 & 0.0081 & 0.1512 & 0.0383 & 0.96 && 0.1002 & 0.0017 & 0.1015 & 0.0170 & 0.95 \\
				& $\nu$        & 2.00  & 2.1282 & 0.0641 & 0.1315 & 0.0329 & 0.93 && 2.0365 & 0.0182 & 0.0696 & 0.0079 & 0.95 && 2.0122 & 0.0061 & 0.0483 & 0.0036 & 0.96 \\
				\bottomrule
			\end{tabular}
		}
		\label{tab5}
	\end{table}
	% vgamma
	\begin{table}[h!]
		\centering
		\caption{SVM-VG model estimation results using HMM approach ($b_{m}=-b_{0}=2.5$) based on $n=300$ replicates of $T=1500, 3000, 6000$.}
		\setlength{\tabcolsep}{2.5pt}
		\scalebox{0.8}{
			\begin{tabular}{l c r *{5}{c} c *{5}{c} c *{5}{c}} 
				\toprule
				$m$ & Param. & {True} & \multicolumn{5}{c}{$T=1500$} & & \multicolumn{5}{c}{$T=3000$} & & \multicolumn{5}{c}{$T=6000$} \\
				\cmidrule(lr){4-8} \cmidrule(lr){10-14} \cmidrule(lr){16-20}
				& & {Value} & {Mean} & {MRB} & {MRAD} & {RMSE} & {CR} & & {Mean} & {MRB} & {MRAD} & {RMSE} & {CR} & & {Mean} & {MRB} & {MRAD} & {RMSE} & {CR} \\
				\midrule
				
				% Bloco m=50
				\multirow{7}{*}{50} 
				& $\beta_{0}$ & 0.20  & 0.2123 & 0.0614 & 0.3264 & 0.1770 & 0.92 && 0.2066 & 0.0332 & 0.2143 & 0.0769 & 0.94 && 0.1989 & -0.0054 & 0.1295 & 0.0261 & 0.97 \\
				& $\beta_{1}$ & 0.07  & 0.0658 & -0.0599 & 0.3317 & 0.1686 & 0.92 && 0.0689 & -0.0158 & 0.2104 & 0.0697 & 0.94 && 0.0695 & -0.0072 & 0.1482 & 0.0353 & 0.95 \\
				& $\beta_{2}$ & -0.18 & -0.1911 & 0.0615 & 0.3388 & 0.2003 & 0.91 && -0.1864 & 0.0357 & 0.2202 & 0.0807 & 0.93 && -0.1798 & -0.0012 & 0.1276 & 0.0260 & 0.96 \\
				& $\mu$        & 0.10  & 0.0875 & -0.1254 & 1.1069 & 1.9638 & 0.90 && 0.0961 & -0.0388 & 0.7842 & 0.9953 & 0.94 && 0.0988 & -0.0124 & 0.5559 & 0.4709 & 0.95 \\
				& $\phi$       & 0.98  & 0.9747 & -0.0054 & 0.0109 & 0.0003 & 0.90 && 0.9775 & -0.0026 & 0.0070 & 0.0001 & 0.95 && 0.9792 & -0.0008 & 0.0039 & 0.0000 & 0.96 \\
				& $\sigma$     & 0.10  & 0.1073 & 0.0726 & 0.2116 & 0.0825 & 0.95 && 0.1014 & 0.0140 & 0.1346 & 0.0308 & 0.95 && 0.1013 & 0.0134 & 0.0931 & 0.0138 & 0.97 \\
				& $\nu$        & 10.00 & 15.5684 & 0.5568 & 0.6815 & 0.9582 & 0.90 && 12.5073 & 0.2507 & 0.3561 & 0.3045 & 0.93 && 11.0762 & 0.1076 & 0.1875 & 0.0767 & 0.96 \\
				\midrule
				
				% Bloco m=100
				\multirow{7}{*}{100} 
				& $\beta_{0}$ & 0.20  & 0.2109 & 0.0544 & 0.3272 & 0.1878 & 0.93 && 0.2063 & 0.0314 & 0.1961 & 0.0674 & 0.94 && 0.1988 & -0.0059 & 0.1357 & 0.0294 & 0.93 \\
				& $\beta_{1}$ & 0.07  & 0.0683 & -0.0244 & 0.2968 & 0.1479 & 0.92 && 0.0676 & -0.0347 & 0.2008 & 0.0658 & 0.94 && 0.0685 & -0.0209 & 0.1463 & 0.0352 & 0.95 \\
				& $\beta_{2}$ & -0.18 & -0.1921 & 0.0673 & 0.3333 & 0.1962 & 0.94 && -0.1869 & 0.0383 & 0.1983 & 0.0691 & 0.93 && -0.1798 & -0.0014 & 0.1320 & 0.0289 & 0.95 \\
				& $\mu$        & 0.10  & 0.0942 & -0.0578 & 1.1690 & 2.0742 & 0.92 && 0.0980 & -0.0203 & 0.7598 & 0.9484 & 0.93 && 0.1014 & 0.0145 & 0.5438 & 0.4822 & 0.96 \\
				& $\phi$       & 0.98  & 0.9758 & -0.0043 & 0.0106 & 0.0002 & 0.92 && 0.9777 & -0.0024 & 0.0065 & 0.0001 & 0.94 && 0.9794 & -0.0006 & 0.0043 & 0.0000 & 0.95 \\
				& $\sigma$     & 0.10  & 0.1008 & 0.0083 & 0.2088 & 0.0744 & 0.95 && 0.1033 & 0.0330 & 0.1548 & 0.0422 & 0.92 && 0.1012 & 0.0115 & 0.1027 & 0.0166 & 0.94 \\
				& $\nu$        & 10.00 & 15.9939 & 0.5994 & 0.7177 & 0.9858 & 0.88 && 13.1098 & 0.3110 & 0.4043 & 0.3717 & 0.92 && 11.2122 & 0.1212 & 0.2092 & 0.0886 & 0.93 \\
				\midrule
				
				% Bloco m=150
				\multirow{7}{*}{150} 
				& $\beta_{0}$ & 0.20  & 0.2113 & 0.0566 & 0.3145 & 0.1695 & 0.95 && 0.2023 & 0.0115 & 0.1941 & 0.0609 & 0.96 && 0.2030 & 0.0152 & 0.1405 & 0.0293 & 0.97 \\
				& $\beta_{1}$ & 0.07  & 0.0660 & -0.0565 & 0.2870 & 0.1339 & 0.94 && 0.0703 & 0.0049 & 0.2217 & 0.0753 & 0.95 && 0.0693 & -0.0095 & 0.1430 & 0.0334 & 0.94 \\
				& $\beta_{2}$ & -0.18 & -0.1918 & 0.0654 & 0.3122 & 0.1686 & 0.94 && -0.1833 & 0.0184 & 0.1973 & 0.0649 & 0.95 && -0.1838 & 0.0210 & 0.1440 & 0.0301 & 0.96 \\
				& $\mu$        & 0.10  & 0.0921 & -0.0795 & 1.1288 & 2.0167 & 0.94 && 0.0870 & -0.1300 & 0.7605 & 0.9050 & 0.95 && 0.0965 & -0.0347 & 0.5253 & 0.4312 & 0.95 \\
				& $\phi$       & 0.98  & 0.9757 & -0.0044 & 0.0104 & 0.0003 & 0.92 && 0.9791 & -0.0009 & 0.0062 & 0.0001 & 0.93 && 0.9794 & -0.0007 & 0.0040 & 0.0000 & 0.95 \\
				& $\sigma$     & 0.10  & 0.1036 & 0.0355 & 0.2410 & 0.1040 & 0.92 && 0.1010 & 0.0096 & 0.1443 & 0.0361 & 0.92 && 0.1007 & 0.0074 & 0.1070 & 0.0182 & 0.92 \\
				& $\nu$        & 10.00 & 15.4596 & 0.5460 & 0.6694 & 0.9236 & 0.91 && 13.1251 & 0.3125 & 0.4141 & 0.4068 & 0.90 && 11.2107 & 0.1211 & 0.2126 & 0.1035 & 0.91 \\
				\midrule
				
				% Bloco m=200
				\multirow{7}{*}{200} 
				& $\beta_{0}$ & 0.20  & 0.2097 & 0.0483 & 0.3566 & 0.1971 & 0.93 && 0.2080 & 0.0401 & 0.2137 & 0.0731 & 0.95 && 0.2057 & 0.0283 & 0.1406 & 0.0322 & 0.94 \\
				& $\beta_{1}$ & 0.07  & 0.0668 & -0.0453 & 0.2916 & 0.1426 & 0.90 && 0.0686 & -0.0206 & 0.2131 & 0.0692 & 0.95 && 0.0692 & -0.0114 & 0.1429 & 0.0331 & 0.93 \\
				& $\beta_{2}$ & -0.18 & -0.1891 & 0.0506 & 0.3206 & 0.1653 & 0.94 && -0.1890 & 0.0499 & 0.2122 & 0.0727 & 0.94 && -0.1862 & 0.0347 & 0.1378 & 0.0316 & 0.93 \\
				& $\mu$        & 0.10  & 0.0777 & -0.2226 & 1.1926 & 2.4307 & 0.93 && 0.1002 & 0.0024 & 0.7851 & 0.9756 & 0.93 && 0.0932 & -0.0678 & 0.5430 & 0.4603 & 0.93 \\
				& $\phi$       & 0.98  & 0.9753 & -0.0048 & 0.0101 & 0.0002 & 0.96 && 0.9783 & -0.0017 & 0.0064 & 0.0001 & 0.92 && 0.9787 & -0.0013 & 0.0044 & 0.0000 & 0.93 \\
				& $\sigma$     & 0.10  & 0.1064 & 0.0644 & 0.2332 & 0.1031 & 0.93 && 0.1015 & 0.0148 & 0.1564 & 0.0422 & 0.91 && 0.1019 & 0.0189 & 0.1018 & 0.0166 & 0.96 \\
				& $\nu$        & 10.00 & 16.6274 & 0.6627 & 0.7698 & 1.0706 & 0.91 && 12.9065 & 0.2907 & 0.3908 & 0.3474 & 0.90 && 11.1835 & 0.1184 & 0.2066 & 0.0943 & 0.93 \\
				\bottomrule
			\end{tabular}
		}
		\label{tab6}
	\end{table}
	
	Regarding the SVM-\textit{t} model, across all sample sizes, the posterior means are very close to the true parameter values. Furthermore, MRB, MRAD, and RMSE remain close to zero for nearly all parameters and values of $m$. The only exception is the parameter $\mu$, which exhibits a modest bias. However, this bias has little practical impact on estimation or prediction because $\mu$ merely represents the unconditional level of the latent volatility process. In addition, the empirical coverage rates remain close to the nominal $95\%$ level. As expected, increasing the sample size further improves estimation accuracy, with the error measures approaching zero and the coverage probabilities becoming increasingly aligned with their nominal values.
	
	The results obtained for the SVM-S and SVM-VG models were qualitatively similar to those reported for the SVM-\textit{t} model and are presented in Tables \ref{tab5} and \ref{tab6}, respectively. Overall, these findings demonstrate that the proposed HMM-based approach provides accurate approximations to the likelihood function of SVM-SMN models while maintaining a high level of computational efficiency.

\section{Real Data Application}\label{Appl}
	
	In this section, we analyze the daily returns of four of the world's main indices: S\&P 500 (EUA), NIKKEI 225 (Japan), DAX 30 (Germany) and IBOVESPA (Brazil). Data were obtained from Yahoo Finance (\url{http://finance.yahoo.com}) considering the period from January 1, 2002, until April 6, 2022. Returns are computed as compounded percentage returns, defined as
	$y_t = 100 \times ( \log P_t - \log P_{t-1}  )$, where $P_t$ denotes the (adjusted) closing price on day $t$, for $t=1,\dots,T$. Table~\ref{tab7} shows the statistics of the observations considered for adjustment. 
	% Summary
	\begin{table}[h!]
		\caption{Summary statistics of the return indexes.}
		\centering
		\begin{tabular}{l c c c c c c c}
			\toprule
			Index   & T    & Mean & SD & Min. & Max. & Skewness & Kurtosis  \\
			\midrule
			S\&P 500        & 5100 & 0.027 & 1.226 & -12.765 & 10.957   & -0.445   & 15.285\\
			NIKKEI 225      & 4959 & 0.019 & 1.458 & -12.111 & 13.235   & -0.409   & 9.882\\
			DAX 30          & 5141 & 0.020 & 1.455 & -13.055 & 10.797   & -0.138   & 9.447 \\
			IBOVESPA        & 5013 &0.043  &1.758  &-15.993  &13.677    & -0.397   &10.540 \\
			\bottomrule
		\end{tabular}
		\label{tab7}
	\end{table}
	
	We can also note that the indices present kurtosis greater than 3, this departure from normality being a well-known stylized fact of this type of series. Also, all indices considered exhibit negative skewness, with S\&P500 and NIKKEI being the most negative, while DAX 30 is the closest to zero. 
	
	In our analysis, we compare the SVM-N, SVM-\textit{t}, SVM-S and SVM-VG models for each one of the series described earlier using the methodology described in Section \ref{SVMmod} setting $m=200$ and $b_m = -b_0 = 5$. For comparison purposes, the models were also fitted using MCMC methods based on HMC and RMHMC, as described in the Supplementary Material. Due to the model’s complexity, 100000 iterations were performed, with the first 50000 discarded as burn-in. Subsequently, thinning was applied with a jump size of 25 to mitigate autocorrelation effects, yielding a final sample of 2000 draws. The results for the S\&P 500, NIKKEI 225, DAX 30, and IBOVESPA returns are summarized in Tables \ref{tab8} and \ref{tab9}, respectively.
	%sp500 e nikkei
	\begin{table}[h!]
		\caption{Estimation of the SVM Model using HMC and HMM machinery with Importance Sampling for S\&P 500 and NIKKEI 225 returns.}
		\centering
		\setlength{\tabcolsep}{2.5pt}
		\scalebox{0.85}{
			\begin{tabular}{l rrrr rrrr}
				\toprule 
				& \multicolumn{4}{c}{S\&P 500} & \multicolumn{4}{c}{NIKKEI 225}\\
				\cmidrule(l){2-5} \cmidrule(l){6-9}
				& \multicolumn{2}{c}{HMM}  & \multicolumn{2}{c}{MCMC} & \multicolumn{2}{c}{HMM}  & \multicolumn{2}{c}{MCMC} \\
				\cmidrule(l){2-3} \cmidrule(l){4-5} \cmidrule(l){6-7} \cmidrule(l){8-9}
				Param. & Mean & $95\%$ Interval & Mean & $95\%$ Interval & Mean & $95\%$ Interval & Mean & $95\%$ Interval \\
				\midrule
				\textbf{SVM-N} \\
				$\beta_{0}$ & 0.1170 & (0.0896, 0.1413) & 0.1155 & (0.0914, 0.1400) & 0.1561 & (0.1132, 0.2024) & 0.1551 & (0.1084, 0.2023) \\
				$\beta_{1}$ & -0.0715 & (-0.0993, -0.0438) & -0.0722 & (-0.1009, -0.0447) & -0.0266 & (-0.0548, 0.0017) & -0.0264 & (-0.0565, 0.0018) \\
				$\beta_{2}$ & -0.0612 & (-0.0918, -0.0308) & -0.0598 & (-0.0871, -0.0336) & -0.0672 & (-0.0996, -0.0343) & -0.0667 & (-0.0970, -0.0355) \\
				$\mu$       & -0.3438 & (-0.5948, -0.0979) & -0.3325 & (-0.6228, -0.0400) & 0.3439 & (0.1563, 0.5474) & 0.3549 & (0.1526, 0.5490) \\
				$\phi$      & 0.9772 & (0.9693, 0.9840) & 0.9797 & (0.9724, 0.9866) & 0.9710 & (0.9593, 0.9804) & 0.9745 & (0.9655, 0.9825) \\
				$\sigma$    & 0.2285 & (0.2012, 0.2587) & 0.2142 & (0.1855, 0.2435) & 0.1946 & (0.1640, 0.2284) & 0.1792 & (0.1582, 0.2057) \\
				\midrule
				\textbf{SVM-\textit{t}} \\
				$\beta_{0}$ & 0.1152 & (0.0931, 0.1432) & 0.1152 & (0.0906, 0.1396) & 0.1486 & (0.1028, 0.1975) & 0.1517 & (0.1018, 0.1991) \\
				$\beta_{1}$ & -0.0708 & (-0.0994, -0.0428) & -0.0716 & (-0.0991, -0.0450) & -0.0294 & (-0.0582, 0.0001) & -0.0291 & (-0.0580, -0.0011) \\
				$\beta_{2}$ & -0.0624 & (-0.0992, -0.0322) & -0.0647 & (-0.0997, -0.0324) & -0.0687 & (-0.1012, -0.0349) & -0.0707 & (-0.1059, -0.0358) \\
				$\mu$       & -0.4596 & (-0.7255, -0.1411) & -0.4481 & (-0.7767, -0.1301) & 0.2453 & (0.0228, 0.4631) & 0.2432 & (0.0114, 0.4584) \\
				$\phi$      & 0.9813 & (0.9734, 0.9905) & 0.9814 & (0.9735, 0.9877) & 0.9772 & (0.9665, 0.9852) & 0.9778 & (0.9680, 0.9863) \\
				$\sigma$    & 0.2054 & (0.1769, 0.2381) & 0.2042 & (0.1780, 0.2371) & 0.1665 & (0.1389, 0.1988) & 0.1661 & (0.1387, 0.2001) \\
				$\nu$       & 16.2488 & (10.5827, 25.2093) & 15.5855 & (9.1236, 25.3906) & 18.1541 & (11.0635, 30.0730) & 17.6655 & (10.0706, 30.4644) \\
				\midrule
				\textbf{SVM-S} \\
				$\beta_{0}$ & 0.1124 & (0.0847, 0.1401) & 0.1158 & (0.0917, 0.1411) & 0.1541 & (0.1008, 0.2062) & 0.1521 & (0.1053, 0.2012) \\
				$\beta_{1}$ & -0.0718 & (-0.0958, -0.0432) & -0.0714 & (-0.0984, -0.0428) & -0.0300 & (-0.0610, -0.0004) & -0.0284 & (-0.0568, -0.0001) \\
				$\beta_{2}$ & -0.0729 & (-0.1150, -0.0326) & -0.0789 & (-0.1196, -0.0394) & -0.0897 & (-0.1348, -0.0385) & -0.0889 & (-0.1347, -0.0457) \\
				$\mu$       & -0.6096 & (-0.9592, -0.3352) & -0.6343 & (-0.9818, -0.3163) & 0.0602 & (-0.1115, 0.2953) & 0.0231 & (-0.2279, 0.2653) \\
				$\phi$      & 0.9799 & (0.9735, 0.9863) & 0.9801 & (0.9725, 0.9867) & 0.9768 & (0.9661, 0.9829) & 0.9791 & (0.9694, 0.9873) \\
				$\sigma$    & 0.2113 & (0.1846, 0.2424) & 0.2127 & (0.1857, 0.2416) & 0.1689 & (0.1395, 0.2023) & 0.1593 & (0.1314, 0.1924) \\
				$\nu$       & 3.9426 & (2.7736, 6.1262) & 3.7982 & (2.7212, 6.4310) & 3.6745 & (2.7608, 5.4425) & 3.4691 & (2.5972, 4.9541) \\
				\midrule
				\textbf{SVM-VG} \\
				$\beta_{0}$ & 0.1149 &  (0.0910,  0.1402) & 0.1140 & (0.0907, 0.1380) & 0.1476 & (0.0993, 0.1917) & 0.1505 & (0.1036, 0.1984) \\
				$\beta_{1}$ &-0.0709 & (-0.0992, -0.0427) & -0.0711 & (-0.0997, -0.0450) & -0.0294 & (-0.0587, 0.0006) & -0.0302 & (-0.0581, -0.0019) \\
				$\beta_{2}$ &-0.0541 & (-0.0869, -0.0278) & -0.0553 & (-0.0862, -0.0265) & -0.0606 & (-0.0929, -0.0280) & -0.0619 & (-0.0948, -0.0304) \\
				$\mu$       &-0.2891 & (-0.5856,  0.0104) & -0.3025 & (-0.6096, 0.0362) & 0.3683 & (0.1579, 0.5645) & 0.3714 & (0.1616, 0.5885) \\
				$\phi$      & 0.9820 &  (0.9746,  0.9879) & 0.9821 & (0.9753, 0.9885) & 0.9771 & (0.9658, 0.9853) & 0.9773 & (0.9676, 0.9866) \\
				$\sigma$    & 0.2026 &  (0.1768,  0.2325) & 0.1994 & (0.1720, 0.2245) & 0.1689 & (0.1407, 0.2015) & 0.1672 & (0.1367, 0.1981) \\
				$\nu$       &11.7969 &  (7.1832, 20.1223) & 12.1705 & (6.6778, 22.7024) & 15.2236 & (8.7519, 25.2884) & 14.8197 & (8.1442, 25.7477) \\
				\bottomrule
			\end{tabular}
		}
		\label{tab8}
	\end{table}
	%dax e ibovespa
	\begin{table}[h!]
		\caption{Estimation of the SVM Model using HMC and HMM machinery with Importance Sampling for DAX 30 and IBOVESPA returns.}
		\centering
		\setlength{\tabcolsep}{2.5pt}
		\scalebox{0.85}{
			\begin{tabular}{l rrrr rrrr}
				\toprule 
				& \multicolumn{4}{c}{DAX 30} & \multicolumn{4}{c}{IBOVESPA}\\
				\cmidrule(l){2-5} \cmidrule(l){6-9}
				& \multicolumn{2}{c}{HMM}  & \multicolumn{2}{c}{MCMC} & \multicolumn{2}{c}{HMM}  & \multicolumn{2}{c}{MCMC} \\
				\cmidrule(l){2-3} \cmidrule(l){4-5} \cmidrule(l){6-7} \cmidrule(l){8-9}
				Param. & Mean & $95\%$ Interval & Mean & $95\%$ Interval & Mean & $95\%$ Interval & Mean & $95\%$ Interval \\
				\midrule
				\textbf{SVM-N} \\
				$\beta_{0}$ & 0.1457 & (0.1114, 0.1778) & 0.1437 & (0.1081, 0.1790) & 0.1821 & (0.1153, 0.2541) & 0.1820 & (0.1131, 0.2546) \\
				$\beta_{1}$ & -0.0275 & (-0.0540, 0.0020) & -0.0293 & (-0.0569, -0.0008) & -0.0223 & (-0.0485, 0.0072) & -0.0228 & (-0.0529, 0.0056) \\
				$\beta_{2}$ & -0.0591 & (-0.0842, -0.0327) & -0.0593 & (-0.0850, -0.0346) & -0.0475 & (-0.0778, -0.0192) & -0.0474 & (-0.0773, -0.0194) \\
				$\mu$       & 0.2109  & (-0.0402, 0.4859) & 0.2178 & (-0.0780, 0.5019) & 0.7764  & (0.5953, 0.9543) & 0.7809 & (0.5801, 0.9760) \\
				$\phi$      & 0.9786  & (0.9697, 0.9857) & 0.9813 & (0.9737, 0.9881) & 0.9770  & (0.9680, 0.9843) & 0.9787 & (0.9696, 0.9862) \\
				$\sigma$    & 0.2010  & (0.1710, 0.2376) & 0.1847 & (0.1585, 0.2120) & 0.1468  & (0.1241, 0.1747) & 0.1396 & (0.1174, 0.1640) \\
				\midrule
				\textbf{SVM-\textit{t}} \\
				$\beta_{0}$ & 0.1434 & (0.1097, 0.1817) & 0.1428 & (0.1092, 0.1775) & 0.1820 & (0.1063, 0.2645) & 0.1815 & (0.1131, 0.2501) \\
				$\beta_{1}$ & -0.0289 & (-0.0560, -0.0011) & -0.0284 & (-0.0561, -0.0013) & -0.0234 & (-0.0534, 0.0045) & -0.0231 & (-0.0500, 0.0057) \\
				$\beta_{2}$ & -0.0690 & (-0.1025, -0.0349) & -0.0687 & (-0.1002, -0.0388) & -0.0502 & (-0.0933, -0.0137) & -0.0503 & (-0.0831, -0.0186) \\
				$\mu$       & 0.0878  & (-0.2029, 0.3932) & 0.0576 & (-0.2719, 0.3821) & 0.7089  & (0.5152, 0.8733) & 0.7019 & (0.4921, 0.9106) \\
				$\phi$      & 0.9843  & (0.9776, 0.9898) & 0.9847 & (0.9773, 0.9914) & 0.9800  & (0.9719, 0.9869) & 0.9802 & (0.9718, 0.9875) \\
				$\sigma$    & 0.1655  & (0.1399, 0.1963) & 0.1653 & (0.1336, 0.1982) & 0.1333  & (0.1106, 0.1562) & 0.1327 & (0.1100, 0.1531) \\
				$\nu$       & 11.6877 & (8.2895, 16.6556) & 11.7709 & (7.7560, 18.4397) & 24.2335 & (15.8754, 34.6290) & 23.6160 & (14.5021, 33.8353) \\
				\midrule
				\textbf{SVM-S} \\
				$\beta_{0}$ & 0.1426 & (0.1059, 0.1747) & 0.1419 & (0.1063, 0.1779) & 0.1816 & (0.1096, 0.2469) & 0.1822 & (0.1096, 0.2539) \\
				$\beta_{1}$ & -0.0261 & (-0.0542, 0.0035) & -0.0268 & (-0.0552, 0.0001) & -0.0245 & (-0.0575, 0.0037) & -0.0236 & (-0.0507, 0.0040) \\
				$\beta_{2}$ & -0.0876 & (-0.1259, -0.0455) & -0.0871 & (-0.1297, -0.0462) & -0.0573 & (-0.0970, -0.0216) & -0.0587 & (-0.0970, -0.0214) \\
				$\mu$       & -0.1596 & (-0.4624, 0.1339) & -0.1783 & (-0.5014, 0.1445) & 0.5743  & (0.3319, 0.8702) & 0.5546 & (0.3113, 0.7869) \\
				$\phi$      & 0.9836  & (0.9763, 0.9893) & 0.9843 & (0.9772, 0.9908) & 0.9797  & (0.9703, 0.9874) & 0.9804 & (0.9714, 0.9880) \\
				$\sigma$    & 0.1723  & (0.1426, 0.2038) & 0.1690 & (0.1410, 0.1985) & 0.1361  & (0.1102, 0.1635) & 0.1327 & (0.1071, 0.1607) \\
				$\nu$       & 2.9597  & (2.3054, 3.8507) & 2.8913 & (2.3029, 3.7691) & 5.4271  & (3.5119, 10.0480) & 5.3848 & (3.2888, 11.4683) \\
				\midrule
				\textbf{SVM-VG} \\
				$\beta_{0}$ & 0.1429 & (0.0981, 0.1759) & 0.1430 & (0.1083, 0.1794) & 0.1842 & (0.1105, 0.2549) & 0.1846 & (0.1161, 0.2589) \\
				$\beta_{1}$ & -0.0303 & (-0.0583, -0.0055) & -0.0298 & (-0.0562, -0.0031) & -0.0242 & (-0.0523, 0.0037) & -0.0233 & (-0.0510, 0.0036) \\
				$\beta_{2}$ & -0.0548 & (-0.0821, -0.0282) & -0.0554 & (-0.0807, -0.0294) & -0.0470 & (-0.0802, -0.0182) & -0.0471 & (-0.0780, -0.0174) \\
				$\mu$       & 0.2469  & (-0.0677, 0.5282) & 0.2562 & (-0.0730, 0.5935) & 0.7910  & (0.5843, 0.9923) & 0.7957 & (0.6104, 0.9890) \\
				$\phi$      & 0.9852  & (0.9786, 0.9912) & 0.9856 & (0.9791, 0.9915) & 0.9803  & (0.9710, 0.9882) & 0.9800 & (0.9713, 0.9875) \\
				$\sigma$    & 0.1600  & (0.1324, 0.1896) & 0.1598 & (0.1351, 0.1870) & 0.1335  & (0.1110, 0.1580) & 0.1338 & (0.1117, 0.1574) \\
				$\nu$       & 7.8058  & (5.4705, 11.1987) & 8.0856 & (5.5496, 12.6144) & 23.3856 & (14.4321, 32.2793) & 21.3955 & (11.8300, 33.1472) \\
				\bottomrule
			\end{tabular}
		}
		\label{tab9}
	\end{table}
	In general, the values obtained are consistent with those commonly reported in the literature. Moreover, it should be noted that the results of both methods are very similar in all applications, in terms of both posterior means and 95\% credibility intervals. This finding underscores the reliability of the estimates produced by the HMM method.
	
	To assess the performance in-sample of the models under the HMM machinery, we computed the Log Predictive Score \citep[LPS;][]{Delatola2011} and the Deviance Information Criterion \citep[DIC;][]{Spiegelhalter2002}. In both cases, the preferred model corresponds to the one with the smallest LPS (DIC) value. In addition to the SVM models mentioned earlier, we also fitted SV models using HMM machinery.
	% info criteria
	\begin{table}[h!]
		\caption{Model selection criteria comparison between SV and SVM models. Values in bold indicate the preferred specification according to the LPS and DIC criteria based on HMM estimation.}
		\centering
		\scalebox{1.0}{
			\begin{tabular}{ l l l l l l l l l }
				\toprule
				& \multicolumn{2}{c}{S\&P 500} & \multicolumn{2}{c}{NIKKEI 225} & \multicolumn{2}{c}{DAX 30} & \multicolumn{2}{c}{IBOVESPA} \\
				\cmidrule(l){2-3} \cmidrule(l){4-5} \cmidrule(l){6-7} \cmidrule(l){8-9}
				Model & DIC & LPS & DIC & LPS & DIC & LPS & DIC & LPS\\
				\midrule
				SV-N    & 44047.90 & 4.3178 & 46062.36 & 4.6437 & 47104.78 & 4.5807 & 48592.92 & 4.8461 \\
				SV-t    & 44037.39 & 4.3166 & 46051.48 & 4.6425 & 47073.46 & 4.5775 & 48591.35 & 4.8458 \\
				SV-S    & 44040.35 & 4.3171 & 46052.08 & 4.6426 & 47081.64 & 4.5783 & 48588.30 & 4.8455 \\
				SV-VG   & 44032.16 & 4.3162 & 46049.98 & 4.6424 & 47066.19 & 4.5767 & 48592.43 & 4.8459 \\
				\midrule
				SVM-N   & 43985.22 & 4.3111 & 46033.34 & 4.6402 & 47061.97 & 4.5759 & 48581.46 & 4.8443 \\
				SVM-\textit{t}   & 43974.63 & 4.3099 & 46025.28 & 4.6393 & 47031.57 & 4.5728 & 48580.28 & 4.8441 \\
				SVM-S   & 43978.83 & 4.3103 & 46026.64 & 4.6393 & 47040.87 & 4.5737 & \textbf{48577.27} & \textbf{4.8437} \\
				SVM-VG  & \textbf{43970.51} & \textbf{4.3095} & \textbf{46024.11} & \textbf{4.6392} & \textbf{47023.59} & \textbf{4.5720} & 48581.14 & 4.8442 \\
				\bottomrule
			\end{tabular}
		}
		\label{tab10}
	\end{table}
	From Table \ref{tab10}, we observe that the SVM-VG model provides the best fit for most of the considered series, according to both criteria. An exception is the IBOVESPA series, for which the SVM-S model provides a superior fit.
	
	Next, we illustrate the efficiency of the HMM approach relative to classical MCMC, as shown in the applications. Table \ref{tab11} reports the computational times for both methods, measured on the same machine used in the first simulation study.
	\begin{table}[h!]
		\centering
		\caption{Elapsed time for model fitting (minutes).}
		\begin{tabular}{ l l l l l l l l l }
			\toprule
			& \multicolumn{2}{c}{S\&P 500}  & \multicolumn{2}{c}{NIKKEI 225} & \multicolumn{2}{c}{DAX 30}  & \multicolumn{2}{c}{IBOVESPA}\\
			\cmidrule(l){2-3} \cmidrule(l){4-5} \cmidrule(l){6-7} \cmidrule(l){8-9}
			Model & HMM & MCMC & HMM & MCMC & HMM & MCMC & HMM & MCMC\\
			\midrule
			SVM-N   & 2.04  & 23.91 & 2.28  & 23.36  & 2.13  & 23.97 & 2.14  & 23.72  \\
			SVM-\textit{t}   & 2.78  & 33.00 & 2.72  & 32.14  & 2.61  & 32.99 & 2.22  & 32.38  \\
			SVM-S   & 2.98  & 38.52 & 2.78  & 37.46  & 2.75  & 38.46 & 2.98  & 38.18  \\
			SVM-VG  & 3.05  & 41.53 & 3.25  & 40.92  & 3.02  & 42.76 & 3.29  & 41.89  \\
			\bottomrule
		\end{tabular}
		\label{tab11}
	\end{table}
	Overall, the HMM method is approximately ten times faster than the MCMC approach in all applications. From a practical perspective, this highlights the considerable advantage of adopting the HMM framework as an alternative to traditional MCMC schemes.
	
\begin{figure}[h!]
		\centering
		\includegraphics[width=1.0\textwidth]{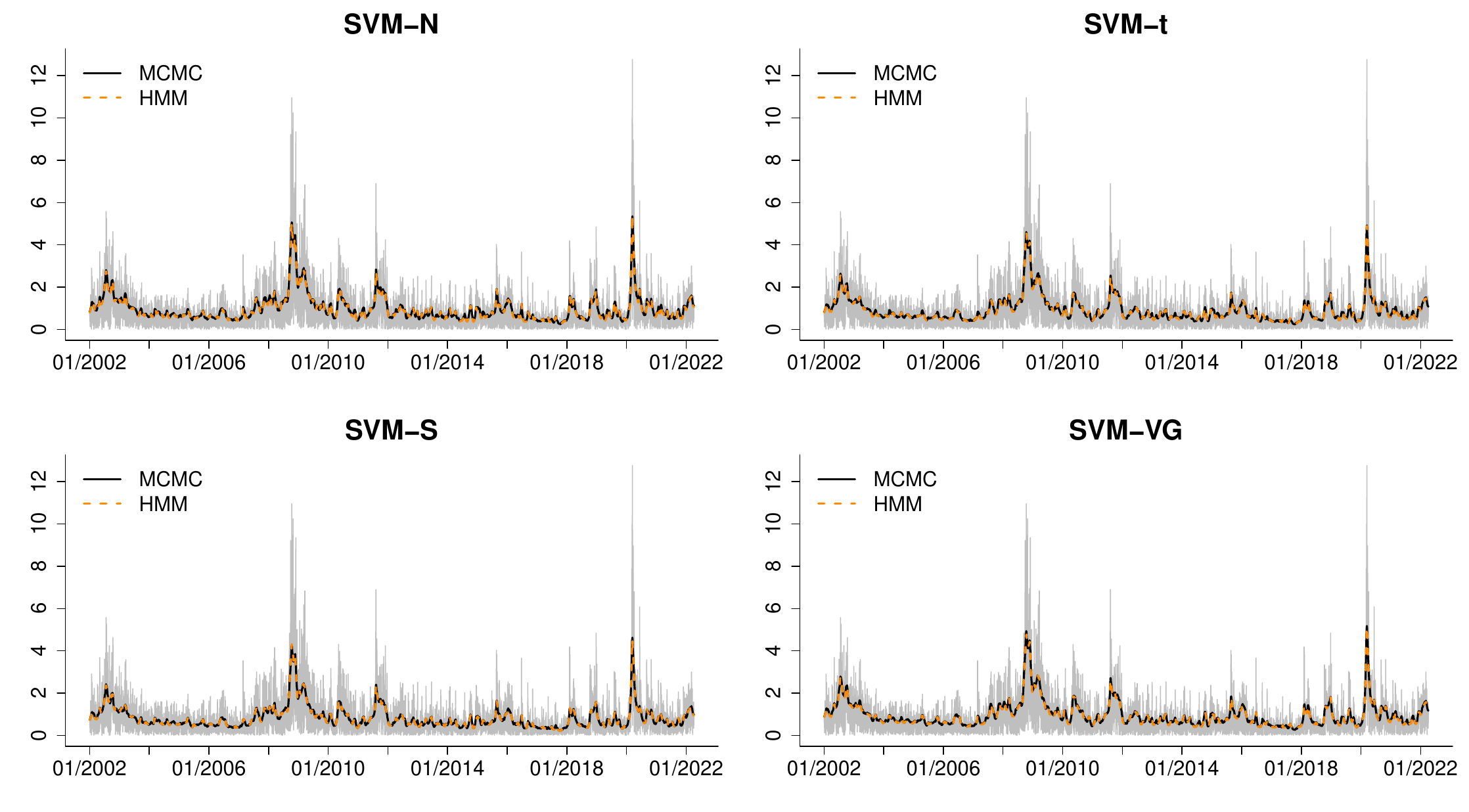}
		\caption{Absolute returns (full gray lines) of the S\&P 500 indices and the corresponding $\exp(h_t/2)$ estimates derived from the MCMC (full black lines) and HMM (dotted orange lines) methods.}
		\label{fig1}
	\end{figure}

	Figure \ref{fig1} shows, for the analyzed S\&P 500 series, the absolute returns (solid gray line), the estimates of $\exp(h_t/2)$ from the SVM models obtained via the HMM approximation (dotted orange lines), and the smoothed means of $\exp(h_t/2)$ from the SVM models obtained using the MCMC sampler (solid black lines).
	% Volatilities
	% sp500
	For the HMM-based approach, the latent log-volatility process is estimated via global decoding using the Viterbi algorithm, which identifies the most probable sequence of hidden states jointly over the entire sample. Further details are provided in the Supplementary Material. The figures show that the volatilities estimated using the HMM method closely overlap with those obtained via MCMC across all models considered. Similar results are observed for the remaining applications; however, they are omitted here due to space constraints. This further demonstrates the accuracy of the approximations provided by the proposed approach.
	
	% Forecasting
	\subsection{Out-of-sample Analysis}
	
	Finally, we conduct an out-of-sample analysis to evaluate the forecasting performance of the SVM models considered, following the procedure described in the Supplementary Material. The validation period spans from April 7, 2022, to January 1, 2025. The Jarque–Bera (JB) and Ljung–Box (LB) tests were applied to the one-step-ahead forecast pseudo-residuals obtained for the different indices and SVM specifications, in order to assess, respectively, the assumptions of normality and serial independence. The corresponding p-values are reported in Table \ref{tab12}.
	% Jarque-Bera test
	\begin{table}[h!]
		\centering
		\caption{The p-values of Jarque-Bera (JB) and Ljung-Box (LB) tests applied to one-step ahead forecast pseudo-residuals (LB test with 10 lags). Bold numbers indicate p-values $<0.05$.}
		\scalebox{1.0}{
			\begin{tabular}{ l cc cc cc cc }
				\toprule
				& \multicolumn{2}{c}{SVM-N} & \multicolumn{2}{c}{SVM-\textit{t}} & \multicolumn{2}{c}{SVM-S} & \multicolumn{2}{c}{SVM-VG} \\
				\cmidrule(lr){2-3} \cmidrule(lr){4-5} \cmidrule(lr){6-7} \cmidrule(lr){8-9}
				Index & JB & LB & JB & LB & JB & LB & JB & LB \\
				\midrule
				S\&P 500    & 0.10 & 0.48 & 0.08 & 0.44 & 0.09 & 0.44 & 0.07 & 0.43 \\
				NIKKEI 225  & \textbf{0.02} & 0.96 & \textbf{0.04} & 0.96 & 0.06 & 0.97 & \textbf{0.04} & 0.96 \\
				DAX 30      & 0.07 & 0.85 & 0.11 & 0.83 & 0.12 & 0.83 & 0.09 & 0.83 \\
				IBOVESPA    & 0.69 & 0.35 & 0.44 & 0.35 & 0.60 & 0.34 & 0.44 & 0.34 \\
				\bottomrule
			\end{tabular}
		}
		\label{tab12}
	\end{table}
	Overall, both tests are not rejected at the 5\% significance level for most series and model specifications, with the exception of the normality test for the NIKKEI 225 under the SVM-N, SVM-\textit{t}, and SVM-VG models. 
	
\begin{figure}[h!]
		\centering
		\begin{subfigure}[b]{1.0\textwidth}
			\centering\includegraphics[width=1.0\textwidth]{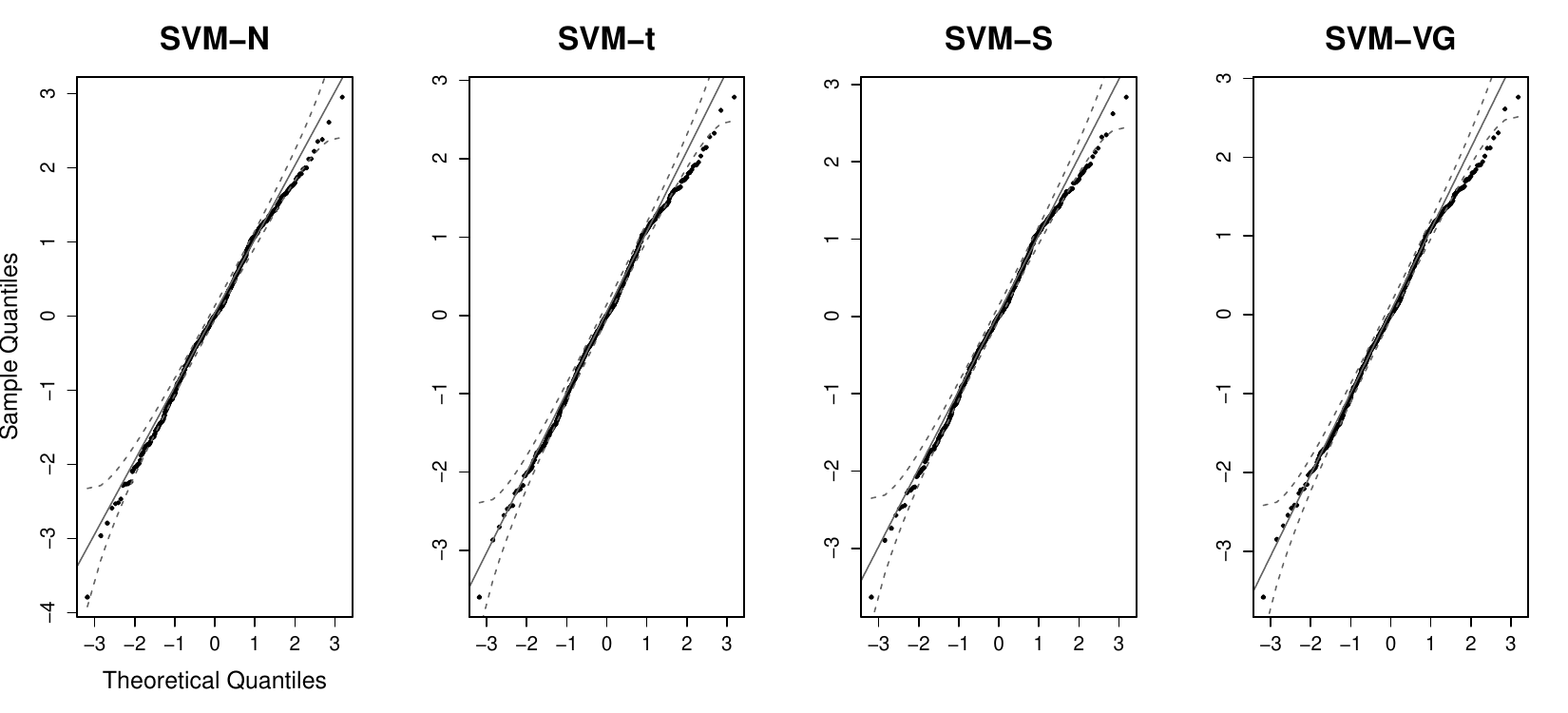}
			\caption{S\&P 500 residuals.}
		\end{subfigure}
		\\
		\begin{subfigure}[b]{1\textwidth}
			\centering
			\includegraphics[width=1.0\textwidth]{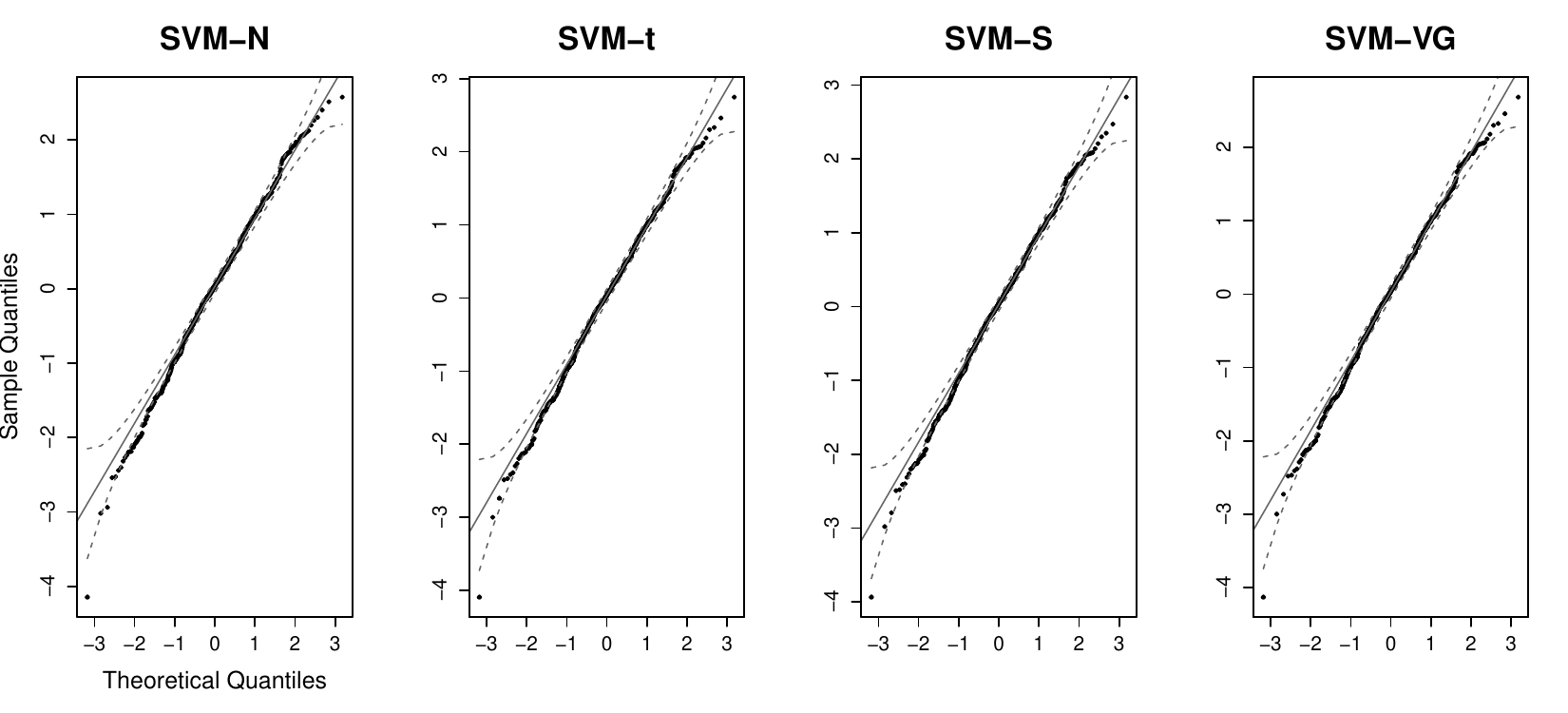}
			\caption{NIKKEI 225 residuals.}
		\end{subfigure}
		\caption{QQ-plots of the forecast pseudo-residuals for the S\&P 500 and NIKKEI 225 indices. Dashed lines denote the 95\% simulated confidence envelopes.}
		\label{fig2}
	\end{figure}

\begin{figure}[h!]
		\centering
		\begin{subfigure}[b]{1\textwidth}
			\centering
			\includegraphics[width=1\textwidth]{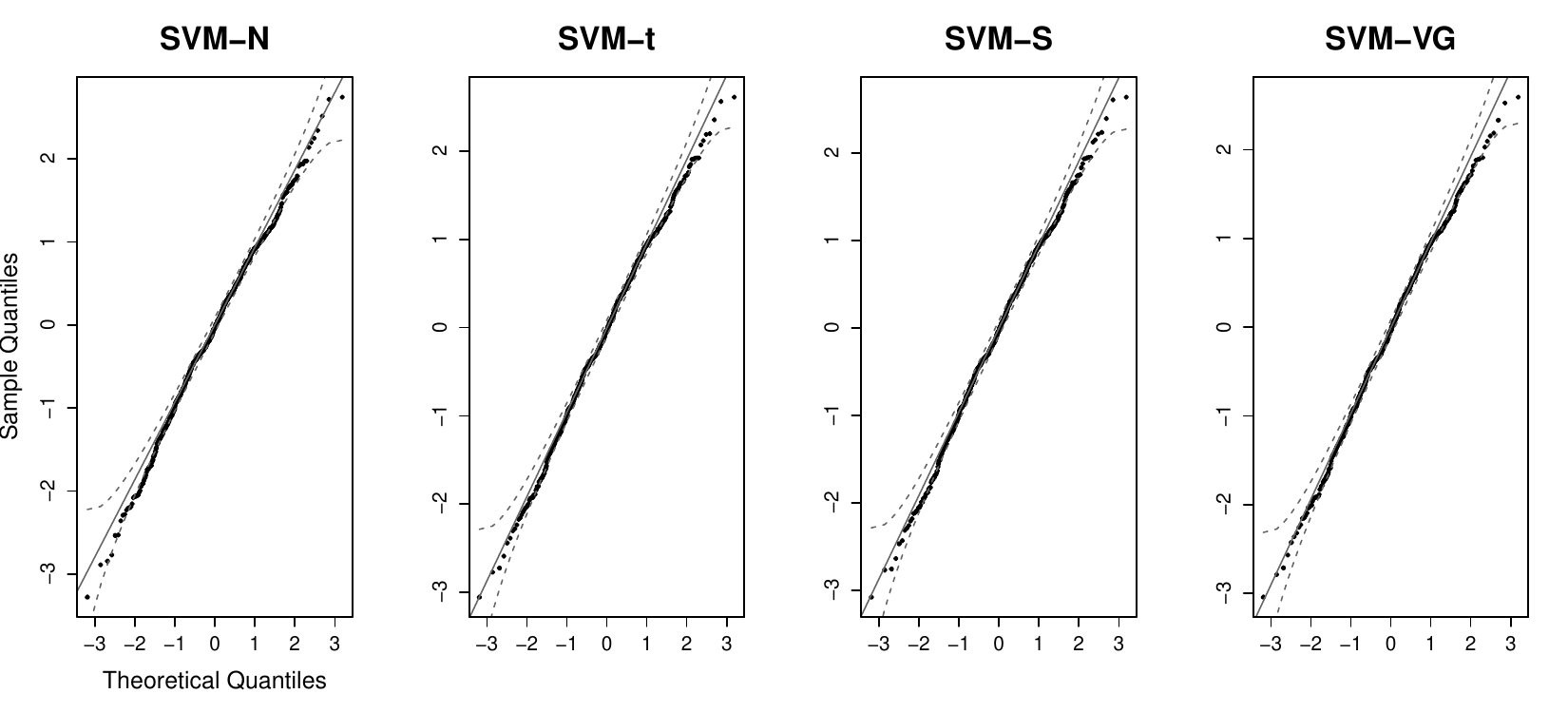}
			\caption{DAX 30 residuals.}
		\end{subfigure}
		\\
		\begin{subfigure}[b]{1\textwidth}
			\centering
			\includegraphics[width=1\textwidth]{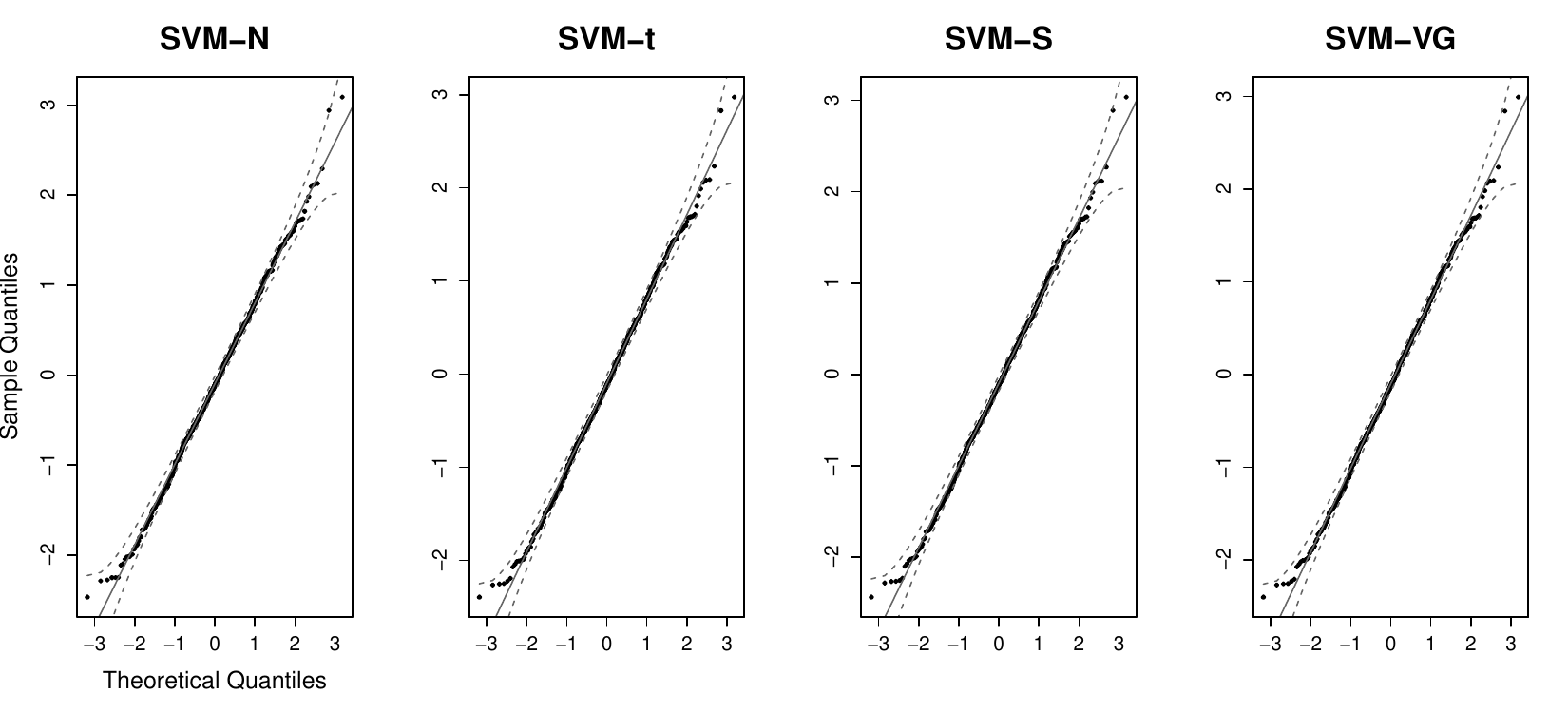}
			\caption{IBOVESPA residuals.}
		\end{subfigure}
		\caption{QQ-plots of the forecast pseudo-residuals for the DAX 30 and IBOVESPA indices. Dashed lines denote the 95\% simulated confidence envelopes.}
		\label{fig3}
	\end{figure}

	% QQ-plots
	The QQ-plots for the four return series under the SVM-N, SVM-\textit{t}, SVM-S, and SVM-VG models are shown in Figures \ref{fig2} and \ref{fig3}, respectively. The plots incorporate 95\% simulated confidence envelopes.
	% Forecasts Residuals
	% sp500 and nikkei
	% dax and ibovespa
	Figure \ref{fig2} reveals that for the S\&P500, the residuals under SVM-\textit{t}, SVM-S and SVM-VG systematically diverge from the upper-tail boundaries, suggesting a specification error likely arising from the unmodeled leverage effect (correlation between return innovations and volatility shocks). A similar failure to capture tail dynamics is observed in the lower tail of all NIKKEI 225 models. In contrast, Figure~\ref{fig3} indicates that the DAX 30 and IBOVESPA series are well accommodated, with virtually all observations falling within the simulated envelopes across the four model specifications.
	
	When assessing the risk associated with a financial asset or index, particular attention must be given to the extreme left tail of the predictive distribution. A widely used tool for this purpose is the Value at Risk (VaR), a standard measure in risk management. VaR summarizes the potential loss of a financial position resulting from adverse market movements and represents the worst expected loss over a given time horizon $N$ at a specified confidence level $\alpha \in (0,1)$. 
	
	To evaluate the accuracy of VaR forecasts, we employ the unconditional coverage (UC) test of \citet{Kupiec1995}, a standard approach in risk model assessment. This likelihood ratio test verifies whether the empirical violation rate matches the nominal level $\alpha$. Specifically, define a violation of the left-tail as $V^{\alpha}_t = 1$, if $y_t < \widehat{VaR}^{\alpha}_t$ and $V^{\alpha}_t = 0$ otherwise. The total number of violations is then given by $n_0 = \sum_{t=1}^{N} V^{\alpha}_t$. Under the null hypothesis of unconditional coverage, the test statistic
	\begin{equation*}
		LR_{UC} = -2 \log \left\{  (1- \alpha)^{N-n_0} \alpha^{n_0}  \right\} - 2 \log \left\{ (1- \hat{\alpha})^{N-n_0} \hat{\alpha}^{n_0} \right \},
	\end{equation*}
	follows a $\chi^2_1$ distribution, where $\hat{\alpha} = n_0/N$ denotes the empirical violation rate. 
	
	In addition, we examine the independence of violations using the test proposed by \citet{christoffersen1998evaluating}, which assesses whether violations exhibit temporal clustering. Let $n_{ij}$ denote the number of occurrences of outcome $j \in \{0,1\}$ conditional on outcome $i \in \{0,1\}$ on the previous day. Under the null hypothesis of independence, the test statistic,
	\begin{equation*}
		LR_{IND} = -2 \log \left\{ (1-\pi)^{n_{00}+n_{10}} \pi^{n_{01}+n_{11}}  \right\} - 2 \log \left \{(1-\pi_{0})^{n_{00}+n_{10}} \pi_{0}^{n_{01}} \pi_{1}^{n_{11}} \right\},
	\end{equation*}
	is distributed as $\chi^2_1$. In this framework, $n_{ij}$ is the number of times where outcome $j$, i.e a violation or no violation, occurred on a day conditional on outcome $i$ occurring on the previous day,  $\pi_{0}=n_{01}/(n_{00}+n_{01})$, $\pi_{1}=n_{11}/(n_{10}+n_{00})$ and $\pi=(n_{01}+n_{11})/(n_{00}+n\_{01}+n_{10}+n_{11})$.
	
	Finally, to jointly assess unconditional coverage and independence, we consider the conditional coverage (CC) test defined as $LR_{CC} = LR_{UC} + LR_{IND}$, which follows a $\chi^2_2$ distribution under the null hypothesis.
	
	\begin{table}[ht]
		\centering
		\caption{Empirical violation rate $\hat{\alpha}$, p-values of the unconditional coverage ($LR_{UC}$), independence ($LR_{IND}$) and conditional coverage ($LR_{CC}$) tests. Bold numbers indicate p-values $<0.05$.}
		\scalebox{1.0}{
			\begin{tabular}{llccccccccc}
				\toprule
				&&&\multicolumn{4}{c}{$\alpha=0.01$} &\multicolumn{4}{c}{$\alpha=0.05$}\\
				\cmidrule(l){4-7} \cmidrule(l){8-11}
				\textbf{Index} & \textbf{Model} &N& $\hat{\alpha}$ & $LR_{UC}$ & $LR_{IND}$ & $LR_{CC}$ & $\hat{\alpha}$ & $LR_{UC}$ & $LR_{IND}$ & $LR_{CC}$\\
				\midrule
				\multirow{4}{*}{S\&P 500}   
				& SVM-N  & 687 & 0.01 & 0.96 & 0.70 & 0.93 & 0.06 & 0.42 & 0.60 & 0.63 \\
				& SVM-\textit{t}  &  & 0.01 & 0.96 & 0.70 & 0.93 & 0.06 & 0.25 & 0.72 & 0.49 \\
				& SVM-S  &  & 0.01 & 0.96 & 0.70 & 0.93 & 0.06 & 0.19 & 0.78 & 0.41 \\
				& SVM-VG &  & 0.01 & 0.96 & 0.70 & 0.93 & 0.06 & 0.25 & 0.72 & 0.49 \\
				\midrule
				\multirow{4}{*}{NIKKEI 225} 
				& SVM-N  & 671 & 0.01 & 0.91 & 0.06 & 0.16 & 0.05 & 0.78 & 0.63 & 0.86 \\
				& SVM-\textit{t}  &  & 0.01 & 0.91 & 0.06 & 0.16 & 0.05 & 0.94 & 0.83 & 0.97 \\
				& SVM-S  &  & 0.01 & 0.91 & 0.06 & 0.16 & 0.05 & 0.94 & 0.83 & 0.97 \\
				& SVM-VG &  & 0.01 & 0.91 & 0.06 & 0.16 & 0.05 & 0.94 & 0.83 & 0.97 \\
				\midrule
				\multirow{4}{*}{DAX 30}     
				& SVM-N  & 698 & 0.01 & 0.99 & 0.71 & 0.93 & 0.06 & 0.30 & 0.77 & 0.56 \\
				& SVM-\textit{t}  &  & 0.01 & 0.70 & 0.75 & 0.88 & 0.06 & 0.23 & 0.76 & 0.47 \\
				& SVM-S  &  & 0.01 & 0.70 & 0.75 & 0.88 & 0.06 & 0.23 & 0.76 & 0.47 \\
				& SVM-VG &  & 0.01 & 0.70 & 0.75 & 0.88 & 0.06 & 0.30 & 0.70 & 0.54 \\
				\midrule
				\multirow{4}{*}{IBOVESPA}   
				& SVM-N  & 682 & 0.00 & \textbf{0.01} & 0.96 & \textbf{0.02} & 0.05 & 0.71 & 0.65 & 0.84 \\
				& SVM-\textit{t}  &     & 0.00 & \textbf{0.01} & 0.96 & \textbf{0.02} & 0.05 & 0.85 & 0.60 & 0.85 \\
				& SVM-S  &     & 0.00 & \textbf{0.01} & 0.96 & \textbf{0.02} & 0.05 & 0.71 & 0.65 & 0.84 \\
				& SVM-VG &     & 0.00 & \textbf{0.01} & 0.96 & \textbf{0.02} & 0.05 & 0.85 & 0.60 & 0.85 \\
				\bottomrule
			\end{tabular}
		}
		\label{tab13}
	\end{table}
	
Table \ref{tab13} reports the empirical violations rates and the p-values of the UC, IND, and CC backtest procedures for the VaR forecasts at the significance levels of 1\% and 5\%. Overall, the results indicate satisfactory VaR performance in most indices and model specifications, with empirical violations generally close to their nominal levels. For S\&P 500, NIKKEI 225, and DAX 30, none of the unconditional or conditional coverage tests are rejected at the significance level of 5\%, suggesting that the proposed models adequately capture both the frequency and temporal dependence of extreme losses. In contrast, for the IBOVESPA at the 1\% level, the UC and CC tests are rejected across all models, reflecting an overestimation of extreme tail risk in this case. At the 5\% level, however, all models perform satisfactorily for the IBOVESPA, with no evidence against correct coverage or independence of violations.

Finally, it should be noted that although the VaR backtests reject the models for the IBOVESPA at the 1\% risk level, the calibration test (see Table~\ref{tab12}) and the QQ-plots (see Figure~\ref{fig3}) indicate a satisfactory global calibration for the predictive distributions. Consequently, rather than a structural misspecification, this result suggests a localized difficulty in capturing challenging extreme events, since at the 5\% level the models remain well-calibrated. This localized difficulty can be explained by the omission of the leverage effect, which is an alternative to the feedback effect for modeling the asymmetry that impacts the tail of the predictive distribution.

\section{Conclusions}\label{discuss}

This paper makes three complementary contributions to the modeling of financial time series using stochastic volatility models. First, by allowing the observation distribution to belong to the Scale Mixture of Normals (SMN) family, we extend the HMM-based methodology proposed by \citet{Abanto2024} to a broad class of heavy-tailed stochastic volatility-in-mean models. Second, we embed the resulting HMM approximation within an approximate Bayesian inference framework based on importance sampling, providing a computationally efficient alternative to conventional MCMC-based methods. Third, by combining parallel computing techniques with analytical representations based on special functions, we obtain substantial improvements in numerical stability and computational efficiency, leading to more reliable estimation and significantly reduced computing times.

The empirical evidence presented in this paper indicates that the proposed HMM approximation yields posterior summaries that closely agree with those obtained using exact MCMC methods, while requiring only a fraction of the computational effort---approximately one-tenth of the execution time in the applications considered. Moreover, the simulation experiments suggest that the approximate likelihood stabilizes rapidly as the discretization grid is refined. From a theoretical perspective, the HMM likelihood can be interpreted as a quadrature approximation to the exact likelihood, and a heuristic justification of its convergence properties is provided in the Supplementary Material. Taken together, these findings suggest that the proposed methodology offers an attractive compromise between computational efficiency and inferential accuracy for the Bayesian analysis of SVM models.

To facilitate reproducibility and encourage further methodological developments, all computational routines developed in this study have been made publicly available through the development version of the \texttt{R} package \texttt{svHMM}, which can be accessed from the GitHub repository: \url{https://github.com/svHMM/svHMM.git}. We believe that the availability of efficient and user-friendly software is particularly valuable for practitioners and applied researchers, who are often interested in obtaining reliable results within a reasonable computational time.

Several avenues for future research remain open. One natural extension is to incorporate the leverage effect into the proposed framework, allowing for a more realistic representation of the asymmetric relationship between returns and volatility, as discussed by \citet{hiraki2025stochastic}. Another promising direction is the systematic comparison of the proposed HMM-based approximate Bayesian methodology with alternative approximate inference techniques, such as Variational Inference, in order to assess their relative performance in terms of approximation accuracy, robustness, and computational efficiency.

\section*{Acknowledgments}

The authors would like to thank the EULER cluster team of the CeMEAI project (FAPESP grant 2013/07375-0) for their support, which enabled part of the computational simulations conducted in this study.

\section*{Funding Statement}
	
The research of Bruno E. Holtz was financed by the Coordenação de Aperfeiçoamento de Pessoal de Nível Superior - Brasil (CAPES) - Finance Code 001. The second author acknowledges partial ﬁnancial support from Fundação Carlos Chagas Filho de Amparo à Pesquisa do Estado do Rio de Janeiro (FAPERJ). Ricardo Ehlers acknowledges financial support from the São Paulo Research Foundation (FAPESP), under grant 2025/06569-2 and from the National Council for Scientific and Technological Development (CNPq), under grant 409630/2024-1.

\section*{Data Availability Statement}

The data used in this work were sourced from Yahoo Finance and are available on the GitHub repository: \url{https://github.com/svHMM/Data}.
        
\section*{Competing Interests}
The authors declare that they have no potential conflict of interest.

\bibliographystyle{unsrtnat}
\bibliography{references}  %%% Uncomment this line and comment out the ``thebibliography'' section below to use the external .bib file (using bibtex) .





% \end{thebibliography}

\end{document}

\maketitle

\begin{abstract}
	This supplementary material provides extended technical documentation and numerical evidence supporting the findings of the main manuscript. It reviews the theoretical background on volatility asymmetry and details the recursive techniques used in HMMs. Additionally, it includes analytical derivations for the probability density functions within the SMN family. Finally, this document offers a comprehensive description of the MCMC algorithm implemented for model comparison, accompanied by detailed convergence diagnostics.
\end{abstract}

% keywords can be removed
\keywords{Stochastic Volatility in Mean \and Scale Mixture of Normals \and Hidden Markov Models \and Bayesian Estimation \and Parallel Computing.}

\section*{Theoretical Background on Volatility Asymmetry}
	
	%Asymmetry
	The asymmetric link between equity returns and market volatility has been a focal point in financial econometric research \citep{Black1976,Campbell1992,Bekaert2000}. 
	This relationship describes a phenomenon where the magnitude of volatility response depends on the sign of the preceding return. Specifically, empirical evidence suggests that negative shocks are associated with significantly larger increases in volatility than positive returns of an equivalent magnitude.
	This asymmetry is of critical importance for several reasons. First, it captures an essential feature of volatility behavior and has direct consequences for asset pricing, as it reflects systematic risk components that are priced by the market. Second, accounting for volatility asymmetry is vital in practical domains such as risk management, derivative pricing, and the development of effective hedging strategies. Third, the presence of asymmetries helps explain the negative skewness often observed in asset return distributions, shedding light on the mechanisms behind extreme market downturns.
	
	% Feedback and Leverage effects
	In order to explain the asymmetry, extensive empirical research has also focused on the link between expected returns and volatility. 
	% Feedbach effect
	Although it is intuitively sound that the relationship between expected returns (expected risk premium) and expected volatility (\textit{ex-ante} volatility) is positive, since rational risk-averse investors require higher expected returns during more
	volatile periods, empirical evidence has remained largely inconsistent. Instead, a consistent finding across the literature is the negative correlation between unexpected returns and unexpected volatility shocks. \cite{French1987} interpreted this as indirect evidence supporting a positive link between the expected risk premium and \textit{ex ante} volatility. They argue that large and unexpected shocks in the return process, which can be caused by good or bad news, induce greater expected volatility for future periods. Assuming that expected volatility and expected returns are positively related and that future cash flows are unaffected, this expectation of increased risk necessitates an immediate reduction in current prices to accommodate the higher required risk premium, in a context of persistent volatility. Consequently, this mechanism generates a contemporaneous negative correlation between returns and volatility. On the other hand, small shocks in the return process lead to an increase in contemporary stock index prices. This phenomenon is commonly referred to as the volatility feedback effect.
	% Leverage effect
	An alternative explanation, known as the leverage effect and originally proposed by \cite{Black1976}, attributes this asymmetry to changes in a company's financial structure. When there is bad news that lowers the price and therefore increases the debt-to-equity ratio (i.e., financial leverage), this makes the company riskier and tends to increase expected future volatility. As a result, the leverage effect should correspond to a negative relationship between volatility and price.
	
	The volatility feedback and leverage effects offer competing explanations for the direction of causality regarding the asymmetry. While the volatility feedback theory attributes the decline in prices to an increase in expected volatility, the leverage effect posits that the initial fall in prices triggers a subsequent rise in future volatility. Empirical analyses by \cite{French1987} and \cite{Schwert1989} found that leverage alone is insufficient to explain the extent of the observed asymmetry. Consequently, \cite{Campbell1992} provides support for the joint presence of both effects. Whereas \cite{Bekaert2000} and \cite{Wu2001} argue that the volatility feedback effect empirically dominates, other seminal studies, such as \cite{Nelson1991}, \cite{Glosten1993}, and \cite{Engle1993}, observe that volatility responds more strongly to negative shocks than to positive ones. Furthermore, these authors find the interplay between expected returns and volatility to be statistically insignificant.
	
	\subsection*{SVM Framework for Capturing the Volatility Feedback Effect}
	
	As mentioned in the main manuscript, $\beta_2$ accounts for both the ex-ante relationship between returns and volatility and the volatility feedback effect. This dual role becomes evident by rewriting the observational equation as
	\begin{equation*}
		y_{t} = \beta_{0} + \beta_{1}y_{t-1} + \beta_{2}e^{h_{t \mid t-1}} + \beta_{2}(e^{h_t} - e^{h_{t \mid t-1}}) + e^{\frac{h_{t}}{2}}\epsilon_{t}.
	\end{equation*}
	This decomposition reveals that $\beta_2$ scales two distinct elements: the expected volatility component, $e^{h_{t \mid t-1}}$, which represents the conditional variance given information up to $t-1$; and the unexpected component, $e^{h_t} - e^{h_{t \mid t-1}}$, which reflects the stochastic shock to the volatility process independent of predictable factors. Theoretically, the former is expected to be positive (rational investor hypothesis), whereas the latter, the volatility feedback effect, is typically expected to be negative. This negative relationship arises because large shocks (assuming they are independent of the sign of $\epsilon_t$) increase contemporaneous volatility through $\eta_t$, thereby inducing a drop in current prices. This return adjustment is further amplified when initial returns are strongly negative, yet it is attenuated when they are strongly positive or only slightly negative. In contrast, smaller shocks tend to reduce volatility, exerting a positive impact on current prices. These price movements are amplified in the presence of slightly positive initial returns, but are similarly attenuated when initial returns are either very positive or slightly negative.
	
	\section*{Hidden Markov Models}
	\label{sec:HMM}
	
	Hidden Markov Models (HMM) are one of the most successful statistical modeling ideas to emerge in recent years. The use of unobservable (or hidden) states makes the model generic enough to handle a variety of complex real-world time series, while the relatively simple prior dependency structure (grounded in the Markov property) still allows the use of efficient computational procedures. This section briefly introduces HMMs, as well as some of their computational techniques. For a more detailed reading on HMM, we recommend the references \cite{cappe2005inference} and \cite{Zucchini2016}.
	
	A HMM is a bivariate discrete-time process $\{X_t, Y_t\}_{t \geq 0}$, where $\{X_t\}$ is a Markov chain with discrete state space and, conditional on $\{X_t\}$, $\{Y_t\}$ is a sequence of independent random variables such that the conditional distribution of $Y_t$ depends only on $X_t$, i.e.,
	\begin{eqnarray}
		P \left(Y_t \mid \bm{Y}_{t-1}, \bm{X}_{t} \right) &=& P(Y_t \mid X_t) \nonumber \\
		P \left( X_t \mid \bm{X}_{t-1} \right) &=& P(X_t \mid X_{ t-1 }),
		\label{hmmeqs}
	\end{eqnarray}
	where $\bm{Y}_{t}=(Y_1, \dots, Y_t)'$ and $\bm{X}_{t}=(X_1, \dots, X_t)'$, for $t \in \mathbb{N}_{+}$.
	
	If the Markov chain $\{X_t\}$ has $m$ states, the transition matrix $\bm{\Gamma} = (\gamma_{ij})$ is defined by $\gamma_{ij} = P(X_t=j \mid X_{t-1}=i)$. The vector $\bm{\delta}$ represents the initial distribution of the chain, i.e., the distribution of $X_0$. Furthermore, consider the conditional probability function
	\begin{equation*}
		p_i(y) = P(Y_t = y \mid X_t = i),
	\end{equation*}
	where $i, j = 1, \dots, m$. Together, these elements characterize an HMM. It should be noted that the probabilities in (\ref{hmmeqs}) may depend on a parameter vector $\bm{\theta}$ (i.e., $P = P_{\bm{\theta}}$). To avoid overloading the notation, this dependency will be suppressed unless an explicit declaration is required.
	
	\subsection*{The HMM Likelihood Function}
	
	Consider an observed series $\bm{y}_{T} = (y_1, \dots, y_T)'$ originating from a HMM. The likelihood of the model can be expressed as
	\begin{equation}
		\mathcal{L}_T(\bm{y}_T; \bm{\theta}) = \bm{\delta} \bm{P}(y_1) \bm{\Gamma} \bm{P}(y_2) \cdots \bm{\Gamma} \bm{P}(y_T) \bm{1}',
		\label{hmmlik_new}
	\end{equation} 
	where $\bm{P}(y_t)$ represents a diagonal matrix of dimension $m \times m$ containing the conditional probabilities $p_i(y_t)$ and ${\bm 1}'$ is a column vector of ones. This matrix formulation underlies the so-called "forward algorithm," a recursive procedure essential for the efficient calculation of likelihood. The importance of this recursion extends beyond parameter estimation, being indispensable for inference tasks such as state prediction, decoding, and model checking. 
	
	Unlike the exhaustive summation over all possible state paths, the recursive approach drastically reduces the computational load. When defining the forward probability vector $\bm{\alpha}_t$, we have 
	\begin{equation}
		\bm{\alpha}_t = \bm{\delta} \bm{P}(y_1) \bm{\Gamma} \bm{P}(y_2) \cdots \bm{\Gamma} \bm{P}(y_t).
		\label{forward}
	\end{equation}
	In this way, the total likelihood is obtained by $\mathcal{L}_T(\bm{y}_T; \bm{\theta}) = \bm{\alpha}_T \bm{1}'$, where the vector update follows the relation $\bm{\alpha}_t = \bm{\alpha}_{t-1} \bm{\Gamma} \bm{P}(y_t)$ for $t = 1, \dots, T$, starting from the initial condition $\bm{\alpha}_0 = \bm{\delta}$. 
	
	The computational cost to evaluate the equation (\ref{hmmlik_new}) is on the order of $\mathcal{O}(Tm^2)$, which makes maximization strategies viable even on large datasets. In practice, numerical maximization of parameters requires attention to challenges such as numerical underflow. To mitigate these limitations, a scaled version of the forward algorithm is routinely used. Technical details on these implementations can be found at \cite{Zucchini2016}.

\subsection*{A heuristic justification for the HMM approximation}

The HMM approximation employed in this work can be interpreted as a multidimensional
quadrature approximation to the exact likelihood of the stochastic volatility model.
Let
\[
L(\bm{\theta})
=
\int_{\mathbb{R}^{T}}
p(\bm{y}_{T}\mid \bm{h}_{T},\bm{\theta})
p(\bm{h}_{T}\mid\bm{\theta})
\,d\bm{h}_{T}
\]
denote the exact likelihood, where
$\bm{h}_{T}=(h_{1},\ldots,h_{T})'$.
The HMM methodology replaces the continuous latent state space by a finite partition
of the interval $[-A,A]$ into $m$ subintervals, yielding the approximate likelihood
$L_{m,A}(\bm{\theta})$ evaluated through the forward algorithm.

The approximation error can be decomposed as
\[
|L(\bm{\theta})-L_{m,A}(\bm{\theta})|
\leq
|L(\bm{\theta})-L_{A}(\bm{\theta})|
+
|L_{A}(\bm{\theta})-L_{m,A}(\bm{\theta})|,
\]
where $L_{A}(\bm{\theta})$ denotes the likelihood obtained by restricting the integration
to the compact region $[-A,A]^{T}$. The first term corresponds to the truncation error,
which vanishes as $A\rightarrow\infty$ under the integrability of the complete-data
density. The second term is the quadrature error associated with replacing the continuous
integral over $[-A,A]^{T}$ by a finite-state approximation. Under standard regularity
conditions, namely continuity of the integrand and boundedness on compact sets, this
error converges to zero as the mesh size of the partition tends to zero (equivalently,
as $m\rightarrow\infty$).

Consequently, for any fixed sample size $T$ and parameter vector $\bm{\theta}$,
\[
L_{m,A}(\bm{\theta})
\longrightarrow
L(\bm{\theta}),
\]
as $A\rightarrow\infty$ and $m\rightarrow\infty$. Therefore, the HMM approximation
may be viewed as a consistent quadrature approximation of the exact likelihood.
This interpretation is closely related to the discretization approach developed by
\cite{Langrock2012} for hidden Markov approximations of state-space models.

	\subsection*{Forecasting, Model Checking, and Decoding}The predictive distribution of an HMM for a forecast horizon $h \geq 0$ can be expressed as
	\begin{equation}
		P\left(Y_{t+h} = y \mid \bm{y_{t-1}}\right) = \sum_{i=1}^{m} \xi_i(h)p_i(y),
		\label{forecast}
	\end{equation}
	where $\xi_i(h)$ is the $i$-th entry of the vector $\bm\phi_{t-1} \bm{\Gamma}^{h}$, with $\bm\phi_{t-1} = \bm{\alpha}_{t-1} / \bm{\alpha}_{t-1}\bm{1}'$ as defined in (\ref{forward}). Consequently, model checking can be performed through residual analysis based on the forecast distribution in equation (\ref{forecast}) \citep{Kim1998}. Specifically, the one-step-ahead forecast pseudo-residuals (quantile residuals) are defined as
	\begin{equation*}
		r_t = \Phi^{-1}\left\{P \left(Y_t \leq y_t \mid \bm{y_{t-1}} \right) \right\}
	\end{equation*}
	for $t = 1, \dots, T$. Under the assumption that the model is correctly specified, these residuals $r_t$ should follow independent realizations of a standard normal distribution \citep{Smith1985, Kim1998}. This property enables pseudo-residuals to serve as an effective diagnostic tool for identifying outliers and verifying the overall model fit. In practice, validation is conducted through visual inspection of QQ-plots or by applying formal tests for normality and independence.
	
	In many practical applications, the primary interest lies in identifying the hidden state trajectory that best explains the observed data. Formally, the goal is to determine the sequence of states $\bm{X}_{T} = (i_1, \dots, i_T)'$ that maximizes the conditional probability
	\begin{equation*}
		P \left(\bm{X}_{T} \mid \bm{Y}_{T} \right), \quad i_j \in \{1, \dots, m\}.
		\label{viterbi}
	\end{equation*}
	This procedure, known as global decoding, poses a significant combinatorial challenge, since an exhaustive search of all $m^T$ possible state configurations becomes computationally prohibitive as $T$ increases. To address this, the Viterbi algorithm is employed, a dynamic programming technique that efficiently solves the optimization problem. Comprehensive details on the implementation and derivation of this algorithm can be found in \cite{Zucchini2016}.
		
	\section*{Technical Details of the SMN Densities}
	
	As described in the main manuscript, if $Y \sim \mathcal{SMN}(\mu, \sigma, \nu)$, then
	\begin{equation*}
		f(y|\mu, \sigma, \nu) = \int_{0}^{\infty} \mathcal{N}(y; \mu, \sigma/ \lambda)h(\lambda|\nu)d\lambda.
	\end{equation*}
	where $\lambda$ is the mixing variable, which depends on $\nu>0$.
	
	In this section, we present key considerations that ensure the robustness of the computational procedures employed in this work. Similarly to the case of the Student’s t-distribution, whose marginal density can be expressed using the gamma function, $\Gamma(\cdot)$, we show that the Slash and Variance Gamma distributions can also be represented through special functions. Without loss of generality, we set $\mu = 0$ and $\sigma = 1$ in the following derivations:
	\begin{itemize}
		\item \textbf{Slash Distribution.} Suppose that $Y$ follows a Slash distribution with tail parameter $\nu$. Its density is given by\begin{equation*}
			f(y \mid \nu) = \frac{\nu}{\sqrt{2\pi}} \int_{0}^{1} \lambda^{\nu-1/2} \exp \left\{ - \frac{\lambda y^{2}}{2} \right\} d\lambda.\end{equation*}By applying the change of variable $t = \lambda y^{2}/2$, we obtain:\begin{align*}
			f(y \mid \nu) &= \frac{\nu}{\sqrt{2\pi}} \left( \frac{2}{y^{2}} \right)^{\nu + 1/2} \int_{0}^{y^{2}/2} t^{\nu+1/2 - 1} e^{-t} dt \\
			&= \frac{\nu}{\sqrt{2\pi}} \left( \frac{2}{y^{2}} \right)^{\nu + 1/2} \gamma\left( \nu+\frac{1}{2}, \frac{y^{2}}{2}\right), \quad \text{for } y \neq 0,
		\end{align*}
		where $\gamma(s, x) = \int_0^{x} t^{s-1} e^{-t} dt$ is the lower incomplete gamma function. To handle the singularity at $y = 0$, let $s = \nu + 1/2$ and $x = y^{2}/2$. The limit as $y \to 0$ is given by:
		\begin{equation*}
			\lim_{y \to 0} f(y \mid \nu) = \lim_{x \to 0^{+}} \frac{\nu}{\sqrt{2\pi}} \frac{\gamma(s, x)}{x^{s}}.
		\end{equation*}
		Since $\lim_{x \to 0^{+}} \gamma(s, x) = 0$ and $\lim_{x \to 0^{+}} x^{s} = 0$, applying L'Hôpital's rule and the Fundamental Theorem of Calculus yields:
		\begin{equation*}
			\lim_{x \to 0^{+}} \frac{\gamma(s, x)}{x^{s}} = \lim_{x \to 0^{+}} \frac{x^{s-1} e^{-x}}{s x^{s-1}} = \frac{1}{s}.
		\end{equation*}
		Therefore, the density of the slash distribution can be computed as:
		\begin{equation*}
			f(y \mid \nu) = \begin{cases}
				\frac{\nu}{\sqrt{2\pi}} \left( \frac{2}{y^{2}} \right)^{\nu + 1/2} \gamma\left( \nu+\frac{1}{2}, \frac{y^{2}}{2}\right), & y \neq 0, \\
				\frac{\nu}{\sqrt{2\pi} (\nu + 1/2)}, & y \rightarrow 0.
			\end{cases}
		\end{equation*}
		
		\item \textbf{Variance Gamma Distribution.} \cite{fischer2025variance} provides a comprehensive review of the literature on the Variance Gamma distribution. In particular, they present a hierarchical representation based on the gamma distribution, $S \sim \mathcal{G}(r/2, 1/2)$ for $r>0$, such that $Y = \sigma_r \sqrt{S} X \sim VG(r, \sigma_r)$ when $X \sim N(0,1)$ e $\sigma_r > 0$. By leveraging the relationship between the Gamma and Inverse Gamma distributions, this hierarchical representation can be reconciled with the Variance Gamma distribution parameterization by mapping $r = \nu$ and $\sigma_r = 1/\nu$. Consequently, for $\nu >1$, we obtain
		\begin{equation*}
			f(y \mid \nu) = \begin{cases}
				\sqrt{ \frac{\nu}{\pi} } \frac{1}{2^{\frac{\nu-1}{2}} \Gamma(\frac{\nu}{2})} \left( \sqrt{\nu} |y| \right)^{\frac{\nu-1}{2}} K_{\frac{\nu-1}{2}}(\sqrt{\nu}|y|), & y \neq 0, \\
				\frac{1}{2} \sqrt{ \frac{\nu}{\pi} } \frac{\Gamma(\frac{\nu-1}{2})}{\Gamma(\frac{\nu}{2})}, & y \rightarrow 0,
			\end{cases}
		\end{equation*}
		where $K_{\nu}(x)$ is a modified Bessel function of the second kind.
		
	\end{itemize}
	
	It is worth highlighting that, in addition to the parallelization strategy, a significant portion of the computational efficiency achieved in this work is due to the use of the special functions presented in this section. Rather than relying on generic numerical integration methods, the evaluation of the Slash and Variance Gamma densities was performed using the lower incomplete gamma function, $\gamma(\cdot, \cdot)$, and the modified Bessel function of the second kind, $K_{\nu}(\cdot)$, respectively. These functions are highly optimized and provided by the \textit{GNU Scientific Library} (GSL) via the \texttt{RcppGSL} package, enabling fast and numerically stable likelihood evaluations.

	\section*{Technical Details of the MCMC Sampler}
	
	This section provides technical details regarding the MCMC inference procedures and additional empirical results for the SVM-SMN models discussed in the main manuscript. The methodology presented here serves as an extension of the framework proposed in \citet{Abanto2021}. While the aforementioned work introduced the use of HMC and RMHMC for Stochastic Volatility in Mean (SVM) models with Gaussian innovations, the current study extends this approach to the Scale Mixture of Normals (SMN) class. This extension allows for greater flexibility in capturing stylized facts of financial time series, such as heavy tails, through the implementation of $t$-Student, Slash, and Variance Gamma distributions within the Hamiltonian Monte Carlo framework. The following sections detail the joint posterior distribution, the block-sampling MCMC algorithm, prior specifications, and full posterior summaries for all empirical applications.
	
	\subsection*{Dynamic Hamiltonian}
	\label{HMC}
	
	Shortly after the introduction of the stochastic particle simulation method by \citet{metropolis1953equation}, \citet{alder1959studies} described a deterministic approach based on Newton’s laws of motion. These two paradigms were later unified by \citet{duane1987hybrid} into what was initially termed 'Hybrid Monte Carlo'. In subsequent statistical literature, \citet{neal1996} rebranded the method as 'Hamiltonian Monte Carlo' (HMC) to emphasize its foundation in Hamiltonian dynamics.To illustrate the mechanics of HMC, consider the simulation of a univariate continuous random variable $\theta \in \mathbb{R}$ with probability density $p(\theta)$. Let the potential energy be defined by the negative log-density, $U(\theta) = -\log p(\theta)$. Conceptually, imagine a frictionless particle moving along the surface defined by $U(\theta)$. Starting from a region of low potential (high probability), the particle moves toward higher elevations with a velocity $v=p/m$, where $m$ denotes the mass and $p$ represents the auxiliary linear momentum. As the particle ascends the 'potential well', its kinetic energy $K(p)$ is converted into potential energy $U(\theta)$ until its velocity vanishes. At this vertex, the particle reverses direction, accelerating back toward the high-density regions as potential energy is reconverted into kinetic energy, thus maintaining the total Hamiltonian (energy) of the system.This physical analogy demonstrates how the system dynamics are exploited to efficiently explore the parameter space. By utilizing the gradient of the log-density, $\nabla_{\theta} \log p(\theta)$, the HMC method systematically directs the sampler toward regions of high probability mass. Consequently, even when the particle visits low-density areas, the trajectory is informed by the geometry of the distribution, allowing for distant proposals that maintain high acceptance rates. 
	
	Hamiltonian dynamics provides a powerful reformulation of classical mechanics, offering a profound perspective on the geometric structures inherent in closed and conservative systems \citep{leimkuhler2004simulating}. Let $\bm{\theta} \in \mathbb{R}^{d}$ denote the position of a particle and $\mathbf{p} \in \mathbb{R}^{d}$ its corresponding linear momentum. The state of the system is thus defined within the phase space $(\bm{\theta}, \mathbf{p}) \in \mathbb{R}^{d} \times \mathbb{R}^{d}$. For such a system, the Hamiltonian function $H(\bm{\theta}, \mathbf{p})$, representing the total energy, is defined as
	\begin{equation}
		H(\bm{\theta}, \mathbf{p}) = U(\bm{\theta}) + K(\mathbf{p}),
	\end{equation}
	where $U(\bm{\theta})$ is the potential energy and $K(\mathbf{p})$ is the kinetic energy. Typically, the kinetic energy is specified as a quadratic form
	\begin{equation}
		K(\mathbf{p}) = \frac{1}{2} \mathbf{p}^{\top} \mathbf{M}^{-1} \mathbf{p},
	\end{equation}
	where $\mathbf{M}$ is the mass matrix (often referred to as the metric in statistical contexts). The temporal evolution of the system is governed by Hamilton's equations
	\begin{equation}
		\frac{d \bm{\theta}}{d t} = \frac{\partial H}{\partial \mathbf{p}} = \nabla_{\mathbf{p}} K(\mathbf{p}), \quad \frac{d \mathbf{p}}{d t} = - \frac{\partial H}{\partial \bm{\theta}} = - \nabla_{\bm{\theta}} U(\bm{\theta}).
	\end{equation}
	These equations define a volume-preserving and reversible flow $T_s$ that maps the state $(\bm{\theta}(t), \mathbf{p}(t))$ to $(\bm{\theta}(t+s), \mathbf{p}(t+s))$ for any time interval $s > 0$ \citep{neal2011mcmc}.
	
	%Discretization: The Leapfrog Method
	
	In practice, particularly when $\bm{\theta}$ and $\mathbf{p}$ are high-dimensional, the analytical solutions to Hamilton's equations are rarely available, necessitating numerical integration. Although Euler's method is a standard approach for discretizing time into $L$ steps of size $\epsilon$ (where $s = L\epsilon$), it fails to preserve the Hamiltonian structure over long periods. A superior alternative for conservative systems is the Störmer-Verlet method, commonly known as the leapfrog integrator. This method is favored in MCMC applications because it is both time-reversible and symplectic (volume-preserving), properties that are fundamental to ensure the detailed balance of the resulting Markov chain \citep{leimkuhler2004simulating}. Given an initial state at $t_0$ and a step size $\epsilon > 0$, a single leapfrog iteration is described by
	\begin{align}
		\mathbf{p}(t_0 + \epsilon/2) &= \mathbf{p}(t_0) - \frac{\epsilon}{2} \nabla_{\bm{\theta}} U(\bm{\theta}(t_0)), \label{eqlf1} \\
		\bm{\theta}(t_0 + \epsilon) &= \bm{\theta}(t_0) + \epsilon \mathbf{M}^{-1} \mathbf{p}(t_0 + \epsilon/2), \label{eqlf2} \\
		\mathbf{p}(t_0 + \epsilon) &= \mathbf{p}(t_0 + \epsilon/2) - \frac{\epsilon}{2} \nabla_{\bm{\theta}} U(\bm{\theta}(t_0 + \epsilon)). 
		\label{eqlf3}
	\end{align}
	Note that in (\ref{eqlf2}), the update to position $\bm{\theta}$ uses the momentum at the half-step, creating the leapfrog effect that maintains numerical stability and energy conservation.
	
	\subsection*{The Hamiltonian Monte Carlo Method}
	
	The core intuition of HMC is to integrate concepts from Hamiltonian physics into statistical sampling, utilizing the analogy of a particle moving within a potential energy field to explore the parameter space. This approach effectively mitigates the limitations of traditional MCMC algorithms, such as the Metropolis-Hastings random walk, which often struggles in high-dimensional settings \citep{betancourt2017conceptual}.
	
	Let $\bm{\theta} \in \mathbb{R}^{d}$ be the vector of parameters of interest with posterior density $p(\bm{\theta} | \mathbf{x})$. We introduce an auxiliary momentum variable $\mathbf{p} \sim N_{d}(\mathbf{0}, \mathbf{M})$, independent of $\bm{\theta}$, where $\mathbf{M}$ is a user-defined mass matrix. The joint density $p(\bm{\theta}, \mathbf{p})$ is then defined as
	\begin{equation}
		p(\bm{\theta}, \mathbf{p}) \propto \exp(-H(\bm{\theta}, \mathbf{p})),
	\end{equation}
	where the Hamiltonian energy function $H(\bm{\theta}, \mathbf{p})$ is given by
	\begin{equation}
		H(\bm{\theta}, \mathbf{p}) = -\mathcal{L}(\bm{\theta}) + \frac{1}{2} \log\left\{ (2\pi)^{d} |\mathbf{M}| \right\} + \frac{1}{2} \mathbf{p}^{\top}\mathbf{M}^{-1}\mathbf{p},
	\end{equation}
	and $\mathcal{L}(\bm{\theta}) = \log p(\bm{\theta} | \mathbf{x})$ represents the log-posterior. As demonstrated by \citet{neal2011mcmc}, a Markov chain with invariant density $p(\bm{\theta} | \mathbf{x})$ can be constructed using these dynamics. Starting from $(\bm{\theta}^{(i)}, \mathbf{p}^{(i)})$, a candidate $(\bm{\theta}^{*}, \mathbf{p}^{*})$ is proposed via the leapfrog integrator (Equations \ref{eqlf1}-\ref{eqlf3}) over $L$ steps of size $\epsilon$. Although the leapfrog method is a discretization, it leaves the Hamiltonian approximately invariant ($H^{(i)} - H^{*} \approx 0$). Consequently, the Metropolis acceptance step,
	\begin{equation}
		\alpha = \min \left\{ 1, \exp\left( H(\bm{\theta}^{(i)}, \mathbf{p}^{(i)}) - H(\bm{\theta}^{*}, \mathbf{p}^{*})\right)\right\},
	\end{equation}
	is used to correct the discretization bias \citep{girolami2011riemann}. 
	
	A significant advantage of HMC is that it eliminates the need for a manual proposal distribution $q(\cdot)$ and suppresses random-walk behavior. By appropriately tuning $L$ and $\epsilon$, the chain can traverse high-probability regions efficiently, leading to faster convergence. However, identifying the optimal parameters remains a non-trivial task \citep{nugroho2015estimation}. While perfect simulation would yield an acceptance rate of $1$, the use of approximations makes the acceptance rate sensitive to $\epsilon$, with the global error being $O(\epsilon^2)$. While small $\epsilon$ values optimize acceptance, they increase the computational cost by requiring a larger $L$ to reach distant points. Conversely, an excessively large $L$ may lead to 'u-turn' trajectories that return to the starting point, while an $L$ that is too small fails to escape the local neighborhood, reintroducing unwanted random-walk characteristics. Finally, the choice of the mass matrix $\mathbf{M}$ (or covariance) is critical. $\mathbf{M}$ scales the momentum and dictates the geometry of the exploration; an ill-conditioned matrix can render the sampler highly inefficient. Given the vast space of possible configurations for $\mathbf{M}$, finding an optimal choice for a specific target distribution is often impractical without more advanced techniques, such as those provided by Riemannian geometry \citep{betancourt2017conceptual}.

	\subsection*{The Hamiltonian Monte Carlo Method on Riemannian Manifolds}
	
	To address the limitations of a fixed mass matrix, \citet{girolami2011riemann} introduced a significant refinement to the HMC framework by exploiting the Riemannian geometry of the parameter space. By allowing the mass matrix $\mathbf{M}(\bm{\theta})$ to vary with the position, the RMHMC method accounts for the local curvature of the target density. This adaptation is particularly effective for distributions with complex geometries, such as those exhibiting varying scales or strong correlations. The fundamental reformulation of the Hamiltonian function is given by
	\begin{equation}
		H(\bm{\theta}, \mathbf{p}) = -\mathcal{L}(\bm{\theta}) + \frac{1}{2} \log\left\{ (2\pi)^{d} |\mathbf{M}(\bm{\theta})| \right\} + \frac{1}{2} \mathbf{p}^{\top}\mathbf{M}(\bm{\theta})^{-1}\mathbf{p},
	\end{equation}where $\mathcal{L}(\bm{\theta})$ denotes the log-posterior density.Unlike the standard HMC, the Hamiltonian is no longer separable, meaning the kinetic energy now depends on the position $\bm{\theta}$. Consequently, Hamilton's equations for the evolution of the system become:
	\begin{equation}
		\frac{d \bm{\theta}}{d t} = \frac{\partial H}{\partial \mathbf{p}} = \mathbf{M}(\bm{\theta})^{-1}\mathbf{p},
	\end{equation}
	and
	\begin{equation}\frac{d \mathbf{p}}{d t} = -\frac{\partial H}{\partial \bm{\theta}} = \nabla_{\bm{\theta}}\mathcal{L}(\bm{\theta}) - \frac{1}{2} \text{tr}\left[\mathbf{M}(\bm{\theta})^{-1}\frac{\partial \mathbf{M}(\bm{\theta})}{\partial \bm{\theta}}\right] + \frac{1}{2} \mathbf{p}^{\top} \mathbf{M}(\bm{\theta})^{-1} \frac{\partial \mathbf{M}(\bm{\theta})}{\partial \bm{\theta}} \mathbf{M}(\bm{\theta})^{-1}\mathbf{p}.
		\label{gleapfrog_refined}
	\end{equation}
	
	As specified in Equation (\ref{gleapfrog_refined}), the dynamics involve the calculation of higher-order derivatives, which significantly increases the computational demand per iteration. Furthermore, the non-separability of the Hamiltonian requires the use of an implicit numerical integrator to maintain symplecticity. The Generalized Leapfrog method is typically employed, requiring fixed-point iterations to solve the implicit updates for both position and momentum \citep{leimkuhler2004simulating}. Despite the higher cost per step, this sampling strategy yields a highly efficient, reversible, and ergodic Markov chain. A primary advantage of RMHMC is the elimination of manual tuning for the mass matrix $\mathbf{M}$, as the metric is automatically adjusted at each step based on the expected Fisher Information or a similar geometric construct. This allows the sampler to adaptively navigate the "warped" regions of the posterior, significantly improving mixing and convergence rates in challenging models \citep{girolami2011riemann}.
	
	Under the Bayesian framework adopted in this study, the metric matrix $\mathbf{M}(\bm{\theta})$ is constructed to incorporate the curvature of the joint log-probability of the data and the parameters, $p(\mathbf{y}, \bm{\theta})$. Formally, $\mathbf{M}(\bm{\theta})$ is defined by the expected Hessian of the negative log-joint density
	\begin{align}
		\mathbf{M}(\bm{\theta}) &= -\mathbb{E}_{\mathbf{y}|\bm{\theta}} \left[\frac{\partial^2}{\partial \bm{\theta}^2} \log p(\mathbf{y}, \bm{\theta}) \right] \nonumber \\
		&= -\mathbb{E}_{\mathbf{y}|\bm{\theta}} \left[\frac{\partial^2}{\partial \bm{\theta}^2} \left( \log p(\mathbf{y} | \bm{\theta}) + \log p(\bm{\theta}) \right) \right] \nonumber \\
		&= \mathbf{I}(\bm{\theta}) - \mathbf{H}_{\bm{\theta}},
		\label{Metric_refined}
	\end{align}
	where $\mathbf{I}(\bm{\theta}) = -\mathbb{E}_{\mathbf{y}|\bm{\theta}} \left[\nabla_{\bm{\theta}}^2 \log p(\mathbf{y}|\bm{\theta})\right]$ represents the Expected Fisher Information matrix, and $\mathbf{H}_{\bm{\theta}} = \nabla_{\bm{\theta}}^2 \log p(\bm{\theta})$ is the Hessian of the log-prior. 
	
	In the RMHMC framework, $\mathbf{M}(\bm{\theta})$ functions as a Riemannian metric tensor. For the Hamiltonian dynamics to be well-defined, $\mathbf{M}(\bm{\theta})$ must be positive definite for all $\bm{\theta}$ in the parameter space. Since the Fisher Information matrix is, by construction, positive semi-definite (and typically positive definite for identifiable models), a sufficient condition for the positive definiteness of $\mathbf{M}(\bm{\theta})$ is that the log-prior be concave, ensuring that $-\mathbf{H}_{\bm{\theta}}$ is positive definite. The sum of these two matrices then guarantees a valid metric for the manifold. To enhance sampling efficiency, HMC-based methods are ideally implemented over unconstrained parameter spaces, $\bm{\theta} \in \mathbb{R}^d$. Consequently, parameters with restricted supports—such as the volatility scale $\sigma > 0$—are transformed using appropriate reparameterizations, such as $\gamma = \log(\sigma)$. A notable property of this geometric approach is that the Fisher Information matrix is equivariant under such coordinate transformations. Specifically, if $\mathbf{I}(\bm{\theta})$ is positive definite, the transformed metric $\mathbf{I}(\bm{\theta}(\phi))$ will also preserve this property, ensuring the stability of the RMHMC sampler across different parameterizations \citep{amari2000methods}. To ensure the numerical stability and proper functioning of the RMHMC, it is essential that the transformed Hessian, $-\mathbf{H}_{\bm{\theta}(\phi)}$, remains positive definite throughout the sampling process. 
	
	Consider $\bm{\theta} \in \mathbb{R}^{d}$ as the parameter vector of interest with a prior density $p(\bm{\theta})$. Assuming the components of $\bm{\theta}$ are independent, the joint prior is given by
	\begin{equation}p(\bm{\theta}) = \prod_{i=1}^{d} p_{i}(\theta_{i}).
	\end{equation}
	Under this independence assumption, the Hessian of the log-prior, $\mathbf{H}_{\bm{\theta}}$, simplifies to a diagonal matrix
	\begin{align}
		-\mathbf{H}_{\bm{\theta}} &= \left[ -\frac{\partial^{2}}{\partial \theta{j} \partial \theta_{i}} \sum_{k=1}^{d} \log p_{k}(\theta_{k}) \right]_{(ij)} \nonumber \\
		&= -\text{diag}\left( \frac{\partial^{2}}{\partial \theta_{1}^{2}}\log p_{1}(\theta_{1}), \dots, \frac{\partial^{2}}{\partial \theta_{d}^{2}}\log p_{d}(\theta_{d}) \right).
		\label{hessian_diagonal}
	\end{align}
	The positive definiteness of such a matrix can be verified using Sylvester’s Criterion: A symmetric $n \times n$ matrix $\mathbf{A}$ is positive definite if and only if all its leading principal minors (the determinants of the upper-left $r \times r$ submatrices $\mathbf{A}_{r}$) are strictly positive for $r = 1, \dots, n$ \citep{seber2012linear}. Given the diagonal structure established in Equation (\ref{hessian_diagonal}), the determinant of each principal minor $\mathbf{H}_{r}$ is simply the product of the first $r$ diagonal elements
	\begin{equation*}
		\text{det}(\mathbf{H}_{r}) = \prod_{i=1}^{r} \left\{ -\frac{\partial^{2}}{\partial \theta_{i}^{2}}\log p_{i}(\theta_{i}) \right\}, 
	\end{equation*}
	for $r=1, \dots, d$.
	
	According to Sylvester's Criterion, $-\mathbf{H}_{\bm{\theta}}$ is positive definite if and only if all diagonal elements are strictly positive. Statistically, this condition is satisfied whenever the prior densities $p_{i}(\theta_{i})$ are log-concave. This observation is crucial when selecting priors or performing reparameterizations ($-\mathbf{H}_{\bm{\theta}(\phi)}$), as it guarantees that the prior information contributes positively to the metric tensor's stability, preventing the Hamiltonian integrator from encountering non-invertible mass matrices.

	\subsection*{Stochastic Volatility in Mean Model with SMN Innovations}
	\label{SVMSMN}
	
	The Stochastic Volatility in Mean model with SMN innovations (SVM-SMN) is defined by
	\begin{align}
		&y_{t} = \beta_{0} + \beta_{1}y_{t-1} + \beta_{2}e^{h_{t}} + e^{h_{t}/2} \lambda_{t}^{-1/2} \epsilon_{t}, \quad \epsilon_{t} \sim N(0,1), \label{eq.1svm_smn} \\ 
		&h_{t+1} = \mu + \phi(h_{t} - \mu) + \sigma_{h}\eta_{t}, \quad \eta_{t} \sim N(0,1), \label{eq.2svm_smn} \\    
		&h_{1} \sim N\left( \mu, \frac{\sigma_{h}^{2}}{1-\phi^{2}} \right), \label{eq.3svm_smn} \\
		&\lambda_{t} \sim p(\lambda_{t} | \nu) \label{eq.4svm_smn}, 
	\end{align}
	where $y_{t}$ is the compound return and $h_{t}$ is the log-volatility at time $t$. The parameters $\mu$ and $\sigma$ represent the mean and standard deviation of the process defined by equation (\ref{eq.2svm_smn}), respectively. For this process to be stationary, it is assumed that the modulus of the persistence parameter, $\phi$, is strictly less than 1, that is, $|\phi| < 1$. The innovations $\epsilon_{t}$ and $\eta_{t}$ are considered serially and mutually independent. The parameter $\beta_{0}$ represents the mean of the series, $\beta_{1}$ the first-order autocorrelation of the returns, and $\beta_{2}$ measures the relationship of the returns with the expected and unexpected volatilities, as well as the feedback effect.

	\subsection*{Parameter Estimation}
	\label{section:estimacao}
	
	Let $\bm{\theta} = (\beta_0, \beta_1, \beta_2, \mu, \phi, \sigma, \nu)'$ be the parameter vector of the SVM-SMN model, $\bm{h}_{1:T} = (h_{1}, ..., h_{T})'$ be the log-volatilities vector, $\bm{\lambda_{1:T}}$ be the mixture variables and $\bm{y}_{0:T} = (y_{0}, ..., y_{T} )'$ be the information available up to time $T$. The joint posterior distribution of the parameters and latent variables can be written as:
	\begin{align}
		p(\bm{h}_{1:T}, \bm{\lambda_{1:T}}, \bm{\theta} | \bm{y}_{0:T}) &\propto p(\bm{y}_{1:T} | y_{0}, \bm{h}_{1:T}, \bm{\lambda_{1:T}}, \bm{\theta})p(\bm{h}_{1:T}| \bm{\theta})p(\bm{\lambda_{1:T}}|\bm{\theta})p(\bm{\theta}) \nonumber \\
		&= \prod_{t=1}^{T} p(y_{t}|y_{t-1}, \beta_{0}, \beta_{1}, \beta_{2}, h_{t}, \lambda_{t}) p(\lambda_{t}|\nu) \times \nonumber \\ 
		& \times p(h_{1}|\mu, \phi, \sigma^{2})\prod_{t=2}^{T}p(h_{t}| h_{t-1}, \mu, \phi, \sigma^{2})p(\bm{\theta}), \label{poster}
	\end{align}
	where $p(y_{t}|y_{t-1}, \beta_{0}, \beta_{1}, \beta_{2}, h_{t}, \lambda_{t})$, $p(h_{1}| \mu, \phi, \sigma^{2})$ and $p(h_{t}| h_{t-1}, \mu, \phi, \sigma^{2})$ are given by (\ref{eq.1svm_smn}), (\ref{eq.2svm_smn}) and (\ref{eq.3svm_smn}), respectively. $p(\lambda_{t}|\nu)$ is the density of the mixture variable and $p(\bm{\theta})$ is the prior of the model.
	
	To conduct the Bayesian inference for the SVM-SMN model, we generate posterior realizations from the joint distribution (\ref{poster}) using the MCMC methods detailed in Section HMC. Given that the joint posterior does not belong to a known parametric family and exhibits high-dimensional dependencies, we employ a sampling scheme that alternates between the model parameters and the latent variables. Specifically, we utilize a Metropolis-within-Gibbs strategy where the parameters $\bm{\theta}$ and the log-volatility vector $\bm{h}_{1:T}$ are updated via Hamiltonian dynamics (HMC/RMHMC) to ensure efficient exploration of the manifold, while the mixture variables $\bm{\lambda}_{1:T}$ are sampled from their respective full conditional distributions. This integrated procedure is summarized in Algorithm \ref{alg_svmsmn}.
	\begin{algorithm}
		\caption{SMN-SMN Sampler}
		\begin{algorithmic}[1]
			\State Take $i=0$ and initialize $\bm{h}_{1:T}^{(0)}, \bm{\lambda_{1:T}}^{(0)}, \bm{\theta}^{(0)}$ 
			\State Simulate $\left(\mu, \phi, \sigma \right)^{(i+1)} \sim p\left(\mu, \phi, \sigma | \bm{h}_{1:T}^{(i)} \right)$ 
			\State Simulate $(\beta_0, \beta_1, \beta_2)^{(i+1)} \sim p\left(\beta_0, \beta_1, \beta_2 | \bm{h}_{1:T}^{(i)}, \bm{\lambda_{1:T}}^{(i)}, \bm{y}_{0:T} \right)$ 
			\State Simulate $\bm{h}_{1:T}^{(i+1)} \sim p\left(\bm{h}_{1:T} | (\mu, \phi, \sigma)^{(i+1)}, (\beta_0, \beta_1, \beta_2)^{(i+1)}, \bm{\lambda_{1:T}}^{(i)}, \bm{y}_{0:T} \right)$ 
			\State Simulate $\bm{\lambda}_{1:T}^{(i+1)} \sim p\left(\bm{\lambda_{1:T}} | (\beta_0, \beta_1, \beta_2)^{(i+1)}, \bm{h}_{1:T}^{(i+1)}, \nu^{(i)}, \bm{y}_{0:T} \right)$
			\State Simulate $\nu^{(i+1)} \sim p\left(\nu | \bm{\lambda_{1:T}}^{(i+1)} \right)$ 
			\State Set $i = i+1$ and return to step $2$ until convergence is achieved.
		\end{algorithmic}
		\label{alg_svmsmn}
	\end{algorithm}
	
	To complete the Bayesian specification, we define a joint prior distribution for the parameter vector $\bm{\theta}$. Assuming prior independence among the components, the joint density is factorized as
	\begin{equation*}
		p(\bm{\theta}) = p(\beta_0)p(\beta_1)p(\beta_2)p(\mu)p(\phi)p(\sigma)p(\nu).
	\end{equation*}
	
	For the mean equation parameters, we assign weakly informative Gaussian priors to the intercept and the exogenous coefficient, $\beta_0, \beta_2 \sim N(0, 10)$. For the autoregressive coefficient $\beta_1$, we employ a transformed Beta distribution, $\frac{\beta_1 + 1}{2} \sim Be(5, 1.5)$, which restricts $\beta_1$ to the interval $(-1, 1)$ to ensure the stationarity of the return process.Similarly, for the log-volatility process, we specify $\mu \sim N(0, 10)$ and use a strongly informative transformed Beta prior for the persistence parameter, $\frac{\phi + 1}{2} \sim Be(20, 1.5)$. This choice reflects the empirical expectation of high persistence ($\phi$ near 1) typically observed in financial volatility series. The scale parameter of the volatility is assigned an Inverse-Gamma prior, $\sigma^{2} \sim IG(2.5, 0.025)$. The prior for the tail-thickness parameter $\nu$ is model-specific, as it dictates the behavior of the mixture variables $\lambda_t$ in the SVM-t, SVM-S, and SVM-VG specifications. These cases are detailed in the following subsections.
	
	\subsection*{Full Conditional Distribution}
	
	\subsubsection*{\textbf{Complete Conditional Distribution of $(\beta_0, \beta_1, \beta_2)'$.}}
	Since $|\beta_{1}|<1$, we use the transformations: $\delta = \text{arctanh}(\beta_{1})$ to ensure that the parameters vary on the line. Let $\bm{\theta}_{1} = (\beta_0, \delta, \beta_2)'$, since the joint distribution $ p( \bm{y}_{0:T}| \bm{\theta}_{1} )$ belongs to the exponential family, the Fisher information matrix can be expressed as
	$$
	\mathbf{I}(\bm{\theta}_{1}) = -\mathbf{E}_{\bm{y}|\bm{\theta}_{1}} \left[\frac{\partial^2 }{\partial \bm{\theta}_{1}^2} log \left\{ p \left( \bm{y}_{0:T}| \bm{\theta}_{1} \right) \right\} \right].
	$$
	Let $\mathcal{L}(\bm{\theta}_{1}) = log \left \{ p \left( \bm{y}_{0:T}| \bm{\theta}_{1} \right)p(\bm{\theta}_{1}) \right\}$, then
	\begin{align}
		\mathcal{L}(\bm{\theta}_{1}) &= k - \frac{1}{2}\sum_{t=1}^{T}\frac{\lambda_{t}}{e^{h_{t}}}[y_{t}-\beta_{0}-\beta_{1}y_{t-1}-\beta_{2}e^{h_{t}}]^{2} - \frac{1}{2\sigma_{\beta_{0}}^{2}}(\beta_{0} - \mu_{\beta_{0}})^{2} \nonumber \\
		&+ (a_{\beta_{1}} - 1)log(1 + \beta_{1}) + (b_{\beta_{1}} - 1)log(1-\beta_{1}) - \frac{1}{2\sigma_{\beta_{2}}^{2}}(\beta_{2}-\mu_{\beta_{2}})^{2} \nonumber \\
		&+ log(1 - \beta_{1}^{2}), \text{ where } k \text{ is constant.} \nonumber 
	\end{align}
	Since $\beta_{1} = \text{tanh}(\delta)$, then $\frac{d\beta_{1}}{d\delta} = 1 - \beta_{1}^{2}$. Thus, the gradient is given by
	$$
	\bigtriangledown_{\bm{\theta}_{1}} \mathcal{L}(\bm{\theta}_{1}) = \begin{bmatrix}
		\frac{\partial }{\partial \beta_{0}} \mathcal{L}(\bm{\theta}_{1})\\
		\frac{\partial }{\partial \delta} \mathcal{L}(\bm{\theta}_{1})\\
		\frac{\partial }{\partial \beta_{2}} \mathcal{L}(\bm{\theta}_{1})
	\end{bmatrix}, 
	$$
	in which
	$$
	\frac{\partial }{\partial \beta_{0}}  \mathcal{L}(\bm{\theta}_{1}) = \sum_{t=1}^{T}\frac{\lambda_{t}}{e^{h_{t}}}[y_{t}-\beta_{0}-\beta_{1}y_{t-1}-\beta_{2}e^{h_{t}}] - \frac{1}{\sigma_{\beta_{0}}^{2}}(\beta_{0} - \mu_{\beta_{0}}),
	$$
	\begin{align}
		\frac{\partial }{\partial \delta} \mathcal{L}(\bm{\theta}_{1}) &= (1-\beta_{1}^{2})\sum_{t=1}^{T}\frac{\lambda_{t}}{e^{h_{t}}}[y_{t}-\beta_{0}-\beta_{1}y_{t-1}-\beta_{2}e^{h_{t}}]y_{t-1} \nonumber \\
		& + (a_{\beta_{1}} - 1)(1-\beta_{1}) - (b_{\beta_{1}} - 1)(1+\beta_{1}) - 2\beta_{1}\nonumber ,
	\end{align}
	$$
	\frac{\partial }{\partial \beta_{2}}  \mathcal{L}(\bm{\theta}_{1}) = \sum_{t=1}^{T}\lambda_{t}[y_{t}-\beta_{0}-\beta_{1}y_{t-1}-\beta_{2}e^{h_{t}}] - \frac{1}{\sigma_{\beta_{1}}}(\beta_{2}-\mu_{\beta_{2}})
	$$
	The metric tensor $\mathbf{M}(\bm{\theta}_{1})$ is given by:
	$$
	\mathbf{M}(\bm{\theta}_{1}) = \mathbf{I}(\bm{\theta}_{1}) - \mathbf{H}_{\bm{\theta}_{1}},
	$$
	in which
	$$
	\mathbf{I}(\bm{\theta}_{1}) = \begin{bmatrix}
		\sum_{t=1}^{T}\frac{\lambda_{t}}{e^{h_{t}}} & (1-\beta_{1}^{2})\sum_{t=1}^{T}\frac{\lambda_{t}}{e^{h_{t}}}y_{t-1} & \sum_{t=1}^{T}\lambda_{t} \\
		(1-\beta_{1}^{2})\sum_{t=1}^{T}\frac{\lambda_{t}}{e^{h_{t}}}y_{t-1} & (1-\beta_{1}^{2})^{2}\sum_{t=1}^{T}\frac{\lambda_{t}}{e^{h_{t}}}y_{t-1}^{2} & \sum_{t=1}^{T}\lambda_{t}y_{t-1}^{2} \\
		\sum_{t=1}^{T}\lambda_{t} & \sum_{t=1}^{T}\lambda_{t}y_{t-1}^{2} & \sum_{t=1}^{T}\lambda_{t}e^{h_{t}}
	\end{bmatrix},
	$$
	is the Fisher information matrix and
	\begin{equation}
		-\mathbf{H}_{\bm{\theta}_{1}} = \begin{bmatrix}
			\frac{1}{\sigma_{\beta_{0}}^{2}} & 0 & 0 \\
			0 & (a_{\beta_{1}} + b_{\beta_{1}})(1-\beta_{1}^{2}) & 0 \\
			0 & 0 & \frac{1}{\sigma_{\beta_{2}}^{2}}
		\end{bmatrix},
		\label{hessian1}
	\end{equation}
	is the Hessian of the logarithm of the priors. The partial derivatives of the metric tensor are of the form:
	$$
	\frac{\partial}{\partial \beta_{0}}\mathbf{M}(\bm{\theta}_{1}) = \frac{\partial }{\partial \beta_{2}}\mathbf{M}(\bm{\theta}_{1}) = \mathbf{0}_{3x3},
	$$
	$$
	\frac{\partial}{\partial \delta} \mathbf{M}(\bm{\theta}_{1}) = -2\beta_{1}(1-\beta_{1}^{2})\begin{bmatrix}
		0 & \sum_{t=1}^{T}\frac{\lambda_{t}}{e^{h_{t}}}y_{t-1} & 0 \\
		\sum_{t=1}^{T}\frac{\lambda_{t}}{e^{h_{t}}}y_{t-1} & 2(1-\beta_{1}^{2})\sum_{t=1}^{T}\frac{\lambda_{t}}{e^{h_{t}}}y_{t-1}^{2} + (a_{\beta_{1}}+b_{\beta_{1}})& \sum_{t=1}^{T}\lambda_{t}y_{t-1} \\
		0 & \sum_{t=1}^{T}\lambda_{t}y_{t-1} & 0
	\end{bmatrix}.
	$$
	
	\subsubsection*{\textbf{Complete Conditional Distribution of $\bm{h}_{1:T}$.}}
	
	Let $\mathcal{L}(\bm{h}_{1:T})$ be given by
	\begin{align}
		\mathcal{L}(\bm{h}_{1:T}) &= k - \frac{1}{2}\sum_{t=1}^{T}h_{t} - \frac{1}{2\sigma^{2}}\sum_{t=1}^{T-1}[h_{t+1} - \mu - \phi(h_{t} - \mu)]^{2} - \frac{1-\phi^{2}}{2\sigma^{2}}(h_{1}-\mu)^{2} \nonumber \\
		&-\frac{1}{2}\sum_{t=1}^{T}\frac{\lambda_{t}}{e^{h_{t}}}[y_{t}-\beta_0-\beta_1y_{t-1}-\beta_2e^{h_{t}}]^{2}, \text{ where } k \text{ is constant.} \nonumber 
	\end{align}
	Then the gradient is defined in the form
	$$
	\bigtriangledown_{\bm{h}_{1:T}} \mathcal{L}(\bm{h}_{1:T}) = \begin{bmatrix}
		\frac{\partial }{\partial h_{1}} \mathcal{L}(\bm{h}_{1:T})\\
		\vdots \\
		\frac{\partial }{\partial h_{t}} \mathcal{L}(\bm{h}_{1:T})\\
		\vdots \\
		\frac{\partial }{\partial h_{T}} \mathcal{L}(\bm{h}_{1:T})
	\end{bmatrix},
	$$
	in which
	\begin{align}
		\frac{\partial }{\partial h_{1}} \mathcal{L}(\bm{h}_{1:T}) &= -\frac{1}{2} + \frac{\lambda_{1}}{2e^{h_{1}}}(y_{1}-\beta_0-\beta_1y_{0}-\beta_2e^{h_{1}})^{2} + \beta_2\lambda_{1}(y_{1}-\beta_0-\beta_1y_{0}-\beta_2e^{h_{1}}) \nonumber \\
		&+ \frac{\phi}{\sigma^{2}}(h_{2}-\mu-\phi(h_{1}-\mu)) - \frac{1-\phi^{2}}{\sigma^{2}}(h_{1}-\mu) \nonumber,
	\end{align}
	
	\begin{align}
		\frac{\partial }{\partial h_{t}} \mathcal{L}(\bm{h}_{1:T}) &= -\frac{1}{2} + \frac{\lambda_{t}}{2e^{h_{t}}}(y_{t}-\beta_0-\beta_1y_{t-1}-\beta_2e^{h_{t}})^{2} + \beta_2\lambda_{t}(y_{t}-\beta_0-\beta_1y_{t-1}-\beta_2e^{h_{t}}) \nonumber \\
		&+ \frac{\phi}{\sigma^{2}}(h_{t+1}-\mu-\phi(h_{t}-\mu)) - \frac{1}{\sigma^{2}}(h_{t}-\mu-\phi(h_{t-1}-\mu)) \nonumber, 
	\end{align}
	
	\noindent for $1<t<T$,
	\begin{align}
		\frac{\partial }{\partial h_{T}} \mathcal{L}(\bm{h}_{1:T}) &= -\frac{1}{2} + \frac{\lambda_{T}}{2e^{h_{T}}}(y_{T}-\beta_0-\beta_1y_{T-1}-\beta_2e^{h_{T}})^{2} + \beta_2\lambda_{T}(y_{T}-\beta_0-\beta_1y_{T-1}-\beta_2e^{h_{T}}) \nonumber \\
		&- \frac{1}{\sigma^{2}}(h_{T}-\mu-\phi(h_{T-1}-\mu)) \nonumber.
	\end{align}
	
	When applying the RMHMC method to sample the log-volatility vector $\bm{h}_{1:T}$, we must derive the metric matrix $\mathbf{M}(\bm{h}_{1:T})$ and its partial derivatives. Based on the Expected Fisher Information for the SVM-SMN model, the metric matrix is formulated as
	\begin{equation*}
		\mathbf{M}(\bm{h}_{1:T}) = \frac{1}{2} \mathbf{I} + \beta_2^{2} \mathbf{G} + \mathbf{C},
	\end{equation*}
	where $\mathbf{I}_T$ is the $T \times T$ identity matrix, $\mathbf{G} = \text{diag}(\lambda_{1} e^{h_{1}}, \dots, \lambda_{T} e^{h_{T}})$ accounts for the local curvature contributed by the observation equation, and $\mathbf{C}$ is a tridiagonal matrix arising from the Gaussian state equation. Specifically, $\mathbf{C}$ has diagonal elements $(1+\phi^{2})/\sigma^{2}$ (with the first and last elements being $1/\sigma^{2}$) and off-diagonal elements equal to $-\phi/\sigma^{2}$. The explicit dependence of $\mathbf{M}(\bm{h}_{1:T})$ on the position vector $\bm{h}_{1:T}$ necessitates the use of the generalized leapfrog integrator. This requirement significantly amplifies the computational burden, as it involves fixed-point iterations and the inversion of a $T \times T$ matrix at each internal step of the trajectory. As the sample size $T$ grows, the $\mathcal{O}(T^3)$ cost of matrix inversion becomes a bottleneck for the algorithm's scalability. To mitigate this computational overhead, we adopt a simplified approach by setting $\mathbf{M}(\bm{h}_{1:T}) = \mathbf{I}_T$, effectively reverting to the standard HMC method with a pre-conditioned identity metric. This choice allows for the use of the explicit leapfrog integrator, which is significantly faster.

	\subsubsection*{\textbf{Complete Conditional Distribution of $\bm{\lambda}_{1:T}$.}}
	
	\begin{itemize}
		\item SVM-t: Since $\lambda_{t}|\nu \sim G \left( \frac{\nu}{2}, \frac{\nu}{2} \right)$, the complete conditional distribution of $\lambda_{t}$ is given by \\
		\begin{equation*}
			\lambda_{t} | {h}_{t}, \bm{\theta}, y_{t},y_{t-1} \sim G \left(\frac{\nu+1}{2}, \frac{\{y_{t}-\beta_0-\beta_1y_{t-1}-\beta_2e^{h_{t}}\}^{2}e^{-h_{t}} + \nu }{2}\right),
		\end{equation*}
		for $t=1,...,T$.
		
		\item SVM-S: Where $\lambda_{t}|\nu \sim Be(\nu, 1)$, the complete conditional distribution of $\lambda_{t}$ is given by
		\begin{equation*}
			\lambda_{t} | {h}_{t}, \bm{\theta}, y_{t},y_{t-1} \sim G_{(\lambda_{t}<1)}\left(\nu+\frac{1}{2}, \frac{1}{2} \{y_{t}-\beta_0-\beta_1y_{t-1}-\beta_2e^{h_{t}}\}^{2} e^{-h_{t}} \right),
		\end{equation*}
		for $t=1,...,T$, where $G_{(\lambda_{t}<1)}$ is the right-truncated gamma distribution.
		
		To ensure numerical stability during the sampling process, we implemented a custom simulation method for truncated Gamma random variables rather than relying on the standard \texttt{truncdist} package \citep{nadarajah2006r}. Our approach leverages the probability integral transform (inverse transform sampling). Specifically, since the quantile function \texttt{qgamma} in \texttt{R} is the inverse of the cumulative distribution function (CDF), realizations from a truncated distribution can be obtained by mapping uniform draws into the restricted range of the CDF. However, a significant numerical challenge arises when sampling from $G(a,b)$ truncated to the interval $(0,1)$ when the rate parameter $b$ is near zero. In such cases, the normalizing constant—the denominator required to scale the CDF to the truncated support—may become infinitesimally small, leading to division-by-zero errors or numerical underflow. To mitigate this issue, we perform these calculations on a logarithmic scale. By evaluating the log-CDF and utilizing the \texttt{log.p = TRUE} argument in \texttt{R} functions, we maintain precision in the extreme tails of the distribution, ensuring the robustness of the MCMC scheme even under diffuse prior specifications. 
		
		\item SVM-VG: Given $\lambda_{t}|\nu \sim IG\left( \frac{\nu}{2}, \frac{\nu}{2} \right)$, the complete conditional distribution of $\lambda_{t}$ is given by
		\begin{equation*}
			\lambda_{t} | {h}_{t}, \bm{\theta}, y_{t},y_{t-1} \sim GIG \left( \frac{1-\nu}{2}, \{y_{t}-\beta_0-\beta_1y_{t-1}-\beta_2e^{h_{t}}\}^{2}e^{-h_{t}}, \nu \right)
		\end{equation*}    
		for $t=1,...,T$, where $GIG$ represents the Generalized Inverse Normal distribution, and its simulation is performed using the \texttt{GIGrvg} package, developed by \citet{leydold2017package}.
		
	\end{itemize}
	
	\subsubsection*{\textbf{Complete Conditional Distribution of $\nu$.}}

	As mentioned previously, the specification of the prior for the parameter $\nu$ depends on the membership of the SVM-SMN class. For the SVM-S model, we adopt $\nu \sim G(0.08, 0.04)$, as realized by \citet{abanto2019threshold}. This choice results in a complete posterior with known form, given by
	\begin{equation}
		\nu|\bm{\lambda}_{1:T} \sim G_{\nu>1} \left( T + a_{\nu}, b_{\nu} - \sum_{t=1}^{T} log\lambda_{t} \right), \nonumber
	\end{equation}
	which is a left-truncated gamma distribution.
	
	Considering now the SVM-t and SVM-VG models, we do not find a suitable choice for the prior of $\nu$ such that it has a complete conditional with a known form, like the SVM-S model. Therefore, we chose to employ the RMHMC method, as described in section \ref{HMC}, to perform simulations for these models. Furthermore, we specify $\nu$ varying in bounded intervals. For SVM-t, we consider $\nu \in (2,40)$, while for the SVM-VG model, $\nu$ varies in the interval $(0,40)$.
	
	For efficiency reasons, we consider the transformation
	\begin{equation}
		\xi(\nu) = \frac{1}{b}\text{tanh}^{-1}\left( \frac{\nu - c}{a} \right),
		\label{trans_nu}
	\end{equation}
	where $\nu \in (i, s)$ varies in a generic bounded range, $2a = s - i$, $2b = \alpha$, and $2c = s + i$. Note that the transformation $\xi(\nu)$ takes the bounded interval $(i, s)$, on the real line $\mathbb{R}$, since
	$$
	i < \nu < s \Rightarrow -1 < \left( \frac{\nu - c}{a} \right) < 1,
	$$
	and the image of the function $\text{tanh}^{-1}(x)$, defined on $(-1,1)$, is the line $\mathbb{R}$. It is worth noting that the transformation (\ref{trans_nu}) depends on the lower bound $i$ and upper bound $s$ of the defined intervals. Thus, $i=2$ was taken for SVM-t, and $i=0$ for the SVM-VG model. The upper bound $s=40$ was defined for both models.
	
	For the (\ref{hessian_diagonal}) condition to be met in the univariate case, it is sufficient that the prior of the transformed parameter, $\xi$, be log-concave. Thus, we take
	$$
	\xi \sim \mathcal{N} \left( \mu_{\xi}, \sigma_{\xi}^{2} \right ),
	$$
	which is a log-concave distribution. The hyperparameters $\mu_{\xi}$, $\sigma_{\xi}$, and $\alpha$ were chosen to ensure that the prior distribution $p(\nu)$ has a mean of approximately $\mathbf{E}(\nu) \approx 12$ and a variance of $Var(\nu) \approx 70$ in the SVM-t case. In the case of the SVM-VG model, the values are adjusted to achieve a mean of $\mathbf{E}(\nu) \approx 4$ and a variance of $Var(\nu) \approx 25$. 
	Similar values are found in the priors selected by \citet{abanto2019threshold} for the SVM-t and SVM-VG models.  
	
	Obtaining the moments of $\nu$ is a challenging task due to the difficulty of finding an analytical solution. To see this, note that the inverse of the (\ref{trans_nu}) transformation is of the form
	\begin{equation}
		\nu(\xi) = a \cdot \text{tanh}(b\xi) + c,
		\label{inv_trans_nu}
	\end{equation}
	and the calculation of $E(\nu)$ and $Var(\nu)$ involves solving the integral
	\begin{equation}
		\frac{1}{\sqrt{2\pi} \sigma_{\xi}} \int_{-\infty}^{\infty} \text{tanh}(b\xi) \text{exp} \left( -\frac{1}{2 \sigma_{\xi}^{2}} (\xi - \mu_{\xi})^{2} \right) d\xi, \nonumber
	\end{equation}
	which is difficult to manipulate analytically. Thus, after performing $S = 10000$ simulations of $\xi$ from the distribution $N(\mu_{\xi}, \sigma_{\xi}^{2})$, we estimate $E(\nu)$ and $Var(\nu)$ using the sample mean and variance, respectively, of the sample $\{ \nu_{1}, ..., \nu_{S} \}$, where $\nu_{i} = \nu(\xi_{i})$ and $i=1, ..., S$.
	
	Finally, the log-posteriori $\mathcal{L}(\xi) = \sum_{t=1}^{T} log p\left( \lambda_{t}|\nu(\xi) \right)$ and their derivatives $\nabla_{\xi}\mathcal{L}(\xi)$, are given, respectively, by
	
	%SVM-t
	\begin{equation}
		\mathcal{L}(\xi) = k + \frac{T\nu(\xi)}{2}log \left( \frac{\nu(\xi)}{2} \right) - Tlog\Gamma\left( \frac{\nu(\xi)}{2} \right) + \frac{\nu(\xi)}{2} \sum_{t=1}^{T}( log \lambda_{t} - \lambda_{t}) - \frac{1}{2\sigma_{\xi}^{2}}(\xi - \mu_{\xi})^{2} \nonumber
	\end{equation}
	\noindent and
	\begin{equation}
		\nabla_{\xi}\mathcal{L}(\xi) = \frac{d\nu}{d\xi} \left( \frac{T}{2}log\left(\frac{\nu}{2}\right) - \frac{T}{2}\psi^{(0)}\left(\frac{\nu}{2}\right) + \frac{T}{2} + \frac{1}{2} \sum_{t=1}^{T}( log \lambda_{t} - \lambda_{t})\right) - \frac{1}{\sigma_{\xi}^{2}}(\xi - \mu_{\xi}), \nonumber
	\end{equation}
	for the SVM-t model, while for the SVM-VG model, we have
	% SVM-VG
	\begin{equation}
		\mathcal{L}(\xi) = k + \frac{T\nu(\xi)}{2}log \left( \frac{\nu(\xi)}{2} \right) - Tlog\Gamma\left( \frac{\nu(\xi)}{2} \right) - \frac{\nu(\xi)}{2} \sum_{t=1}^{T}( log \lambda_{t} - \lambda_{t}^{-1}) - \frac{1}{2\sigma_{\xi}^{2}}(\xi - \mu_{\xi})^{2} \nonumber
	\end{equation}
	and
	\begin{equation}
		\nabla_{\xi}\mathcal{L}(\xi) = \frac{d\nu}{d\xi} \left( \frac{T}{2}log\left(\frac{\nu}{2}\right) - \frac{T}{2}\psi^{(0)}\left(\frac{\nu}{2}\right) + \frac{T}{2} - \frac{1}{2} \sum_{t=1}^{T}( log \lambda_{t} - \lambda_{t}^{-1} )\right) - \frac{1}{\sigma_{\xi}^{2}}(\xi - \mu_{\xi}). \nonumber
	\end{equation}
	
	\noindent In these expressions, $\Gamma(\cdot)$ represents the gamma function, $\psi^{(n)}(\cdot) = \frac{d^{n+1}}{dx^{n+1}}log\Gamma(\cdot)$ represents the nth-order polygamma function, and $\frac{d\nu}{d\xi}$ represents the derivative of the inverse transformation (\ref{inv_trans_nu}). The metric tensor $M(\xi)$, for both the SVM-t and SVM-VG models, is defined by
	\begin{equation}
		M(\xi) = - \left( \frac{d\nu}{d\xi} \right)^{2} T \left( \frac{1}{2\nu(\xi)} - \frac{1}{4}\psi^{(1)}\left( \frac{\nu(\xi)}{2}\right) \right), \nonumber
	\end{equation}
	so that its derivative is of the form
	\begin{align}
		\frac{\partial}{\partial \xi} M(\xi) = &-2 \left( \frac{d\nu}{d\xi} \right) \left( \frac{d^{2} \nu}{d \xi^{2}} \right) T \left( \frac{1}{2\nu(\xi)} - \frac{1}{4}\psi^{(1)}\left( \frac{\nu(\xi)}{2}\right) \right) + \left( \frac{d\nu}{d\xi}\right)^{3} T \left( \frac{1}{2\nu(\xi)} + \frac{1}{8} \psi^{(2)} \left( \frac{\nu(\xi)}{2} \right) \right). \nonumber
	\end{align}
		
	\subsection*{Detailed Results for Empirical Applications}
	
	In this section, we provide the full posterior summaries for the models fitted to the four stock market indices: S\&P 500, NIKKEI 225, DAX 30, and IBOVESPA.
	
	The latent volatilities $h_t$ are sampled in a single block using the HMC method, configured with 50 leapfrog steps and a step size of 0.015. The parameter sets $(\mu, \phi, \sigma)$ and $(b_0, b_1, b_2)$ are updated in blocks via the RMHMC algorithm. Specifically, for $(\mu, \phi, \sigma)$, we employ a step size of 0.5 with 20 leapfrog steps and 5 fixed-point iterations. Similarly, the coefficients $(b_0, b_1, b_2)$ are sampled using a step size of 0.1, 20 leapfrog steps, and 5 fixed-point iterations. Finally, the degrees of freedom $\nu$ are also updated using the RMHMC method, adopting a step size of 0.5, 20 leapfrog steps, and 5 fixed-point iterations.
	
	For all models, 100000 MCMC iterations were performed, discarding the first 50,000 as burn-in. To reduce autocorrelation between consecutive iterations, a thinning factor of 25 was applied, resulting in a final sample size of 2000 for each model. Based on these samples, we calculated the posterior mean, 95\% credibility intervals, Geweke’s convergence diagnostic (CD) statistics, and the inefficiency factor (IF).
	
	%sp500 and nikkei
	\begin{table}[h!]
		\centering
		\caption{Posterior summaries for S\&P 500 and Nikkei 225 indices.}
		\label{tab:combined_sp_nikkei}
		\scalebox{0.75}{
			\begin{tabular}{l ccccccc ccccccc}
				\toprule
				& \multicolumn{7}{c}{\textbf{S\&P 500}} & \multicolumn{7}{c}{\textbf{Nikkei 225}} \\
				\cmidrule(lr){2-8} \cmidrule(lr){9-15}
				Param. & Mean & SD & 2.5\% & 50\% & 97.5\% & CD & IF & Mean & SD & 2.5\% & 50\% & 97.5\% & CD & IF \\ 
				\midrule
				\multicolumn{15}{l}{\textit{SVM-N}} \\
				$b_0$    & 0.1155 & 0.0125 & 0.0914 & 0.1156 & 0.1400 & -0.65 & 1.00 & 0.1551 & 0.0239 & 0.1084 & 0.1549 & 0.2023 & 0.50 & 0.97 \\
				$b_1$    & -0.0722 & 0.0144 & -0.1009 & -0.0720 & -0.0447 & -0.12 & 1.00 & -0.0264 & 0.0148 & -0.0565 & -0.0262 & 0.0018 & 0.30 & 1.00 \\
				$b_2$    & -0.0598 & 0.0143 & -0.0871 & -0.0598 & -0.0336 & 1.10 & 0.89 & -0.0667 & 0.0161 & -0.0970 & -0.0665 & -0.0355 & 1.06 & 1.03 \\
				$\mu$    & -0.3325 & 0.1490 & -0.6228 & -0.3348 & -0.0400 & -0.17 & 0.99 & 0.3549 & 0.1021 & 0.1526 & 0.3551 & 0.5490 & -0.27 & 1.00 \\
				$\phi$   & 0.9797 & 0.0037 & 0.9724 & 0.9798 & 0.9866 & 0.54 & 12.86 & 0.9745 & 0.0044 & 0.9655 & 0.9746 & 0.9825 & 0.99 & 9.77 \\
				$\sigma$ & 0.2142 & 0.0146 & 0.1855 & 0.2142 & 0.2435 & -0.16 & 38.04 & 0.1792 & 0.0126 & 0.1582 & 0.1784 & 0.2057 & -0.84 & 27.17 \\ 
				\midrule
				\multicolumn{15}{l}{\textit{SVM-t}} \\
				$b_0$    & 0.1152 & 0.0124 & 0.0906 & 0.1150 & 0.1396 & -1.25 & 1.00 & 0.1517 & 0.0249 & 0.1018 & 0.1516 & 0.1991 & 0.75 & 1.32 \\
				$b_1$    & -0.0716 & 0.0140 & -0.0991 & -0.0715 & -0.0450 & 1.53 & 1.08 & -0.0291 & 0.0145 & -0.0580 & -0.0291 & -0.0011 & 1.34 & 1.00 \\
				$b_2$    & -0.0647 & 0.0172 & -0.0997 & -0.0643 & -0.0324 & 1.87 & 1.11 & -0.0707 & 0.0180 & -0.1059 & -0.0703 & -0.0358 & 0.05 & 1.07 \\
				$\mu$    & -0.4481 & 0.1634 & -0.7767 & -0.4489 & -0.1301 & 0.52 & 1.28 & 0.2432 & 0.1163 & 0.0114 & 0.2479 & 0.4584 & 0.51 & 1.24 \\
				$\phi$   & 0.9814 & 0.0036 & 0.9735 & 0.9816 & 0.9877 & -0.06 & 10.91 & 0.9778 & 0.0046 & 0.9680 & 0.9779 & 0.9863 & 1.24 & 21.44 \\
				$\sigma$ & 0.2042 & 0.0145 & 0.1780 & 0.2038 & 0.2371 & 0.44 & 38.17 & 0.1661 & 0.0156 & 0.1387 & 0.1652 & 0.2001 & -0.85 & 39.27 \\
				$\nu$    & 15.5855 & 4.1875 & 9.1236 & 15.0848 & 25.3906 & -0.85 & 7.72 & 17.6655 & 5.2963 & 10.0706 & 16.6991 & 30.4644 & -0.71 & 9.06 \\ 
				\midrule
				\multicolumn{15}{l}{\textit{SVM-S}} \\
				$b_0$    & 0.1158 & 0.0126 & 0.0917 & 0.1156 & 0.1411 & -0.02 & 1.00 & 0.1521 & 0.0244 & 0.1053 & 0.1521 & 0.2012 & -1.01 & 1.00 \\
				$b_1$    & -0.0714 & 0.0146 & -0.0984 & -0.0712 & -0.0428 & 0.40 & 1.03 & -0.0284 & 0.0146 & -0.0568 & -0.0285 & -0.0001 & -0.57 & 1.08 \\
				$b_2$    & -0.0789 & 0.0207 & -0.1196 & -0.0793 & -0.0394 & 0.16 & 1.07 & -0.0889 & 0.0230 & -0.1347 & -0.0885 & -0.0457 & 0.75 & 1.00 \\
				$\mu$    & -0.6343 & 0.1675 & -0.9818 & -0.6367 & -0.3163 & -0.04 & 3.12 & 0.0231 & 0.1238 & -0.2279 & 0.0262 & 0.2653 & 0.62 & 2.05 \\
				$\phi$   & 0.9801 & 0.0036 & 0.9725 & 0.9802 & 0.9867 & -0.12 & 12.36 & 0.9791 & 0.0046 & 0.9694 & 0.9794 & 0.9873 & 1.46 & 21.68 \\
				$\sigma$ & 0.2127 & 0.0148 & 0.1857 & 0.2120 & 0.2416 & 0.12 & 35.36 & 0.1593 & 0.0159 & 0.1314 & 0.1586 & 0.1924 & -1.50 & 49.38 \\
				$\nu$    & 3.7982 & 0.8904 & 2.7212 & 3.5940 & 6.4310 & -0.41 & 31.78 & 3.4691 & 0.5990 & 2.5972 & 3.3805 & 4.9541 & -0.72 & 14.85 \\ 
				\midrule
				\multicolumn{15}{l}{\textit{SVM-VG}} \\
				$b_0$    & 0.1140 & 0.0122 & 0.0907 & 0.1138 & 0.1380 & 0.42 & 1.00 & 0.1505 & 0.0247 & 0.1036 & 0.1499 & 0.1984 & 1.25 & 0.98 \\
				$b_1$    & -0.0711 & 0.0141 & -0.0997 & -0.0709 & -0.0450 & -1.12 & 1.00 & -0.0302 & 0.0144 & -0.0581 & -0.0304 & -0.0019 & -0.80 & 1.00 \\
				$b_2$    & -0.0553 & 0.0150 & -0.0862 & -0.0555 & -0.0265 & 0.63 & 1.08 & -0.0619 & 0.0163 & -0.0948 & -0.0617 & -0.0304 & -1.49 & 1.00 \\
				$\mu$    & -0.3025 & 0.1645 & -0.6096 & -0.3064 & 0.0362 & 0.02 & 1.00 & 0.3714 & 0.1111 & 0.1616 & 0.3739 & 0.5885 & -0.69 & 1.13 \\
				$\phi$   & 0.9821 & 0.0034 & 0.9753 & 0.9822 & 0.9885 & -0.98 & 8.46 & 0.9773 & 0.0048 & 0.9676 & 0.9776 & 0.9866 & -1.00 & 18.02 \\
				$\sigma$ & 0.1994 & 0.0135 & 0.1720 & 0.1999 & 0.2245 & 1.19 & 24.75 & 0.1672 & 0.0159 & 0.1367 & 0.1674 & 0.1981 & 0.87 & 42.72 \\
				$\nu$    & 12.1705 & 4.1153 & 6.6778 & 11.4197 & 22.7024 & 0.28 & 9.83 & 14.8197 & 4.5777 & 8.1442 & 13.9470 & 25.7477 & 0.82 & 9.96 \\ 
				\bottomrule
			\end{tabular}
		}
		\label{tab5}
	\end{table}
	%dax and ibovespa
	\begin{table}[h!]
		\centering
		\caption{Posterior summaries for DAX 30 and IBOVESPA indices.}
		\label{tab:combined_dax_ibov}
		\scalebox{0.75}{
			\begin{tabular}{l ccccccc ccccccc}
				\toprule
				& \multicolumn{7}{c}{\textbf{DAX 30}} & \multicolumn{7}{c}{\textbf{IBOVESPA}} \\
				\cmidrule(lr){2-8} \cmidrule(lr){9-15}
				Param. & Mean & SD & 2.5\% & 50\% & 97.5\% & CD & IF & Mean & SD & 2.5\% & 50\% & 97.5\% & CD & IF \\ 
				\midrule
				\multicolumn{15}{l}{\textit{SVM-N}} \\
				$b_0$    & 0.1437 & 0.0176 & 0.1081 & 0.1440 & 0.1790 & 0.42 & 1.00 & 0.1820 & 0.0357 & 0.1131 & 0.1819 & 0.2546 & -0.57 & 1.08 \\
				$b_1$    & -0.0293 & 0.0145 & -0.0569 & -0.0294 & -0.0008 & -1.05 & 1.07 & -0.0228 & 0.0150 & -0.0529 & -0.0230 & 0.0056 & 1.92 & 1.00 \\
				$b_2$    & -0.0593 & 0.0128 & -0.0850 & -0.0595 & -0.0346 & 0.22 & 0.81 & -0.0474 & 0.0151 & -0.0773 & -0.0473 & -0.0194 & 0.97 & 1.00 \\
				$\mu$    & 0.2178 & 0.1468 & -0.0780 & 0.2210 & 0.5019 & 0.28 & 1.00 & 0.7809 & 0.1016 & 0.5801 & 0.7822 & 0.9760 & 1.17 & 1.22 \\
				$\phi$   & 0.9813 & 0.0037 & 0.9737 & 0.9814 & 0.9881 & -1.47 & 15.29 & 0.9787 & 0.0043 & 0.9696 & 0.9791 & 0.9862 & 1.00 & 25.17 \\
				$\sigma$ & 0.1847 & 0.0140 & 0.1585 & 0.1847 & 0.2120 & 1.22 & 29.29 & 0.1396 & 0.0115 & 0.1174 & 0.1388 & 0.1640 & -0.55 & 55.32 \\ 
				\midrule
				\multicolumn{15}{l}{\textit{SVM-t}} \\
				$b_0$    & 0.1428 & 0.0176 & 0.1092 & 0.1426 & 0.1775 & -1.41 & 1.00 & 0.1815 & 0.0358 & 0.1131 & 0.1824 & 0.2501 & 1.15 & 1.00 \\
				$b_1$    & -0.0284 & 0.0140 & -0.0561 & -0.0285 & -0.0013 & 0.12 & 1.00 & -0.0231 & 0.0142 & -0.0500 & -0.0232 & 0.0057 & -0.80 & 1.00 \\
				$b_2$    & -0.0687 & 0.0160 & -0.1002 & -0.0688 & -0.0388 & 1.44 & 1.00 & -0.0503 & 0.0166 & -0.0831 & -0.0497 & -0.0186 & -0.78 & 0.77 \\
				$\mu$    & 0.0576 & 0.1658 & -0.2719 & 0.0552 & 0.3821 & 0.51 & 1.58 & 0.7019 & 0.1044 & 0.4921 & 0.7032 & 0.9106 & -0.28 & 1.00 \\
				$\phi$   & 0.9847 & 0.0036 & 0.9773 & 0.9848 & 0.9914 & 1.19 & 20.00 & 0.9802 & 0.0041 & 0.9718 & 0.9803 & 0.9875 & 0.50 & 14.73 \\
				$\sigma$ & 0.1653 & 0.0165 & 0.1336 & 0.1651 & 0.1982 & -1.56 & 48.73 & 0.1327 & 0.0111 & 0.1100 & 0.1327 & 0.1531 & -0.33 & 36.85 \\
				$\nu$    & 11.7709 & 2.8131 & 7.7560 & 11.3201 & 18.4397 & -0.07 & 6.30 & 23.6160 & 5.2457 & 14.5021 & 23.3614 & 33.8353 & 0.04 & 5.88 \\ 
				\midrule
				\multicolumn{15}{l}{\textit{SVM-S}} \\
				$b_0$    & 0.1419 & 0.0182 & 0.1063 & 0.1423 & 0.1779 & 0.30 & 1.00 & 0.1822 & 0.0356 & 0.1096 & 0.1823 & 0.2539 & -0.23 & 0.92 \\
				$b_1$    & -0.0268 & 0.0139 & -0.0552 & -0.0266 & 0.0001 & 1.38 & 1.09 & -0.0236 & 0.0142 & -0.0507 & -0.0233 & 0.0040 & -1.05 & 0.93 \\
				$b_2$    & -0.0871 & 0.0212 & -0.1297 & -0.0865 & -0.0462 & 0.21 & 1.54 & -0.0587 & 0.0195 & -0.0970 & -0.0581 & -0.0214 & -1.07 & 1.67 \\
				$\mu$    & -0.1783 & 0.1664 & -0.5014 & -0.1813 & 0.1445 & -1.31 & 2.34 & 0.5546 & 0.1192 & 0.3113 & 0.5561 & 0.7869 & -1.64 & 12.24 \\
				$\phi$   & 0.9843 & 0.0035 & 0.9772 & 0.9845 & 0.9908 & 0.78 & 12.35 & 0.9804 & 0.0044 & 0.9714 & 0.9807 & 0.9880 & 0.93 & 22.88 \\
				$\sigma$ & 0.1690 & 0.0150 & 0.1410 & 0.1684 & 0.1985 & -0.24 & 38.19 & 0.1327 & 0.0129 & 0.1071 & 0.1321 & 0.1607 & -0.62 & 47.81 \\
				$\nu$    & 2.8913 & 0.3754 & 2.3029 & 2.8260 & 3.7691 & -0.28 & 11.39 & 5.3848 & 2.0287 & 3.2888 & 4.7658 & 11.4683 & -1.55 & 69.06 \\ 
				\midrule
				\multicolumn{15}{l}{\textit{SVM-VG}} \\
				$b_0$    & 0.1430 & 0.0182 & 0.1083 & 0.1424 & 0.1794 & 1.53 & 1.00 & 0.1846 & 0.0362 & 0.1161 & 0.1846 & 0.2589 & -0.42 & 1.00 \\
				$b_1$    & -0.0298 & 0.0139 & -0.0562 & -0.0300 & -0.0031 & -0.38 & 1.00 & -0.0233 & 0.0143 & -0.0510 & -0.0232 & 0.0036 & -0.62 & 1.00 \\
				$b_2$    & -0.0554 & 0.0131 & -0.0807 & -0.0556 & -0.0294 & -1.45 & 1.07 & -0.0471 & 0.0153 & -0.0780 & -0.0467 & -0.0174 & 0.28 & 0.89 \\
				$\mu$    & 0.2562 & 0.1649 & -0.0730 & 0.2538 & 0.5935 & -0.03 & 1.00 & 0.7957 & 0.0995 & 0.6104 & 0.7967 & 0.9890 & -0.11 & 1.00 \\
				$\phi$   & 0.9856 & 0.0032 & 0.9791 & 0.9857 & 0.9915 & -1.49 & 14.85 & 0.9800 & 0.0042 & 0.9713 & 0.9802 & 0.9875 & -0.55 & 19.15 \\
				$\sigma$ & 0.1598 & 0.0131 & 0.1351 & 0.1596 & 0.1870 & 1.29 & 35.94 & 0.1338 & 0.0118 & 0.1117 & 0.1330 & 0.1574 & 0.80 & 41.94 \\
				$\nu$    & 8.0856 & 1.8089 & 5.5496 & 7.7324 & 12.6144 & 0.80 & 7.91 & 21.3955 & 5.7474 & 11.8300 & 20.9457 & 33.1472 & 0.09 & 6.56 \\ 
				\bottomrule
			\end{tabular}
			\label{tab6}
		}
	\end{table}
	
	Tables \ref{tab5} and \ref{tab6} summarize the results. According to these summaries, the null hypothesis of stationarity is accepted at the 5\% significance level for all parameters and models, as the CD statistics remain within the $(-1.96, 1.96)$ interval.

\clearpage
\bibliographystyle{unsrtnat}
\bibliography{references}  %%% Uncomment this line and comment out the ``thebibliography'' section below to use the external .bib file (using bibtex) .





% \end{thebibliography}